\documentclass[preprint,12pt]{elsarticle}

\usepackage{booktabs}
\usepackage{tikz}
\usetikzlibrary{positioning}
\usetikzlibrary{calc}
\usetikzlibrary{shadows.blur}

\usepackage{amssymb}
\usepackage{amsmath}

\newcommand{\Psineut}[1]{\Psi_{#1}^{N+1}}
\newcommand{\Psiscat}[1]{\Psi_{#1}^{(-)}}
\newcommand{\psik}[2]{\psi_{k#1}^{ N+1\mathrm{#2}}}

\newcommand{\kfinalMOL}{\boldsymbol{k}_{f}}

\newcommand{\kfinalLAB}{\boldsymbol{k}_{f}}
\newcommand{\kdirLAB}{\boldsymbol{\hat k}_{f}}

\newcommand{\braopket}[3]{\langle {#1}|{#2}|{#3}\rangle}
\newcommand{\opdipole}{\boldsymbol{d}}

\newcommand{\MTXrotmat}[1]{\boldsymbol{D}^{#1}}
\newcommand{\spharm}[2]{Y_{#1,#2}}
\newcommand{\MTXspharm}[1]{\boldsymbol{Y}_{#1}}

\newcommand{\euler}{\alpha,\beta,\gamma}
\newcommand{\coulombphase}[1]{\sigma_{#1}}

\newcommand{\chnl}[2]{\overline{\Phi}_{#1}^{\Gamma_{#2}}({\mathbf{X}}_{N};{\hat{\mathbf{r}}}_{N+1}\sigma_{N+1})}
\newcommand{\chnlshort}[2]{\overline{\Phi}_{#1}^{\Gamma_{#2}}}

\newcommand{\textprog}[1]{\uppercase{#1}}  
\newcommand{\textcode}[1]{\texttt{#1}}  
\newcommand{\textfile}[1]{\textit{#1}}  

\newcommand{\UKRmol}[0]{UKRmol}    
\newcommand{\Bohr}[0]{\mbox{a$_0$}}

\newcounter{bla}

\journal{Computer Physics Communications}

\begin{document}

\begin{frontmatter}



\title{\UKRmol+: a suite for modelling of electronic processes in molecules interacting with electrons, positrons and photons using the R-matrix method}


\author[a]{Z. Ma\v{s}\'{i}n\corref{author}}
\author[b]{J. Benda}
\author[b]{J. D. Gorfinkiel}
\author[c]{A. G.  Harvey}
\author[d]{Jonathan Tennyson}

\cortext[author] {Corresponding author.\\\textit{E-mail address:} zdenek.masin@utf.mff.cuni.cz}
\address[a]{Institute of Theoretical Physics, Faculty of Mathematics and Physics, Charles University, V Hole\v{s}ovi\v{c}k\'{a}ch 2, 180 00 Prague 8, Czech Republic}
\address[b]{School of Physical Sciences, The Open University, Walton Hall, Milton Keynes, MK7 6AA, United Kingdom}
\address[c]{Max-Born-Institut, Max-Born-Str. 2A, 12489 Berlin, Germany}
\address[d]{Department of Physics and Astronomy, University College London, London WC1E 6BT, UK}
\begin{abstract}

UKRmol+ is a new implementation of the UK R-matrix electron-molecule scattering code. Key features of the implementation are the use of quantum chemistry codes such as Molpro to provide target molecular orbitals; the optional use of mixed Gaussian -- B-spline basis functions to represent the continuum and improved configuration and Hamiltonian generation. The code is described, and examples covering electron collisions from a range of targets, positron collisions and photionisation are presented. The codes are freely available as a tarball from Zenodo.

\end{abstract}


\end{frontmatter}



{\bf PROGRAM SUMMARY}

\begin{small}
\noindent
{\em Program Title:} \UKRmol+\\
{\em Licensing provisions:} GNU GPLv3 \\
{\em Programming language:}  Fortran 95 with use of some Fortran 2003 features  \\             
{\em Program repository available at: https://gitlab.com/UK-AMOR/UKRmol}\\ 
{\em Computers on which the program has been tested:}  Cray XC30 ARCHER,  Lenovo SD530 node (UCL's Myriad), TACC Stampede2, Intel pcs.\\
{\em Number of processors used:} Min: 1, Max tested: program dependent, up to 100 cores for the parallel ones \\
{\em Number of lines in program:} 158178 in UKRmol-in (including GBTOlib) and  79760 in UKRmol-out\\
{\em Distribution format:} Tarball available from Zenodo (https://zenodo.org/)\\
{\em External routines/libraries:}  LAPACK, BLAS; optionally MPI, ScaLAPACK, Arpack, SLEPc \\

{\em Nature of problem:} The computational study of electron and positron scattering from a molecule requires the determination of multicentric time-independent wavefunctions describing the target+projectile system. These wavefunctions can also be used to calculate photoionization cross sections (in this case the free particle is the ionized electron) or provide input for time-dependent calculations of laser-induced ultrafast processes. \\
 
{\em Solution method:} We use the R-matrix method \cite{Burkebook}, that partitions space into an `inner' and an `outer' region. In the inner region (within a few tens of a$_0$ of the nuclei at most) exchange and correlation are taken into account.  In the outer region, where the free particle is distinguishable from the target electrons, a single-centre multipole potential describes its interaction with the molecule. The key computational step is the building and diagonalization of the target + free particle Hamiltonian in the inner region, making use of integrals generated using the GBTOlib library. The eigenpairs obtained are then  used as input to outer region suite programs to determine scattering quantities (K-matrices, etc.) or transition dipole moments and, from them, photoionization cross sections. The suite also generates input data for the R-matrix with time (RMT) suite \cite{RMT-CPC}. \\

{\em Additional comments:} CMake scripts for the configuration, compilation, testing and installation of the suite are provided.
This article describes the release version UKRmol-in 3.0, that uses GBTOlib 2.0, and UKRmol-out 3.0.
  

\end{small}

\section{Introduction}
\label{intro}

The R-matrix method is a form of embedding method which involves the division of space into a (spherical) inner region and an outer region.
It is widely used for theoretical studies in atomic, molecular and optical physics \cite{BB93,Burke}, nuclear physics \cite{10DeBa} and recently ultra-cold
chemistry \cite{jt643}. A feature of the R-matrix method for scattering applications is that the inner region problem is independent of the scattering
energy. This means that solution of the inner region only needs to be performed once and that the energy dependence of the problem is confined
to the physically simpler outer region. This facilitates, for example, the use of fine energy meshes which can be important for finding and characterising resonances (metastable states embedded in the continuum).

The UK molecular R-matrix codes are an implementation of the R-matrix method originally designed for treating electron-molecule
collisions. They have been subsequently generalised to treat other processes such photoionisation, positron molecule collisions \cite{jt491} and studies of diffuse bound states.
Theses codes have been developed over a number of years \cite{jt161,jt204,epjd_ukrmol}. 

The present paper reports the release of a new version of 
the codes, known as \UKRmol+. \UKRmol+ represents a major improvement in functionality, algorithms and parallelisation compared to the previous
version known as \UKRmol. In particular, \UKRmol+ allows the optional use of B-spline basis functions to represent the continuum which facilitates calculations with higher kinetic energies of the free electron and the
use of greatly enlarged inner regions allowing both large targets and targets with more diffuse electronic states to be studied. Previous versions of the \UKRmol{} codes have incorporated a (limited) quantum chemistry capability to
provide target orbitals. In a change from this, \UKRmol+ utilizes external
 electronic structure codes (e.g. Molpro \cite{MOLPRO}) to provide molecular orbitals allowing considerably more
flexibility in the representation of the molecular targets. Algorithmic improvements include use of the new GBTOlib library \cite{GBTOlib} for computing integrals, generating configurations
and constructing the Hamiltonian matrix among others. These new modules are designed to take advantage of MPI, where available, an option not available in the older code.
In addition, \UKRmol+ contains an option to compute photoionisation dipoles and cross sections \cite{hbm14,BHM15} plus these dipoles can also be used as the input for the RMT (R-matrix with time) code \cite{RMT} which can treat molecules in intense, ultrashort, arbitrarily-polarized laser pulses. A number of other
improvements in functionality are discussed below.

We note that some of the works containing results that are referenced throughout the papers actually used the  \UKRmol{} suite.  \UKRmol+  should be able to reproduce virtually all the old results (some functionality has yet to be implemented in the new suite); this has indeed been tested for a number of targets.

This paper is structured as follows. Section~\ref{sec:theory}  introduces the molecular R-matrix theory: a succinct derivation in Section~\ref{sec:derivation} will help those readers interested in a deeper understanding of the background of the method; those interested in how the  quantities generated by the suite combine to solve the scattering/photoionization problem and the R-matrix scattering models used in practice can safely avoid this derivation. Section~\ref{sec:progs} details  the input data required and the capabilities of each program in the suite. Sections~\ref{sec:scatt}, \ref{sec:photo} and \ref{sec:rmt}  describe how the programs in suite are combined to study electron/positron scattering, photoionization and to produce input for the RMT suite respectively. The test suite is described in Section~\ref{sec:test} followed by Section~\ref{sec:results} containing several examples of practical applications which illustrate the current capabilities of the suite.

\section{Overview of the R-matrix approach}
\label{sec:theory}

In the R-matrix method space is divided by a so-called R-matrix sphere of radius $a$. This radius needs to be set large enough to ensure that the wavefunction representing the target can be assumed to have zero amplitude on the boundary (in fact, all target orbitals used should have approximately zero amplitude on the boundary). As explained below, this division allows us to solve the Schr\"{o}dinger equation separately in these two parts and join the solution on the R-matrix sphere. A full exposition of the R-matrix theory can be found in the  monographs~\cite{bnb07,B2011}. In this Section, we start by deriving the fundamental equations of the R-matrix method. Readers interested only
 in the main equations implemented in the suite  can skip this derivation and start with Section~\ref{sec:maineqs}.


\subsection{Derivation of the approach}
\label{sec:derivation}

We wish to find the solution of the multi-electron Schr\"{o}dinger equation in the whole space for a problem where one of the (total) N+1 electrons (or a positron) can be found outside of the R-matrix sphere ($r \geq a$). Here correlation and exchange with the inner region can be neglected and the outer-region particle can be regarded as moving in a generally non-spherical static potential of the molecule (Appendix A.2 in~\cite{RMT} details the form of this potential). The total wavefunction in the outer region can therefore be written using the channel expansion
\begin{eqnarray}\label{eq:psiouter}
\vert \Psi_{out}(E)\rangle = \sum_{p}^{n}\frac{F_{pp'}(r_{N+1})}{r_{N+1}}\chnl{p}{},
\end{eqnarray}
where $\chnl{p}{}$ is the channel wavefunction given by a product of the wavefunction representing a target electronic state and the angular (spherical harmonic) part of the wavefunction of the outer region electron and $n$ is the total number of channels. $\mathbf{X}_{N}$ stands for all spin-space coordinates of the N electrons confined to the inner region and $\hat{\mathbf{r}}_{N+1}\sigma_{N+1}$ are the angular and spin coordinates of the (N+1)th particle (electron/positron) and $r_{N+1}$ is its radial coordinate. From now on we drop the index of the (N+1)th particle when referring to its coordinates. The functions $F_{pp'}(r)$ are the reduced radial wavefunctions of the outer region particle and $p'$ labels the linearly independent solutions of the single-particle Schr\"{o}dinger equation. As the analytic form of these functions is well known in the asymptotic region~\cite{gai76,hbm14}, we can match  outer region solutions with the ones from the inner region, $\vert \Psi_{in}(E)\rangle$:
\begin{eqnarray}
H_{N+1}\vert\Psi_{in}(E)\rangle &=& E\vert\Psi_{in}(E)\rangle,\label{eq:1}\\
\langle\chnlshort{p}{}\mbox{\small $\frac{1}{r}$}\vert \Psi_{in}(E)\rangle\big\vert_{r=a} &=& \langle\chnlshort{p}{}\mbox{\small $\frac{1}{r}$}\vert \Psi_{out}(E)\rangle\big\vert_{r=a}, p=1,\dots,n,\label{eq:bc1a}\\
\frac{d}{dr}\langle\chnlshort{p}{}\mbox{\small $\frac{1}{r}$}\vert \Psi_{in}(E)\rangle\bigg\vert_{r=a} &=& \frac{d}{dr}\langle\chnlshort{p}{}\mbox{\small $\frac{1}{r}$}\vert \Psi_{out}(E)\rangle\bigg\vert_{r=a}, p=1,\dots,n.\label{eq:bc2}
\end{eqnarray}
Here $E$ is the total energy and $H_{N+1}$ is the non-relativistic molecular Hamiltonian in the fixed-nuclei approximation
\begin{eqnarray}
H_{N+1} &=& \sum_{i=1}^{N+1}\left( -\frac{1}{2}\nabla_{i}^{2} + \sum_{i>j}^{N+1}\frac{1}{\vert \mathbf{r}_{i} - \mathbf{r}_{j}\vert} -\sum_{k=1}^{Nuclei}\frac{Z_{k}}{\vert \mathbf{r}_{i} - \mathbf{R}_{k}\vert}\right).
\end{eqnarray}
where $Z_{k}$ is the charge and $\mathbf{R}_{k}$ is the position of the nucleus. Note that equations~(\ref{eq:bc1a}-\ref{eq:bc2}) are equivalent to:
\begin{eqnarray}
F_{pp',in}(a) &=& F_{pp',out}(a),\\
F_{pp',in}^{'}(a) &=& F_{pp',out}^{'}(a).
\end{eqnarray}
The R-matrix method is an equivalent formulation of this boundary value problem which uses the Bloch operator, $L$,  to embed the second (derivative) boundary condition into the solution of the Schr\"{o}dinger equation~(\ref{eq:1})
\begin{eqnarray}
(H_{N+1} + L)\vert\Psi_{in}(E)\rangle &=& E\vert\Psi_{in}(E)\rangle + L\vert\Psi_{out}(E)\rangle,\label{eq:scheq-bloch}\\
\langle\chnlshort{p}{}\mbox{\small $\frac{1}{r}$}\vert \Psi_{in}(E)\rangle\big\vert_{r=a} &=& \langle\chnlshort{p}{}\mbox{\small $\frac{1}{r}$}\vert \Psi_{out}(E)\rangle\big\vert_{r=a}, p=1,\dots,n,\label{eq:bc1}\\
L &=& \sum_{i=1}^{N+1} \frac{1}{2}\delta(r_{i}-a)\frac{d}{dr_{i}}.\label{eq:bloch}
\end{eqnarray}
The Bloch operator ensures that the operator $H_{N+1} + L$ is self-adjoint and that the boundary condition given by Eq.~(\ref{eq:bc2}) is included in Eq.~(\ref{eq:scheq-bloch}). Using Eq.~(\ref{eq:scheq-bloch}) to express $\vert\Psi_{in}(E)\rangle$ we obtain
\begin{eqnarray}
\vert\Psi_{in}(E)\rangle &=& G_{N+1}(E) L\vert\Psi_{out}(E)\rangle,\label{eq:psiin}\\
G_{N+1}(E) &=& (H_{N+1} + L - E)^{-1},
\end{eqnarray}
where $G_{N+1}(E)$ is the Green's operator for the inner region. Next we take advantage of the spectral decomposition of the Green's operator
\begin{eqnarray}\label{eq:green}
G_{N+1}(E) = \sum_{k}\frac{\vert\psik{}{}\rangle\langle\psik{}{}\vert}{E_{k}-E},
\end{eqnarray}
where $\vert\psik{}{}\rangle$ and $E_{k}$ are the so-called R-matrix basis functions and poles respectively:
\begin{eqnarray}
(H_{N+1}+L)\vert\psik{}{}\rangle &=& E_{k}\vert\psik{}{}\rangle,\label{eq:rmatbasis}
\end{eqnarray}
which are defined only in the inner region;  due to the Bloch operator the \textit{exact} eigenvectors have zero derivative on the boundary. We insert Eq.~(\ref{eq:green}) back into Eq.~(\ref{eq:psiin}) obtaining
\begin{eqnarray}
\vert\Psi_{in}(E)\rangle = \sum_{k}\frac{\vert\psik{}{}\rangle\langle\psik{}{}\vert L\vert\Psi_{out}(E)\rangle}{E_{k}-E}.\label{eq:innerwf}
\end{eqnarray}
We can now project this equation on the channel functions and obtain a formula for the corresponding reduced radial wavefunctions:
\begin{eqnarray}
\langle\chnlshort{p}{}\mbox{\small $\frac{1}{r}$}\vert\Psi_{in}(E)\rangle = \sum_{k}\frac{\langle\chnlshort{p}{}\frac{1}{r}\vert\psik{}{}\rangle\langle\psik{}{}\vert L\vert\Psi_{out}(E)\rangle}{E_{k}-E}.
\label{eq:innerproj}
\end{eqnarray}

The next section shows shows how this leads to the definition of the R-matrix.

\subsection{Fundamental equations of the R-matrix approach}
\label{sec:maineqs}

The matrix elements defined by Eq.~(\ref{eq:innerproj}) can be evaluated with  the help of Eqns.~(\ref{eq:psiouter}),~(\ref{eq:bloch}). Applying the boundary condition given by Eq.~(\ref{eq:bc1}) yields the  result
\begin{eqnarray}
\mathbf{F}_{E}(a) &=& \mathbf{R}_{E}(a)\mathbf{F}_{E}^{'}(a),\label{eq:match}\\
\mathbf{R}_{E}(a) &=& \frac{1}{a}\mathbf{w}(a)\mathbf{[E_{k}-E]^{-1}}\mathbf{w^{T}}(a),\label{eq:rmatrix}
\end{eqnarray}
where $\mathbf{R}_{E}(a)$ is the R-matrix in the basis of the channel wavefunctions (the Green's function evaluated on the R-matrix sphere) and $\mathbf{F}_{E}(a)$ is the matrix of the channel reduced radial wavefunctions evaluated at $r=a$. The matrix $\mathbf{[E_{k}-E]^{-1}}$ is diagonal and the matrix of the reduced boundary amplitudes $\mathbf{w}(a)$ is defined as:
\begin{eqnarray}\label{eq:bamps}
w_{pk}(a) = \frac{1}{\sqrt{2}}\bigg\langle\chnlshort{p}{}\frac{1}{r}\bigg\vert\psik{}{}\bigg\rangle\bigg\vert_{r=a} = \frac{1}{\sqrt{2}}\bigg\langle\Phi_{i_p}^{N}\frac{1}{r}X_{l_{p},m_{p}}(\hat{\mathbf{r}})\bigg\vert\psik{}{}\bigg\rangle\bigg\vert_{r=a},
\end{eqnarray}
where $\Phi_{i_p}^{N}=\Phi_{i_p}^{N}(\mathbf{x}_{1},\dots,\mathbf{x}_{N})$ is the N-electron wavefunction representing the target electronic state $i_p$ corresponding to channel $p$ and $X_{l_{p},m_{p}}(\hat{\mathbf{r}})$ is the real spherical harmonic of the outer region particle in that channel. For the full expression of the boundary amplitudes in terms of the raw (single-particle) boundary amplitudes, see~\ref{app:bamps}. From Eqns.~(\ref{eq:match}-\ref{eq:rmatrix}) and the known (asymptotic) form of $\mathbf{F}_{E}(a)$ we can compute the K-matrix and all scattering observables. In practice, especially in the case of scattering calculations, the radius $r=a$ typically does not lie in the asymptotic region. Therefore the R-matrix is first propagated~\cite{bbm82,mor84,pfarm} in the static multipole molecular potential, see Appendix A.2 in~\cite{RMT}, to a large distance $a_p$ (typically 100 a$_0$) where the matching of the radial functions to known asymptotic expressions is performed \cite{nn84,B2011}.

If the inner region wavefunction is required (as in the case of photoionization calculations) it can be determined through Eq.~(\ref{eq:innerwf}) inserting in it the now fully specified outer region wavefunction. The result is\footnote{It is often stated that Eq.~(\ref{eq:inner_wfn_expansion}) is an expansion in the basis of the $\vert\psik{}{}\rangle$ functions. However, rigorously $\vert\psik{}{}\rangle$ do not form a basis in the Hilbert space of the solutions since the equivalent derivative series generally does not converge to the derivative of the wavefunction at $r=a$, see~\cite{szmytkowski1996} for a detailed discussion.}:
\begin{equation}\label{eq:inner_wfn_expansion}
\vert\Psi_{in}(E)\rangle = \sum_{k}A_{k}(E) \vert\psik{}{}\rangle,
\end{equation}
where the form of the coefficients $A_{k}(E)$ depends on the choice of the asymptotic boundary conditions in the outer region (photoionization or scattering)~\cite{hbm14}.

The strength of the R-matrix method lies in the energy factorisation of the inner-region's Green function (see Eqns.~(\ref{eq:green}-\ref{eq:rmatbasis})) which requires, to obtain the $E_k$ and $\psik{}{}$, only one diagonalization of the inner-region Hamiltonian. Consequently, the R-matrix can be constructed easily for an arbitrary grid of energies and the desired solutions  determined  efficiently. Not surprisingly the construction of the R-matrix basis functions $\vert\psik{}{}\rangle$ is typically the most important and the most difficult part of the whole calculation. These wavefunctions are represented by a close-coupling expansion of the form:
\begin{equation}\label{e:rmat}
    \psi_{k}^{N+1} = \hat A \sum_{i, j} c_{ijk} \Phi^{N}_{i}(\mathbf{x}_{1},\dots,\mathbf{x}_{N}) \eta_{ij}(\mathbf{x}_{N+1}) + \sum_{m} b_{m k} \chi^{N+1}_{m}(\mathbf{x}_{1},\dots,\mathbf{x}_{N+1}).
\end{equation}
The first of the terms on the right-hand side of the equation represents the product of the wavefunctions describing the target, $\Phi^{N}_{i}(\mathbf{x}_{1},\dots,\mathbf{x}_{N})$, with continuum orbitals, $\eta_{ij}(\mathbf{x}_{N+1})$ which are non-zero on the boundary;
the anti-symmetriser $\hat A$ ensures that this product obeys the Pauli principle. 

\UKRmol+  allows the use of both Gaussian type orbitals (GTOs) and B-spline orbitals (BTOs) \cite{bdc01} to represent the continuum: options allow $\eta_{ij}(\mathbf{x}_{N+1})$ to be represented by GTOs, a hybrid set of GTOs and BTOs or simply a set of BTOs, see Section~\ref{sec:sintegrals}. The \UKRmol{} code \cite{epjd_ukrmol} used  GTOs only to represent the continuum \cite{gtobas}.

The second, so-called $L^2$, terms in Eq.~(\ref{e:rmat}) comprises configurations where the scattering electron is placed in target orbitals; they describe short range correlation/polarisation. The coefficients $c_{ijk}$ and $b_{m k}$ are determined variationally by constructing
and diagonalising the inner region Hamiltonian matrix using Eq.~(\ref{eq:rmatbasis}). This step normally dominates the computational requirements. 

\subsection{R-matrix scattering models}\label{sec:rmat_models}

Within the framework described above there are a variety of different models and procedures that can be used. Key ones are discussed below, but for more details on these and the use of the molecular R-matrix method in general see the review by Tennyson \cite{jt474}.

{\bf Static exchange} (SE) is the simplest scattering model which uses a single
target wavefunction represented at the Hartree-Fock (HF) level. In this model the $L^2$
configurations
are given simply by placing the scattering electron in unoccupied target (virtual)
orbitals of the appropriate symmetry. For a closed shell target, the N+1 configurations can be written:
\begin{eqnarray}
&(\text{HF})^N&(\text{cont})^1,\nonumber\\
&(\text{HF})^N&(\text{virt})^1,\nonumber
\end{eqnarray}
where HF represents a single Hartree-Fock determinant, virt is an unoccupied target orbital and
cont is a continuum orbital. The SE model is rather crude but does have the
advantage that it is well defined so can be used for benchmarks against other methods
and codes. 

{\bf Static exchange plus polarisation} (SEP) builds on the SE model by also including $L^2$
configurations which involve promoting an electron 
 from the HF target wavefunction to a virtual orbital while also placing the scattering
 electron in a target virtual orbital. For a closed shell target, the SEP model augments the SE configurations
 with configurations of the type
 \begin{center}
 (core)$^{N_{c}}$(valence)$^{(N_{v}-1)}$(virt)$^2$,
 \end{center}
 where $N_{c}$ is the number of electrons in doubly occupied orbitals and $N_{v}$ is the number of electrons in the ``valence" orbitals so that $N = N_{c}+N_{v}$.
 Experience shows that many more virtuals are required to achieve a good description of the scattering for SEP  than SE calculations \cite{jt533}.
 The extra configurations included in the SEP model allow for the inclusion
 of short-range target polarisation effects in the model. The SEP model is still
 relatively simple but is found to provide a good representation of low-lying shape
 resonances which are, in particular, important for providing a gateway for dissociative
 electron recombination and are also involved in dissociative electron attachment.
 
 {\bf Close-coupling} (CC) expansions involve including several target states in 
  Eq.~(\ref{e:rmat}). This model normally uses a complete active spaces (CAS) description of these  states and, when possibe, (state-averaged) CASSCF orbitals. Within a CAS model with $M$ active electrons in the CAS, the N+1 configurations can generally be represented as
 \begin{eqnarray}
 &(\text{core})^{(N-M)}&\text{CAS}^M(\text{cont})^1\nonumber\\
 &(\text{core})^{(N-M)}&\text{CAS}^{(M+1)}\nonumber\\
 &(\text{core})^{(N-M)}&\text{CAS}^M(\text{virt})^1\nonumber\\
 &(\text{core})^{(N-M)}&\text{CAS}^{(M-1)}(\text{virt})^2\nonumber
  \end{eqnarray}
although other models have been used \cite{jt585}. The first and second type of configurations are always used whereas the last two are not (they tend to be needed for targets with large polarizabilities; the last type is actually rarely included). Use of the CC method is essential
 for describing  electronic excitation and is also best for studying Feshbach resonances.
 However CC calculations can be computationally demanding and there are subtle questions
 that need to be addressed over how best to build a model. \cite{jt189,dora_r-matrix_2009}.
 
 {\bf R-matrix with pseudostates} (RMPS) is a generalisation of the CC method. Given
 that there are an infinite number of states below each ionisation threshold, it is 
 not possible to work with complete CC expansion of physical states. The RMPS method \cite{bhs96}
 uses an extra set of target orbitals,
 known as pseudo-continuum orbitals (PCOs), to provide a representation of the discretized
 continuum in the inner region. The molecular implementation of this uses even-tempered GTOs
 \cite{jt341,jt354}. The RMPS model leads to
 fairly complex set of configurations of the
 type:
 \begin{eqnarray}
   &(\text{core})^{(N-M)}&\text{CAS}^M(\text{cont})^1\nonumber\\
 &(\text{core})^{(N-M)}&\text{CAS}^{M-1}(\text{PCO})^1(\text{cont})^1\nonumber\\
 &(\text{core})^{(N-M)}&\text{CAS}^{M+1}\nonumber\\
 &(\text{core})^{(N-M)}&\text{CAS}^M(\text{PCO})^1\nonumber\\
 &(\text{core})^{(N-M)}&\text{CAS}^{M-1}(\text{PCO})^2\nonumber\\
 &(\text{core})^{(N-M)}&\text{CAS}^M(\text{virt})^1\nonumber\\
 &(\text{core})^{(N-M)}&\text{CAS}^{M-1}(\text{PCO})^1(\text{virt})^1\nonumber\\
 &(\text{core})^{(N-M)}&\text{CAS}^{(M-1)}(\text{virt})^2.\nonumber
 \end{eqnarray}
Again, here the first four  types of configurations are always used, whereas the last four are optional. Construction of the configuration set has to be performed with care as the choice of the number of core, CAS, virt and PCO orbitals has to be balanced with computational demands \cite{jt444}. 
 The RMPS approach has very useful properties
 in terms of extending the energy range of the calculations \cite{jt341} and allowing polarisation effects
 to be rigorously converged \cite{jt468}, but are computationally very demanding
 \cite{jt444,jt510} so as yet the RMPS procedure is only rarely used. 

A word on nomenclature: as seen above, the R-matrix method requires the determination of energies and wavefunctions for N- and N+1-electron systems. In scattering calculations, the N-electron system is normally called the target and the N+1 electron wavefunctions are referred to as the scattering wavefunctions. In the context of photoionization (and RMT calculations) the N-electron system is referred to as the residual (molecular) ion and the N+1-electron system as the neutral system. All these names will be used, as appropriate, throughout the paper.

\section{Programs in the suite}\label{sec:progs}

\medskip

The \UKRmol+ suite consists  of about a dozen  computer programs 
written in various versions of Fortran. The programs are provided in two suites: \UKRmol{}-in, containing those necessary for the target and inner region calculations  and \UKRmol{}-out, containing those needed for the outer region scattering calculation and the interface programs (e.g. to produce the input for RMT). The \UKRmol{}-in suite has been almost completely rewritten over the last few years and that is the one we will describe here in detail. The \UKRmol{}-out suite has remained relatively unchanged since Ref. \citet{epjd_ukrmol}, so will not be discussed in detail in this paper. We will, however, detail the existing interface programs.

Each of the programs is responsible for a specific set of tasks within the scattering or photoionization calculation workflow. The execution of the programs is controlled using case-insensitive input namelists, which are either read from the standard input, from disk files with hard-coded names, or from disk files in paths provided on the command line. The programs communicate with each other using intermediate disk files. In most  cases, the files are not standard named files, but Fortran numerical units, represented by most compilers as disk files with name ``\textfile{fort.n}'', where \textit{n} is a number that can be changed via the program's input namelist. Some \UKRmol+ programs are serial, some are multi-threaded, and some are capable of running in MPI (distributed) mode, as detailed below.

\UKRmol+ supports the following Abelian point groups: C$_1$, C$_2$, C$_s$, C$_i$, C$_{2h}$, C$_{2v}$ and D$_{2h}$.  Molecules that belong to other (non-Abelian) points groups (e.g. those belonging to  C$_{\infty v}$ and D$_{\infty h}$) need to be assigned to the closest smaller group, with C$_1$ as the last-resort option. The irreducible representations of these groups are frequently referenced in the input namelists. They are labelled using what is often often referred to as ``$M$-values'', in analogy to  linear molecules (the first molecular implementation of the R-matrix method was for diatomic molecules \cite{jt161}), for which ``$M$-values'' referred to the projection of the angular momentum on the molecular axis. The assignment of $M$-values to individual irreducible representations is given in \ref{app:irr-tables} in Table~\ref{tab:M-values}. 

The \UKRmol+ suite requires as input a file generated by an external Quantum Chemistry suite, containing geometrical information about the molecule as well as the bound orbitals to be used in the description of the process (see next section for more details). The file should be in Molden format \cite{molden}; for most of the tests provided in the test suite included in the release (and the calculations performed so far) the files have been generated using Molpro \cite{MOLPRO}, although Psi4~\cite{Psi4}
 has also been used for some calculations.
 
 The continuum GTO basis sets are generated using two programs in the suite: \textprog{NUMCBAS} and \textprog{GTOBAS} \cite{gtobas}. These programs do not need to be run for each calculation: the basis is generated once for a specific R-matrix radius and  charge of the N-electron system and a   range of kinetic energies of the free electron. Briefly, the exponents of the GTOs are optimized for each angular momentum by fitting to a set of numerical Bessel (if neutral targets are going to be studied) or Coulomb (if charged targets are to be investigate) functions within a specified radial range given by the R-matrix radius to be used. The number of numerical functions to be fitted is given by a selected  maximum wavenumber. 
 
 
 The sections that follow describe all the other programs in the suite and the input they require.  Sections~\ref{sec:scatt} to \ref{sec:rmt} describe how these programs are used in three different types of calculations: electron-scattering ones (including  calculations to determine bound states), photoionization calculations and those to produce input for the RMT suite. 
 Brief summaries on the inputs are provided to illustrate key points; full documentation of the inputs is provided with the release.
 Section~\ref{sec:test} briefly describes the test suite and finally Section~\ref{sec:results} presents some of the results obtained with the latest version of the codes.
 
\subsection{\textprog{SCATCI\_INTEGRALS}}\label{sec:sintegrals}

The program \textprog{SCATCI\_INTEGRALS} performs all the calculations related to basis functions and orbitals: it evaluates all the required 1- and 2-particle integrals for the atomic basis functions, orthogonalizes bound and continuum orbitals, transforms the integrals from the atomic to the molecular basis, etc. 

The program uses a stand-alone library GBTOlib~\cite{GBTOlib} that offers the choice of using centre-of-mass centred
Gaussian-type orbitals (GTOs) and/or  B-spline type orbitals (BTOs) as the single-particle orbitals, as illustrated  in Fig.~\ref{fig:continuum}. The BTOs and GTOs are defined as follows:
\begin{eqnarray}
{\cal B}_{i,l,m}(\mathbf{r}) &=& N_{i}\frac{B_{i}(r)}{r}X_{l,m}(\hat{\mathbf{r}}),\\
G_{\alpha,l,m}(\mathbf{r}-\mathbf{A}) &=& N_{\alpha,l}^{GTO}S_{l,m}(\mathbf{r}-\mathbf{A} )\exp[-\alpha\vert \mathbf{r}-\mathbf{A} \vert^{2}],
\end{eqnarray}
where $B_{i}(r)$ is the radial B-spline with index $i$ and $X_{l,m}(\hat{\mathbf{r}})$ is a real spherical harmonic~\cite{homeier}. $S_{l,m}(\mathbf{r}-\mathbf{A})$ is a real solid harmonic centred on the point $\mathbf{A}$ (atomic centre or the centre-of-mass) and is defined via the real spherical harmonic:
\begin{equation}
 S_{l,m}(\hat{\mathbf{r}})=\sqrt{\frac{4\pi}{2l+1}}r^{l}X_{l,m}(\hat{\mathbf{r}}).   
\end{equation}
In both cases the factors $N_{i}$ and $N_{\alpha,l}^{GTO}$ are chosen to normalize the functions to a unit integral of their modulus squared. If contracted GTOs (linear combinations of primitive GTOs) are used an overall normalization factor is needed to ensure unit integral over modulus squared of the contracted GTO.

The BTOs and GTOs can be used to build three types of bases: atomic (representing the orbitals of the target molecule), continuum (representing the unbound particle) and pseudocontinuum. These three bases  can be included in the calculation in an arbitrary combination. Table~\ref{tab:basis} lists these bases  together with the type of 1-particle orbitals that can be included in each of them and the namelist required to specify the details of each basis.
\begin{table}[htbp]
\caption{Basis sets and 1-particle orbitals supported by \textprog{SCATCI\_INTEGRALS} together with the namelists used to specify the input.}
    \centering
\begin{tabular}{llccc}
        \toprule
                &                              & \multicolumn{3}{l}{1-particle orbital supported} \\
Basis type      & Namelist                     & GTO           & BTO           & GTO+BTO          \\
        \midrule
Target molecule & \textcode{\&target\_data}    & YES           & NO            & NO               \\
Continuum       & \textcode{\&continuum\_data} & YES           & YES           & YES              \\
Pseudocontinuum & \textcode{\&pco\_data}       & YES           & NO            & NO       \\               \bottomrule

\end{tabular}
\label{tab:basis}
\end{table}

The use of the continuum and pseudocontinuum bases is optional (and therefore the code can be used for pure GTO-based bound-state quantum chemistry calculations) but the basis representing the target molecule must always be present. Obviously, the continuum basis is required for \UKRmol+ calculations involving the construction of continuum states of the molecule.

In addition to the namelists \textcode{\&target\_data}, \textcode{\&continuum\_data} and \textcode{\&pco\_data} listed in Table~\ref{tab:basis}, input is provided via the namelist \textcode{\&process\_control}. Below we describe the most relevant input parameters for each of those namelists.

\subsubsection{Namelist \textcode{target\_data}}

 \textprog{SCATCI\_INTEGRALS} reads a formatted file in Molden format~\cite{molden} containing the geometry, the GTO atomic basis and the molecular orbitals, specified in this namelist by \textcode{molden\_file}. This file can be generated by a range of external quantum chemistry software and thus enables the  use in \UKRmol+ of molecular orbitals produced at  various levels of theory (Hartree-Fock, CASSCF, etc.). The namelist, see Table~\ref{tab:sintegrals-input-target}, also contains information on the molecular symmetry of the target molecule  and how many  (externally generated)  molecular orbitals are used for its description.




\begin{table}[htbp]
     \caption{Selected parameters in the \textcode{\&target\_data} namelist in the input for \textprog{SCATCI\_INTEGRALS}}
   \centering
    \begin{tabular}{ll}
        \toprule
        \textcode{\&target\_data} & \\
        \midrule
        \textcode{a} & R-matrix radius (in \(\Bohr\)) (default:$-1.0$)\\
        \textcode{no\_symop} &  Number of symmetry operators required to define \\
        & the point group\\
        \textcode{sym\_op} &  Symmetry operators to be used \\
        \textcode{molden\_file} &  Path and name of  the input file in Molden format \\
        \textcode{nob} & Number of target orbitals of each irreducible representation \\
                     & to be read from the Molden file \\
        \bottomrule
    \end{tabular}
    \label{tab:sintegrals-input-target}
\end{table}

The choice of the R-matrix radius \textcode{a} is, perhaps, counter-intuitively, done in the \textcode{\&target\_data} namelist rather than in the continuum one. The reason is that it is the spatial extent of the target electronic orbitals that determines the size of the R-matrix sphere required. Integrals involving only the target functions are always computed over all space, in agreement with the assumption of the R-matrix method that the electronic density associated to the target molecular states is completely contained inside the R-matrix sphere. If, on input, $a\leq 0$ then all integrals, including those involving the continuum functions, are computed over all space. 

The definition of the symmetry (using the common symmetry operators \textcode{'X'}, \textcode{'Y'}, \textcode{'Z'}, \textcode{'XY'}, \textcode{'YZ'}, \textcode{'XZ'} and \textcode{'XYZ'}) is necessary to perform the transformation of the integrals from the atomic to the (symmetry adapted) molecular basis. The parameter \textcode{nob} is an array with the length of the number of irreducible representations; again, Table~\ref{tab:M-values} indicates the order of these. \textcode{nob}  specifies the number of molecular orbitals to include in the integral calculation. One should pick here both the orbitals that will be used to describe the target and those to be used (if required) for the $L^2$ functions (i.e. the virtual orbitals, see Section~\ref{sec:rmat_models}).

\subsubsection{Namelist \textcode{continuum\_data}}

The important parameters defining the continuum basis are listed in Table~\ref{tab:sintegrals-input-continuum}. They are used to define both the GTO and/or the BTO continuum bases centred on the centre of mass. As noted above both types of functions can be mixed freely to define the continuum orbitals. 

Using a pure GTO continuum is straightforward. The exponents of the GTO continuum basis are generated, as explained, using the programs \textprog{NUMCBAS} and \textprog{GTOBAS} \cite{gtobas} and provided as input, as a list of values for each partial wave $l$ in the array \textcode{exponents(:,$l$)}. Note that if neither \textcode{min\_l} or \textcode{max\_l} are given non-zero values, then no GTO continuum is included in the calculation.

\begin{figure}[htbp]
    \centering
    \includegraphics[angle=-90,width=\textwidth]{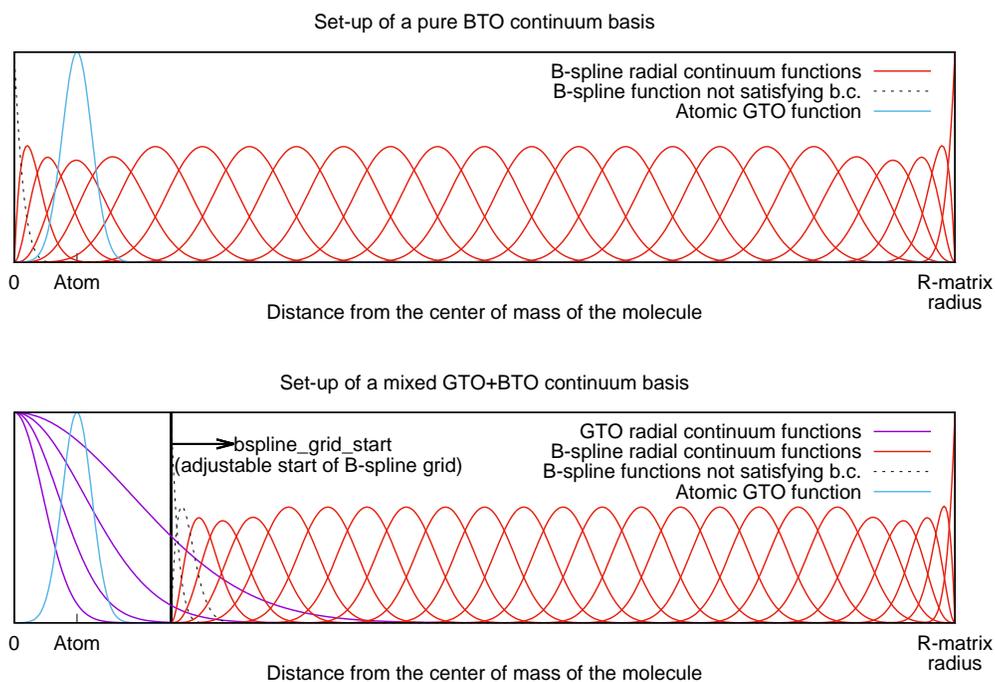}
    \caption{Set-up of the continuum basis in a \UKRmol+ calculation. The GTO and the BTO continuum functions centred on the centre of mass can be mixed freely inside the R-matrix sphere. An additional pseudocontinuum basis built from GTOs centred on the centre of mass can also be included  (not shown). The parameter \textcode{bspline\_grid\_start} sets the radial distance for start of the B-spline basis. The B-spline functions plotted using the dashed lines are those not satisfying the boundary conditions, see text.}
    \label{fig:continuum}
\end{figure}

An illustration of the set-up using either a pure BTO or a mixed GTO + BTO continuum is shown in Fig.~\ref{fig:continuum}. If the BTO basis is used care must be taken to include only those radial B-spline functions which are compliant with the boundary conditions: if the B-spline basis starts at the origin the first B-spline must not be included and if the B-spline basis starts at $r > 0$ then the first two B-splines must not be included.
The values of \textcode{bspline\_indices(1,l)} and \textcode{bspline\_indices(2,l)} determine the starting and the final index of the radial B-spline to include in the BTO basis for partial wave $l$: only if \textcode{bspline\_indices}(1,l) $\leq$ \textcode{bspline\_indices}(2,l) are BTOs for that partial wave included. The radius at which the radial B-spline basis starts  is controlled by the parameter~\textcode{bspline\_grid\_start}. The order of the B-spline \cite{bdc01} to be used is given by \textcode{bspline\_order}; typical values are 8 to 11.

\begin{table}[htbp]
\caption{Selected parameters in the \textcode{\&continuum\_data} namelist in the input for \textprog{SCATCI\_INTEGRALS}. Most default values not indicated are set to $-1$.}
    \centering
    \begin{tabular}{ll}
        \toprule
        \textcode{\&continuum\_data} & \\
        \midrule
        \textcode{min\_l} & The lowest GTO partial wave to include \\
        \textcode{max\_l} & The highest GTO partial wave to include \\
         \textcode{exponents(:,$l$)}            &  Exponents of the continuum GTOs for  \\
         & partial wave $l$ \\
         \textcode{min\_bspline\_l} & The lowest BTO partial wave to include  \\
         \textcode{max\_bspline\_l} & The highest BTO partial wave to include \\
         \textcode{bspline\_grid\_start} & The radial distance from which the BTOs   \\
         & start, see Fig.~\ref{fig:continuum} \\
         \textcode{bspline\_order} &  Order of the B-splines to be included in \\
         &the calculation \\
         \textcode{no\_bspline} &  The number of radial B-splines in the basis  \\
         \textcode{bspline\_indices(1,$l$}) & Indices of the first and the last radial B-spline\\
         \textcode{bspline\_indices(2,$l$}) & to be included for the partial wave $l$\\
\textcode{del\_thresh} &  Deletion thresholds for each irreducible \\
        & representation used in the symmetric\\
        & orthogonalization of the continuum\\
\textcode{run\_free\_scattering} & Logical flag that enables the running of a\\
        & free scattering calculation (default:.false.). \\
        \textcode{min\_energy} &  Minimum scattering energy in Hartree in the \\
        & free scattering calculation  \\
        \textcode{max\_energy} &  Maximum energy in Hartree the free\\
        &scattering calculation\\
        \textcode{nE} & Number of energies between min\_energy\\
        &and max\_energy \\
        \bottomrule
    \end{tabular}
     \label{tab:sintegrals-input-continuum}
\end{table}

The deletion threshold to be used for each irreducible representation in the symmetric orthogonalization is given by \textcode{del\_thresh} (see Section~\ref{sec:orb_ortho}).

If the free scattering calculation is requested using the flag \textcode{run\_free\_scattering} then the program uses the R-matrix methodology (with the R-matrix radius given in the namelist \textcode{\&target\_data}) and the target + continuum orbitals as a basis to solve the 1-particle ``scattering" problem with zero potential (i.e. with Hamiltonian $H = -\frac{\nabla^2}{2}$) and computes the eigenphase sums in each spatial symmetry for a range of energies by matching to spherical Bessel functions at $r=a$. For free scattering, the exact result should be zero eigenphase sums. Therefore the size of the deviations from zero give an idea of the quality of the continuum basis. A rule of thumb for a good continuum representation are eigenphase sums less than or equal to $10^{-2}$~rad. The energy range and grid for this caculations can be adjusted if required using the namelist parameters \textcode{min\_energy}, \textcode{max\_energy} (in Hartree) and \textcode{nE}. If several free scattering calculations are run in order to tune the continuum basis, this can be made more efficient setting the flag \textcode{do\_two\_particle\_integrals = .false.} in the namelist \textcode{\&process\_control}, see below.

\subsubsection{Namelist \textcode{pco\_data}}
\label{sec:pco_data}

The basis for the pseudocontinuum is built from even-tempered GTOs centred on the centre of mass. The pseudocontinuum exponents are generated using the formula:
\begin{eqnarray}\label{eq:pco}
\alpha_{j,l} = \alpha_{0,l}\beta_{l}^{(j-1)},\; \; j=1,\dots n_{l},
\end{eqnarray}
where the parameters $\alpha_{0,l}$, $\beta_{l}$ and $n_{l}$ correspond to the values \textcode{PCO\_alpha0}, \textcode{PCO\_beta} and \textcode{num\_PCOs} respectively in the namelist \&\textcode{pco\_data}, see Table~\ref{tab:sintegrals-input-pco}. Usual values of $\beta_{l}$ range from and 1.1 to 1.5. 

\begin{table}[htbp]
    \caption{Parameters in the \textcode{\&pco\_data} namelist in the input for \textprog{SCATCI\_INTEGRALS}. All default values are set to $-1/-1.0$.}
    \centering
    \begin{tabular}{ll}
        \toprule
        \textcode{\&pco\_data} & \\
        \midrule
        \textcode{PCO\_alpha0} & Parameters $\alpha_{0,l}$ in Eq~(\ref{eq:pco}).  \\
        \textcode{PCO\_beta} & Parameters $\beta_{l}$ in Eq~(\ref{eq:pco}). \\
        \textcode{num\_PCOs} & Parameter $n_{l}$ in Eq~(\ref{eq:pco}): number of pseudocontinuum  \\ 
        &  exponents per partial wave to be generated.\\
        \textcode{min\_PCO\_l} & The lowest angular momentum for PCOs \\
         \textcode{max\_PCO\_l} & The highest angular momentum for PCOs \\
         \textcode{PCO\_gto\_thrs} & Thresholds, per partial wave, for removal of\\
                                 & exponents of continuum GTOs, see text for details.\\
         \textcode{PCO\_del\_thrs} & Deletion thresholds for each irreducible representation\\
        & used in the symmetric  orthogonalization of the\\
        & pseudocontinuum\\
        \bottomrule
    \end{tabular}
    \label{tab:sintegrals-input-pco}
\end{table}

If the GTO continuum is included in the calculation too then the exponents of the continuum GTOs for each partial wave are scanned and those greater than or equal to the smallest PCO exponent (minus the value \textcode{PCO\_gto\_thrs}) in the same partial wave removed. This procedure improves numerical stability of generation of the continuum orbitals. \textcode{PCO\_del\_thrs} are the deletion thresholds for the symmetric orthogonalization of this basis. 

\subsubsection{Namelist \textcode{process\_control}}

\begin{table}[htbp]
    \caption{Selected parameters in the \textcode{\&process\textunderscore control} namelist in the input for \textprog{SCATCI\_INTEGRALS}.}
    \centering
    \begin{tabular}{ll}
        \toprule
        \textcode{\&process\_control} & \\
        \midrule
        \textcode{max\_ijrs\_size} & Maximum size (in MiB) of the auxiliary array  \\                      & used during 2-electron integral transformation. \\
                       & For distributed (MPI) calculation this value \\
                       &  represents memory per MPI task. This value\\
                       & always has to be set by the user.\\
        \textcode{do\_two\_particle\_integrals} & Logical flag requesting calculation of 2-electron \\ 
                       &integrals (useful for tuning of the continuum \\
                       & basis) (default:.true.)\\
        \textcode{two\_p\_continuum} & Flag requesting calculation of 2-electron\\
                                   & integrals with two particles in the continuum.\\
                                   & (default:.false.)\\
        \textcode{mixed\_ints\_method} & Method to use for evaluation of the mixed \\
                                     & GTO/BTO integrals (see documentation)\\
         \textcode{max\_l\_legendre\_1el} & Highest partial wave  in the Legendre expansion \\
                                        & of the mixed GTO/BTO nuclear attraction\\
                                        &  integrals\\
         \textcode{max\_l\_legendre\_2el} & Highest partial wave in the Legendre expansion\\
                                        &  of the mixed GTO/BTO two-electron integrals.\\
         \textcode{calc\_radial\_densities} & Logical flag requesting calculation of the radial\\
                                          & charge densities of the orbitals included\\
                                          &  in the calculation (default: .false.)\\
        \textcode{scratch\_directory} & Path to the scratch directory. Only used\\
                    & in case of mixed integrals. If not set the \\
                     & auxiliary quantities will be kept in memory (default).\\
        \textcode{delta\_r1} & Length in \(\Bohr\) of the elementary  \\
                          & Gauss-Legendre quadrature used for evaluation  \\
                          & of the mixed integrals (default: 0.25)\\
                          
        \bottomrule
    \end{tabular}
    \label{tab:sintegrals-input-process}
\end{table}

This namelist, see Table~\ref{tab:sintegrals-input-process}, controls mainly how the integrals are calculated and how much memory or whether scratch space is requested for some auxiliary quantities needed during the computation of the mixed GTO/BTO integrals. Most importantly, the parameters \textcode{max\_l\_legendre\_1el} and \textcode{max\_l\_legendre\_2el} control the highest partial wave included in the Legendre expansion of the Coulomb potential when calculating the mixed GTO/BTO integrals. The minimum sensible numbers for these quantities is twice the value of the highest (pseudo)continuum angular momentum included in the calculation. Careful convergence checks of the final results with respect to these parameters are recommended. The length of the Legendre expansion  significantly affects both the computational time and the memory requirements of the calculation. Values as high as $75$ have been used in calculations~\cite{Darby_Lewis_2017}. The flag  \textcode{calc\_radial\_densities} is useful for testing
whether the R-matrix radius chosen is sufficiently large.

\subsubsection{Orbital orthogonalization}
\label{sec:orb_ortho}

\begin{figure}[htbp]
    \centering
    \tikzstyle{program}=[rectangle, thick, minimum size=0.5cm, draw=blue!80, fill=blue!20, rounded corners = 5pt, inner sep = 5pt, drop shadow]
\tikzstyle{infile}=[rectangle, thick, minimum size=0.5cm, draw=red!80, fill=red!20, rounded corners = 5pt]
\tikzstyle{imtfile}=[rectangle, thick, minimum size=0.5cm, draw=black!20, fill=white!20, rounded corners = 5pt]
\tikzstyle{outfile}=[rectangle, thick, minimum size=0.5cm, draw=green!80, fill=green!20, rounded corners = 5pt]
\tikzstyle{number}=[circle, inner sep=1pt, minimum size=0mm, draw=blue!80, fill=blue!10]
\tikzstyle{inarrow}=[black!50, line width = 0.5mm, >=latex]
\tikzstyle{outarrow}=[black!50, line width = 0.5mm, >=latex]
\usetikzlibrary{decorations.pathreplacing}
\newcommand*{\connectorV}[4][]{
  \draw[#1] (#3) |- ($(#3) !#2! (#4)$) -| (#4);
}
\newcommand*{\connectorH}[4][]{
  \draw[#1] (#3) -| ($(#3) !#2! (#4)$) |- (#4);
}

\begin{tikzpicture}[auto, outer sep=3pt, node distance=1.5cm]
    \node [program, text width = 3.25cm, align = center, draw=blue!80, fill=blue!20] (GStgt) {\strut{Gram-Schmidt orthogonalization of the target orbitals}};
    \node [number, left = 0.25cm of GStgt] (N1) {\textsf{\small 1}};
    
    \node [program, below = 0.6cm of GStgt, text width = 3.75cm, align = center, draw=red!80, fill=red!20] (GSPCO) {\strut{Individual Gram-Schmidt orthogonalization of the PCOs against target orbitals}};
    \node [number, left = 0cm of GSPCO, draw=red!80, fill=red!20] (N2a) {\textsf{\small 2a}};
    
    \node [program, below = 0.6cm of GSPCO, text width = 3.25cm, align = center, draw=red!80, fill=red!20] (SYMPCO) {\strut{Symmetric orthogonalization of the PCOs}};
    \node [number, left = 0cm of SYMPCO, draw=red!80, fill=red!20] (N2b) {\textsf{\small 2b}};

    \node [program, below = 0.6cm of SYMPCO, text width = 3.25cm, align = center, draw=red!80, fill=red!20] (DELPCO) {\strut{Deletion of linearly-dependent PCOs}};
    \node [number, left = 0cm of DELPCO, draw=red!80, fill=red!20] (N2c) {\textsf{\small 2c}};

    \node [program, right = 0.75cm of GStgt, text width = 3.3cm, align = center, draw=cyan!80, fill=cyan!20] (TGTset) {\strut{Orthonormal set of target orbs. (+PCOs)}};

    \node [program, below = 0.8cm of TGTset, text width = 3.75cm, align = center, draw=green!80, fill=green!20] (GSCO) {\strut{Individual Gram-Schmidt orthogonalization of the COs against target (+PCO) orbs.}};
    \node [number, right = 0.275cm of GSCO, draw=green!80, fill=green!20] (N3a) {\textsf{\small 3a}};

    \node [program, below = 0.5cm of GSCO, text width = 3.25cm, align = center, draw=green!80, fill=green!20] (SYMCO) {\strut{Symmetric orthogonalization of the COs}};
    \node [number, below = 2.20cm of N3a, draw=green!80, fill=green!20] (N3b) {\textsf{\small 3b}};
    
    \node [program, right = 0.75cm of DELPCO, text width = 3.25cm, align = center, draw=green!80, fill=green!20] (DELCO) {\strut{Deletion of linearly-dependent COs}};
    \node [number, below = 1.95cm of N3b, draw=green!80, fill=green!20] (N3c) {\textsf{\small 3c}};

    \node [program, right = 0.75cm of TGTset, text width = 3.2cm, align = center, draw=cyan!80, fill=cyan!20] (FINALset) {\strut{Orthonormal set of target orbs. (+PCOs +COs)}};
    
    \draw [inarrow, ->] (GStgt) --  node { } (GSPCO) ;
    \draw [inarrow, ->] (GSPCO) --  node { } (SYMPCO) ;
    \draw [inarrow, ->] (SYMPCO) --  node { } (DELPCO) ;
    \connectorH[inarrow, ->]{0.50}{DELPCO}{TGTset};
    \draw [inarrow, ->] (TGTset) --  node { } (GSCO) ;
    \draw [inarrow, ->] (GSCO) --  node { } (SYMCO) ;
    \draw [inarrow, ->] (SYMCO) --  node { } (DELCO) ;
    \connectorH[inarrow, ->]{0.5}{DELCO}{FINALset};

    \draw [inarrow, dashed, ->] ([yshift=0.5cm]GStgt) --  node { } (TGTset) ;
    \draw [inarrow, dashed, ->] ([yshift=0.5cm]TGTset) --  node { } (FINALset) ;

\end{tikzpicture}
    \caption{Workflow of the orbital orthogonalization. The steps 2 and 3 are only performed if the basis includes, respectively, the corresponding pseudocontinuum atomic orbitals (PCOs) and continuum orbitals (COs). The dashed arrows represent the workflow in the case PCOs and/or COs are not included in the calculation.}
    \label{fig:orthogonalization}
\end{figure}
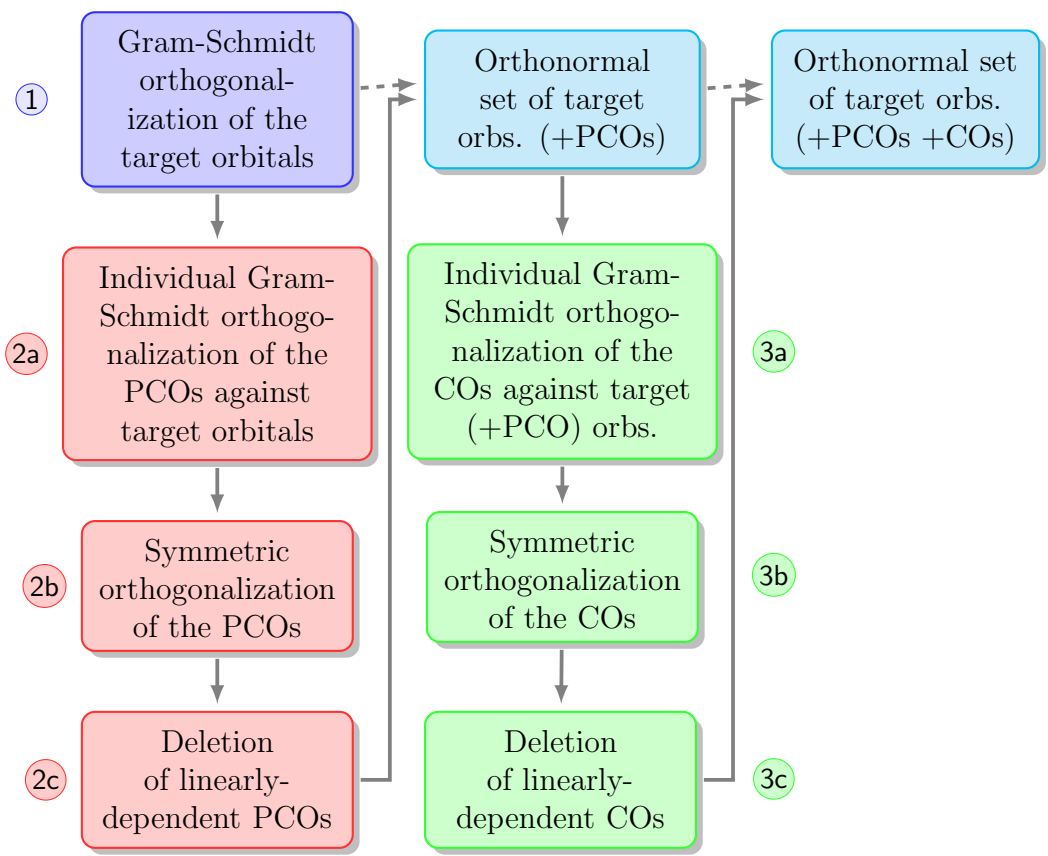

It is a requirement of the  \UKRmol+ suite that all the  orbitals used form a single orthonormal set;  orthogonalization steps are therefore required. The target orbitals are usually orthonormal on input but as the first step they are preemptively reorthogonalized using the Gram-Schmidt method. The flow of the orthogonalization process is shown in or ig.~\ref{fig:orthogonalization}.

In a standard scattering calculation only the target and the continuum bases are included and only steps 1 and 3 from Fig.~\ref{fig:orthogonalization} are performed. In this case  the Gram-Schmidt orthogonalization of the continuum basis with the target orbitals (step 3a) leaves the latter unchanged. The second (step 3b), symmetric orthogonalization~\cite{szabo_ostlund}, ensures that the continuum orbitals (now expanded in a a linear combination of the continuum and atomic bases) are orthogonal among themselves. 

Deletion thresholds for each irreducible representation are provided by the user for the symmetric orthogonalization in the namelist~\textcode{\&continuum\_data}: those continuum orbitals with eigenvalues of the overlap matrix smaller than this threshold are deleted. The appropriate value for this threshold varies depending on the type of continuum basis used and typically lies in the range $10^{-7}$ to $10^{-9}$ for a pure GTO continuum basis. Lowering the deletion threshold corresponds to decreasing the relative precision of the transformed integrals and therefore increasing the risk of running into numerical linear dependence problems when diagonalizing the inner-region Hamiltonian. Increasing the deletion threshold, on the other hand, means more continuum functions are deleted from the basis, lowering the quality of the continuum description. The numerical problems can be avoided (and the deletion thresholds set to a value low enough ensuring no continuum orbitals are deleted) by compiling and running \textprog{SCATCI\_INTEGRALS} in quadruple precision albeit at the expense of a much increased compute time (see below for details). For this reason quad precision calculations have only been performed using pure GTO continuum in which case all integrals have analytic form. In calculations using either a mixed GTO+BTO or a pure BTO continuum the deletion thresholds can be set to a much higher value (typically $10^{-5}$). The higher value of the deletion threshold means that the numerical linear dependence problems are essentially absent in this case (as expected for a B-spline basis) thus removing the need to run the integral calculation in quad precision.

If a pseudocontinuum basis is included in the calculation,  additional orthogonalization steps are required: this basis needs to be orthogonalized to the target orbitals and among itself (step 2) before the continuum is orthogonalized to the joint target + pseudocontinuum basis (step 3). The deletion threshold to be used for the symmetric orthogonalization of the pseudocontinuum orbitals is provided in the namelist~\textcode{\&pco\_data} and is usually set to a relatively high value of about $10^{-4}$;
some pruning of the continuum orbitals may also be necessary in this case to avoid problems with linear dependence.

\subsubsection{Integrals evaluated by \textprog{SCATCI\_INTEGRALS}}
The one-electron integrals are defined as:
\begin{eqnarray}\label{eq:sintegrals-oneel}
\langle a\vert \hat{O} \vert b\rangle =\int d^{3}r a(\mathbf{r}) \hat{O} b(\mathbf{r}),
\end{eqnarray}
where $\hat{O}$ is a one-electron operator and $a$, $b$ are either atomic or molecular orbitals. \textprog{SCATCI\_INTEGRALS} generates by default the following types of 1-electron integrals:
\begin{eqnarray}
\text{overlap:} &\hat{O}& = 1,\\
\text{kinetic energy:} &\hat{O}& = -\frac{\nabla^2}{2} + L,\\
\text{nuclear attraction:} &\hat{O}& = \sum_{k=1}^{Nuclei}-\frac{Z_{k}}{\vert \mathbf{r} - \mathbf{R}_{k}\vert},\\
\text{multipole:} &\hat{O}& = S_{l,m}(\mathbf{r}),\ \  l\leq 2,\\
\text{1-electron Hamiltonian:} &\hat{O}& = -\frac{\nabla^2}{2} + L + \sum_{k=1}^{Nuclei}-\frac{Z_{k}}{\vert \mathbf{r} - \mathbf{R}_{k}\vert},
\end{eqnarray}
where  $L=\frac{1}{2}\delta(r-a)\frac{d}{dr}$, the Bloch operator, is included only if both $a$ and $b$ are continuum functions (the multielectronic form of this operator was introduced in Eq.~(\ref{eq:bloch})). The two-electron (Coulomb) integrals are:
\begin{eqnarray}\label{eq:sintegrals-twoel}
\langle ab\vert\vert bc\rangle =\int d^{3}r a(\mathbf{r}_{1})b(\mathbf{r}_{1})\frac{1}{\vert \mathbf{r}_{1}-\mathbf{r}_{2}\vert}c(\mathbf{r}_{2})d(\mathbf{r}_{2}).
\end{eqnarray}
Depending on the particular combination of the functions $a$, $b$, $c$, $d$ the integral will belong in one of six unique classes:
\begin{eqnarray}
\langle TT\vert\vert TT\rangle,
\langle TC\vert\vert TT\rangle,
\langle TC\vert\vert TC\rangle,
\langle CC\vert\vert TT\rangle,
\langle CC\vert\vert CT\rangle,
\langle CC\vert\vert CC\rangle,\nonumber
\end{eqnarray}
where $T$ and $C$ stand for atomic/molecular orbitals representing the target (or pseudocontinuum) and continuum respectively. The last two classes of integrals (with at least three $C$-type functions) are only needed in calculations where two particles are in the continuum. Since UKRmol+ supports only calculations with one particle in the continuum these integrals are not generated by default. However, they can be generated setting the variable \textcode{two\_p\_continuum = .true.} in the namelist \textcode{\&process\_control}, see Table~\ref{tab:sintegrals-input-process} (Section~\ref{sec:pco_data}). In the current version this option is not available if the basis contains BTOs but it will be implemented in the future.

In the case of the class $\langle CC\vert\vert TT\rangle$ and the one-electron integrals $\langle C\vert C\rangle$, when on input the R-matrix radius is \textcode{a}$> 0$, the atomic integrals are computed only over the interior of the R-matrix sphere with radius \textcode{a}. The other classes of integrals are computed over all space unless the continuum functions are all BTOs.

GBTOlib~\cite{GBTOlib} uses object oriented features from the Fortran~2003 standard and involves distributed and shared-memory parallelization to evaluate the 1-electron and 2-electron integrals defined above. The program proceeds by generating first the 1-electron and 2-electron integrals in the atomic basis and then transforming them into the molecular basis. The whole calculation is performed in core (perhaps with exception of the mixed integrals, see namelist~\textcode{\&process\_control}). Only the final array of the transformed molecular integrals is saved to disk (along with the full basis set information). It is the integrals in the molecular basis which are used by the other programs in the \UKRmol+ suite.

A very useful feature of the GBTOlib is that it can be compiled and run in quad precision: this does not amount to a mere recompilation using a different \textcode{kind} parameter but to actual evaluation of the integrals with increased relative precision (i.e. using more accurate versions of special functions and higher-order quadratures which evaluate to near-full quad precision accuracy). This feature has been used extensively over the last couple of years mostly in photoionization calculations requiring large GTO continuum bases, see e.g.~\cite{masin_2018,brambila_2017} but also in calculations with molecular clusters \cite{SiG17a,SiG17b} and biomolecules \cite{Loupas_alanine}.

In the current release version of GBTOlib distributed evaluation is not allowed when mixed GTO/BTO 2-electron integrals are required: only shared-memory parallelism is available for this type of calculation. However, distributed evaluation of these integrals will be implemented in the next major release of the code along with the option to save the atomic integrals to disk.

\subsection{\textprog{CONGEN}}
\label{sec:congen}

The program \textprog{CONGEN} generates the configuration state functions (CSF) that will be used in building the N or N+1 electron wavefunctions. As such, input to this program defines to a large extent the model that will be used in the description of the scattering/photoionization process; it is in the input required by \textprog{CONGEN} that the SE, SEP, CC and RMPS approximations are most different. This can make the input of \textprog{CONGEN} complex to generate. 

\textprog{CONGEN} does not require any input (data) files generated by other programs. The only information needed from earlier parts of the calculation is the number of target orbitals (specified by the parameter \textcode{nob0}, see Table~\ref{tab:congen-input-state}) belonging to each irreducible representation and the total number of orbitals resulting from the orthogonalization of the continuum to the bound orbitals (again, for each irreducible representation and specified by the parameter \textcode{nob}). This is provided by the user in the input, together with detailed information  specifying the construction of the CSF. When the projectile is a positron, the number of orbitals needs to be doubled: the same orbitals that can be occupied by the electrons will have a differnet index when they are occupied by the positron.

Two namelists are required for \textprog{CONGEN}: \textcode{\&state} and \textcode{\&wfngrp}. Only one occurrence of the former is allowed, but several instances of the second one may be required to define all the CSFs to be generated.   

\begin{table}[htbp]
     \caption{Selected parameters in the \textcode{\&state} namelist in the input of \textprog{CONGEN}. The storage size and memory allocation is indicated in words.}
   \centering
    \begin{tabular}{ll}
        \toprule
        \textcode{\&state} & \\
        \midrule
        \textcode{iscat} & Indicates whether N, if set to 1 (no phase correction) or N+1 \\ 
        & electron calculation, if set to 2 (phase correction) (default:0)\\
        & (see next section for details on the phase correction) \\
        \textcode{qntot} &  Triplet that indicates the space-spin symmetry of  CSFs \\
        & generated\\
        \textcode{nelect} & Total number of electrons  \\
        \textcode{megul} & File unit for the output CSFs (default: 13). \\
        \textcode{nob} &  Total number of orbitals for each irreducible representation \\
          \textcode{nob0} &  NUmber of target orbitals for each irreducible representation \\     
        \textcode{nrefo} & Number of quintets in  \textcode{reforb} \\
        \textcode{reforb} & Quintets that describe the reference determinant \\
        \textcode{lndo} &  Controls assignment of temporary storage (default: 5000) \\
        \textcode{lcdo} & Controls assignment of temporary storage (default: 500) \\
        \textcode{nbmx} &   Memory allocation for CSFs generation (default: 2000000) \\
        \textcode{iposit}&  Lepton charge flag (default; 0, indicates electron; 1, positron)  \\
        \bottomrule
    \end{tabular}
    \label{tab:congen-input-state}
\end{table}

\textprog{CONGEN} generates CSFs as differences from a reference configuration. This configuration, specified using \textcode{nrefo} and \textcode{reforb} in the namelist \textcode{\&state}, should have the correct space-spin symmetry (given by \textcode{qntot}), but does not need to be physically meaningful. However, use of a `sensible' configuration is recommended, as the number of differences that need to be stored will be reduced. The \textcode{iposit} flag in the namelist \textcode{\&state} indicates whether the calculation is for a positron or electron in the continuum.

When \textprog{CONGEN} is used for the N+1 system, the CSFs to be generated are those described by the two terms in Eq.~(\ref{e:rmat}). This means that there are two types of CSFs (a) those which correspond to the product of a target state (whose symmetry must be given by the parameter  \textcode{qntar}, see Table~\ref{tab:congen-input-wfn}) and a continuum orbital and (b) those
L$^2$ CSFS where all electrons occupy target orbitals. There are some rules about the order in which these configurations
must be generated, namely: (1) all CSFs of the target plus continuum type must precede the L$^2$ ones and (2) all CSFs associated with a single target symmetry, and hence value of \textcode{qntar}, must be grouped together and generated in a canonical order
which is used in all parts of the calculation (eg for the target only calculation). As the first step in the \textprog{SCATCI}
algorithm \cite{jt180} is based on the use of Yoshimine's prototype configurations \cite{Y73}, \textprog{CONGEN} only actually generates CSFs of the first type (a) corresponding to the first two continuum orbitals for each scattering symmetry. Note that it is possible to treat CSFs of the form target times virtual orbital in
same fashion as the target times continuum orbitals, see Ref.~\cite{jt189} for a discussion of this.

\begin{table}[htbp]
    \caption{Selected parameters in the \textcode{\&wfngrp} namelist in the input of \textprog{CONGEN}. }
    \centering
    \begin{tabular}{ll}
        \toprule
        \textcode{\&wfngrp} & \\
        \midrule
        \textcode{qntar} & Space-spin symmetry of the target states being coupled to \\
        \textcode{nelecg} &  Number of 'movable' electrons \\
        \textcode{nrefog} &  Number of \textcode{reforg}  quintets needed to describe movable electrons\\
        \textcode{reforg} &  Quintets which describe the movable electrons \\
        \textcode{ndprod} & Number of sets of orbitals into which the electrons will be \\
        & distributed\\
        \textcode{nelecp} &  Number of electrons to be distributed into each set of orbitals \\
        \textcode{nshlp} &  Number of  \textcode{pqn} triplets needed to describe each set of orbitals \\
        \textcode{pqn} &  Triplets that describe each set of orbitals in which the movable \\
		&  electrons will be distributed \\
         \textcode{mshl} &  For each  \textcode{pqn}, the irreducible representation the described \\
         & orbitals belong to\\     
        \bottomrule
    \end{tabular}
    \label{tab:congen-input-wfn}
\end{table}

The namelist \textcode{\&wfngrp} contains the information that specifies the CSFs to be built. The `movable' electrons are those that are going to be promoted from the reference configuration to generate the desired CSFs. 
If all the CSFs to be generated contain the same set of core electrons (in other words, the same set of doubly occupied orbitals), then the number of movable electrons, specified by \textcode{nelecg} in the \textcode{\&wfngrp} namelist, will be smaller than the total number of electrons of the system. In that case, \textcode{nreforg} and \textcode{reforg} will  specify the orbitals in the reference configuration from which electrons are going to be promoted (if \textcode{nelecg}=\textcode{nelect} then   \textcode{reforg} and \textcode{reforb} will specify the same configuration). The CSFs themselves are specified using th parameters \textcode{nelecp},  \textcode{nshlp}, \textcode{pqn} and \textcode{mshl}.

The details of the syntax for the \textcode{qntot} and \textcode{qntar} triplets, \textcode{reforb} and \textcode{reforg} quintets as well as the \textcode{pqn} triplets can be found in the documentation of \textprog{CONGEN}. Special care should be  exercised in the case of positron scattering because of the doubling up of the orbital labels.

\textprog{CONGEN} runs serially and does not require linking to any external libraries for compilation. The program needs to be run for each space-spin symmetry for which CSFs are required. It was initially written in Fortran~66 as part of the Alchemy~I suite \cite{mcl71} and  recently modernized to use Fortran~95 and~2003 features.

\subsection{\textprog{SCATCI} and \textprog{MPI-SCATCI}}

The main task performed by \UKRmol{}-in is the construction and diagonalization of a Hamiltonian matrix in order to obtain the bound and (discretized) continuum wavefunctions describing the molecule + free (N+1) electron system. This task uses
an algorithm \cite{jt180} especially designed  to efficiently construct Hamiltonian matrices with the close-coupling structure  of Eq.~(\ref{e:rmat}); this is carried out by either \textprog{SCATCI} or \textprog{MPI-SCATCI}~\cite{MPI_SCATCI}. The latter is a parallel implementation of a  re-engineered version of the algorithm employed by \textprog{SCATCI} that makes use of modern computer architectures and has added functionality (see Section~\ref{sec:rmt}). 

\begin{table}[htbp]
   \caption{Diagonalization modes and methods of \textprog{SCATCI} and \textprog{MPI-SCATCI} depending on the number of requested eigenvectors and library used. Except for the Davidson diagonalization routine \cite{Davidson}, all other depend on  external libraries: LAPACK, Arpack~\cite{Arpack}, ScaLAPACK~\cite{ScaLAPACK}, or SLEPc~\cite{SLEPc}. The classifications ``few'' and ``many'' is automatically assigned using built-in percentages of requested eigenpairs and the size of the Hamiltonian. The diagonalization method can also be specified manually, in the \textcode{\&cinorn} namelist.}
    \centering
    \begin{tabular}{ccc}
        \toprule
        eigenvectors & \textprog{SCATCI}/serial \textprog{MPI-SCATCI} & parallel \textprog{MPI-SCATCI} \\
        \midrule
        ``few'' & Davidson & Krylov-Shur (SLEPc) \\
        ``many'' & Arnoldi (Arpack) & Krylov-Shur (SLEPc) \\
        all & dense (LAPACK) & dense (ScaLAPACK) \\
        \bottomrule
    \end{tabular}
     \label{tab:diagonalizations}
\end{table}

In typical situations, \textprog{SCATCI}/\textprog{MPI-SCATCI} require two files on input: the CSFs constructed by \textprog{CONGEN} and the molecular integrals file generated by \textprog{SCATCI\_INTEGRALS}. These files must be generated specifying  the same number of target orbitals. The calculated eigenvectors and eigenvalues (eigenpairs) are  either written to a new file as a first ``dataset'', or appended to an already existing file as another ``dataset''.

As with \textprog{CONGEN}, \textprog{SCATCI}/\textprog{MPI-SCATCI} is executed for each space-spin symmetry independently and operates either in target or in scattering mode. In the case of diagonalization of the scattering, N+1-electron Hamiltonian, \textprog{SCATCI} can either calculate the target, N-electron eigenstates needed in Eq.~(\ref{e:rmat}) on its own, or read them from a file produced before by diagonalization of the target Hamiltonian (\textcode{nftg} in namelist \textcode{\&input)}. The former method is sufficient for scattering calculations of integral cross sections. The latter method is needed  to ensure phase consistency \cite{jt195} across all irreducible representations of the $(N+1)$-electron system; the same target eigenstates are supplied to all independent diagonalizations of the scattering Hamiltonians. If this was not done, some of the independently on-the-fly re-calculated target eigenstates for two different scattering irreducible representations might end up with different phases, introducing random relative signs to the obtained coefficients \(c_{ijk}\) in Eq.~(\ref{e:rmat}) and causing errors in the calculation of matrix elements between the (N+1)-electron eigenstates. This is important  in calculations of angular distributions in  electron impact electronic excitation  \cite{jt179,TashiroO2} and  is crucial when using \textprog{CDENPROP} for photoionization calculations (see below).

Depending on the setup of the calculation, \textprog{SCATCI}/\textprog{MPI-SCATCI} may either produce all eigenpairs, or just the requested number of lowest-lying ones. Target calculations usually only request a few low-lying states. Traditional R-matrix calculations require all solutions of the N+1 inner region Hamiltonian; however, when the Hamiltonian is very large the partitioned R-matrix approach provides an alternative which only requires a subset of the solutions \cite{jt332} (this approach, available in the UKRmol suite, will be implemented for scattering calculations in a forthcoming release of UKRmol+). The number of eigenpairs to be calculated determines the choice of diagonalization method,  not all of which are available if the programs are not compiled with all optional libraries;  Table~\ref{tab:diagonalizations} summarizes the choices and availability.

\begin{table}[htbp]
    \caption{Selected members of the \textprog{SCATCI} and \textprog{MPI-SCATCI} \textcode{\&input} namelist. Entries labelled with an asterisk are only used in scattering runs. Detailed information is provided in the program documentation.}
    \centering
    \begin{tabular}{ll}
        \toprule
        \textcode{\&input} & \\
        \midrule
        \textcode{icitg} & Indicates target (= 0, default) or scattering (= 1), i.e. \\
        & with target  contraction, run. \\
        \textcode{iexpc}* & Expand (= 1) continuum in scattering prototype CSFs \\ & (default: 0). \\
        \textcode{idiag} & Hamiltonian evaluation flag (see documentation); \\
        &  often 1 (default) for target, 2 for scattering. \\
        \textcode{nfti} & File unit with molecular integrals (default: 16). \\
        \textcode{nfte} & File unit for Hamiltonian output (default: 26). \\
        \textcode{megul} & File unit with CONGEN generated CSFs (default: 13). \\
        \textcode{ntgsym}* & Number of different target space-spin symmetries    \\
        & included in the Hamiltonian construction  \\
        \textcode{numtgt}* & Number of target eigenstates per \textcode{ntgsym}  \\
        \textcode{mcont}* & $M$-value of continuum orbitals per target \textcode{ntgsym} \\
        \textcode{notgt}* & Number of continuum orbitals used per target \textcode{ntgsym} \\
        \textcode{nftg}* & File unit  for input target eigenstates; if 0, not read \\ & (default: 39) \\
        \textcode{ntgtf}* & For each target state used in \textprog{CONGEN}, index of the set \\
        & in \textcode{nftg} containing it  (default: 0)   \\
        \textcode{ntgts}* & For each target state used in \textprog{CONGEN}, sequence number \\
        & of the state within the ntgtf set\\
        \bottomrule
    \end{tabular}
    \label{tab:scatci-input}
\end{table}

\textprog{SCATCI}/\textprog{MPI-SCATCI} require two namelists in its standard input, called \textcode{\&input} and \textcode{\&cinorn}. The first one defines the construction of the Hamiltonian, while the second one specifies details of the diagonalization method and how and where the produced eigenpairs will be stored. Tables~\ref{tab:scatci-input} and~\ref{tab:scatci-cinorn} summarize the most important members of the two namelists.

When a target calculation is run, the namelist input is rather simple: in \textcode{\&input} one should make sure to indicate the correct unit for the \textprog{CONGEN} output (using  \textcode{megul})
and in \textcode{\&cinorn} one should indicate how many states (eigenpairs) of that particular space-spin symmetry are requested (\textcode{nstat}) and which set they will be stored in (\textcode{nciset}). In the case of an N+1
calculation, the parameters in the \textcode{\&input} namelist will specify how
many target states per space-spin symmetry (\textcode{numtgt}) and how many space-spin symmetries (\textcode{ntgsym})  are being used to build the wavefunctions; \textcode{mcont}
indicates the symmetry of  the continuum orbitals each set of target states is coupled to in order to generate eigenfunctions of the appropriate (N+1) symmetry and  \textcode{notgt} indicates how many of these orbitals there are.

\textprog{MPI-SCATCI} shares most serial capabilities with \textprog{SCATCI} when executed in a single process, and also supports parallel execution based on MPI. In the latter case the construction of the Hamiltonian matrix is distributed among all processes and the diagonalization performed by the appropriate parallelized SLEPc or ScaLAPACK subroutine. On top of that, \textprog{MPI-SCATCI} allows concurrent diagonalization of many Hamiltonians in one run, which is particularly advantageous when only matrix elements between the calculated eigenstates (possibly of different irreducible representations) are desired rather than the states themselves. Concurrent in-core diagonalization then avoids the need of writing the eigensolutions to disk and enables immediate fast in-memory processing once all diagonalizations finish.

\begin{table}[htbp]
    \caption{Selected members of the \textprog{SCATCI} and \textprog{MPI-SCATCI} \textcode{\&cinorn} namelist.}
    \centering
    \begin{tabular}{ll}
        \toprule
        \textcode{\&cinorn} & \\
        \midrule
        \textcode{igh} & Force iterative (= 0 Davidson, =-1 Arpack) or dense (= 1) \\
        & diagonalization method. \\
        \textcode{nstat} & Number of eigenvectors to be calculated (default: 0 = all). \\
        \textcode{nftw} & File unit for output of eigenvectors (default: 25). \\
        \textcode{nciset} & Output dataset index in \textcode{nftw} (default: 1). \\
        \textcode{vecstore} &  Indicates whether all coefficients of the  expansion, Eq.~(\ref{e:rmat}),  \\
         &   are to  be saved (=0, default) or only those that multiply \\ 
         &  configurations of the type 'target state + continuum' (=1). \\
        \bottomrule
    \end{tabular}
    \label{tab:scatci-cinorn}
\end{table}

\begin{table}[htbp]
     \caption{Members of \textprog{MPI-SCATCI} optional input namelist \textcode{\&outer\_interface}.}
   \centering
    \begin{tabular}{ll}
        \toprule
        \textcode{\&outer\_interface} & \\
        \midrule
        \textcode{write\_amp} & If \textcode{.true.}, produce boundary amplitudes file (unit 21). \\
        \textcode{write\_dip} & If \textcode{.true.}, produce transition dipoles file (unit 624). \\
        \textcode{write\_rmt} & If \textcode{.true.}, produce RMT input file \textfile{molecular\textunderscore data}. \\
        \textcode{rmatr} & R-matrix radius, for evaluation of boundary amplitudes. \\
        \textcode{ntarg} & (See \textprog{SWINTERF}, Table~\ref{tab:swinterf-input}.) \\
        \textcode{idtarg} & (See \textprog{SWINTERF}, Table~\ref{tab:swinterf-input}.) \\        
        \textcode{delta\_r} & (See \textprog{RMT\_INTERFACE}, Table~\ref{tab:rmt-interface}.) \\
        \textcode{nfdm} & (See \textprog{RMT\_INTERFACE}, Table~\ref{tab:rmt-interface}.) \\
        \bottomrule
    \end{tabular}
    \label{tab:mpi-scatci-outer-interface}
\end{table}

To this end, \textprog{MPI-SCATCI} features some advanced wave function post-processing capabilities  which can be used as a simplified alternative to \textprog{CDENPROP\_ALL}, \textprog{SWINTERF} and \textprog{RMT\_INTERFACE} described below: it can directly evaluate the permanent and transition dipole moments between the eigenstates or produce the input for \UKRmol{}-out. These additional stages are controlled by an optional namelist \textcode{\&outer\_interface} summarized in Table~\ref{tab:mpi-scatci-outer-interface}. The outer interface built into \textprog{MPI-SCATCI} takes advantage of MPI parallelization and carries out all matrix multiplications needed to evaluate wave function properties and amplitudes in parallel, using the ScaLAPACK subroutines. This also  has the advantage of evenly distributing the memory requirements, which might otherwise become insurmountable for the serial codes.

\subsection{\textprog{DENPROP}}

The program \textprog{DENPROP} calculates the permanent and transition dipole and quadropole moments between states of the target (N-electron system). In order to do this, the program  calculates density matrices. In addition, the program can determine the spherical  polarizability of the target molecule \cite{jt468} using the perturbative sum-over-states formula; this quantity can be used to determine approximately how well polarization effects are being modelled in the calculation \cite{jt341,jt354}. \textprog{DENPROP} produces as its main output a target property file that is used by the interface programs. This file contains all the permanent and transition moments as well as the energy of the target states (read in from the output of \textprog{SCATCI}/\textprog{MPI-SCATCI}).

\begin{table}[htbp]
    \caption{Selected parameters in the \textprog{DENPROP} \textcode{\&input} namelist.}
   \centering
    \begin{tabular}{ll}
        \toprule
        \textcode{\&input} & \\
        \midrule
		\textcode{nftg} & File unit number for the output of \textprog{SCATCI}/\textprog{MPI-SCATCI} \\
		& (default: 25)\\
        \textcode{ntgt} &  Number of different space-spin symmetries to be read in \\
        & from \textcode{nftg}\\
        \textcode{ntgtf} & Set number in unit \textcode{nftg} in which the eigenvectors for a  \\
        &  specific space-spin symmetry are found  (default: 1,2,3, ...) \\
        \textcode{ntgs} & Sequence number, within a specific \textcode{ntgtf}, of the first   \\
        & eigenvector to be used (default: 1 for all \textcode{ntgtf})   \\
        \textcode{ntgtl} & Sequence number, within a specific \textcode{ntgtf}, of the last  \\
        & eigenvector to be used (default: 1 for all \textcode{ntgtf})   \\
        \textcode{nftsor} & File unit numbers of  the \textprog{CONGEN} output files for each \\
        & space-spin symmetry \\
        \textcode{ipol} & Flag for the calculation of the polarizability (default: 1, calculate) \\
        \textcode{zlast} &  Set to .true. to save CPU time  (default: .false.) \\
        \bottomrule
    \end{tabular}
     \label{tab:denprop-input}
\end{table}

\textprog{DENPROP} requires as input the property integrals determined by \textprog{SCATCI\_INTEGRALS}, the target CSFs generated by \textprog{CONGEN} and the wavefunctions obtained from \textprog{SCATCI} or \textprog{MPI-SCATCI}, respectively. Further input to \textprog{DENPROP} is provided via the namelist \textcode{\&input}; the main parameters are given in Table~\ref{tab:denprop-input}. This namelist  provides information on which target states and of which symmetry the  moments are to be determined for, as well as which intermediate files contain the relevant input data.

The program \textprog{CDENPROP\_ALL} (described in he next section) can be used instead of \textprog{DENPROP}, providing  the same input. It is expected that, in the medium term, \textprog{CDENPROP} will replace \textprog{DENPROP}, possibly with the exception of very large target calculations for which the algorithms used by \textprog{DENPROP} are more suitable. 


\subsection{\textprog{CDENPROP} and \textprog{CDENPROP}}


The program \textprog{CDENPROP} constructs the 1-particle reduced density matrix represented in the close-coupling basis of Eq.~(\ref{e:rmat}).
For a concise notation we rewrite Eq.~(\ref{e:rmat}) in the more compact form
\begin{equation}
\Psi_k = \sum_i c_{ki} \Phi_i
\end{equation}
where \( \Phi_i \in \{A\Phi^N_m\eta_n,\Phi^{(N+1)}_p  \}\). The expressions \(A\Phi^N_m\eta_n \) and \(\Phi^{(N+1)}_p\) indicate the two types of terms used in Eq.~(\ref{e:rmat}). The 1-particle density matrix is then
\begin{equation}
    \rho_{ij}(\mathbf{x}_{N+1}) = (N+1)\int {\Phi}_i {\Phi}^*_j \, \mathrm{d}^3\mathbf{x}_1 \dots \mathrm{d}^3\mathbf{x}_N
\end{equation}
and can be used to calculate the transition moment matrix \(\mathcal{M}\),
\begin{equation}
    \mathcal{M}_{ij} = \int \mu \rho_{ij} \, \mathrm{d}^3\mathbf{x} \,,
\end{equation}
and finally the transition moments, i.e. elements of the multipole (dipole,\(\mu\), quadrupole, etc.) operators, between the inner region wavefunctions,
\begin{equation}
    M_{kl} = \langle \Psi_k | \mu | \Psi_l \rangle = \sum_{kl} c_{ki} c_{lj} \mathcal{M}_{ij} \,.
\end{equation}
When the coefficients for the bound \((N+1)\)-electron system are supplied on input ($A_k(E)$ in  Eq.~(\ref{eq:inner_wfn_expansion})), \textprog{CDENPROP} can also evaluate the Dyson orbitals (overlaps of neutral and ionic bound state wave functions)
\begin{equation}
    D_{ij}(\mathbf{x}_{N+1}) = \sqrt{N+1} \int \Phi_i^{N} \Psi_j^{N+1*} \, \mathrm{d}^3\mathbf{x}_1 \dots \mathrm{d}^3\mathbf{x}_N
\end{equation}
and write them to disk in the Molden format. For a detailed description of the code and algorithms used see~\cite{CDENPROP}.

The \UKRmol+ suite contains two programs with similar functionality that are based on \textprog{CDENPROP} code: \textprog{CDENPROP} itself and \textprog{CDENPROP\_ALL}. The first one calculates  transition dipole and quadrupole moments between a chosen number of N+1-electron states belonging to two identical or distinct space-spin symmetries. This is useful for photoionization, where transition moments are usually calculated between  the initial ground state of the neutral molecule and states from only those symmetries coupled to it by the dipole operator. Table~\ref{tab:cdenprop-input} contains the most frequently used namelist parameters for CDENPROP. The first of the two numbers assigned to the parameters \textcode{lucsf}, \textcode{lucivec}, \textcode{nstat} and \textcode{nciset} correspond to the ground (``ket'') state, while the second one corresponds to the final (``bra'') states. See also the Section~\ref{sec:photo} a for detailed description of the photoionization calculations.

\begin{table}[htbp]
    \caption{Selected parameters in the \textprog{CDENPROP} input namelist.}
    \centering
    \begin{tabular}{ll}
        \toprule
        \textcode{\&deninp} & \\
        \midrule
        \textcode{luprop} & Input file unit with the molecular integrals (default: 17). \\
        \textcode{lutdip} & Input file unit with target properties (default: 24). \\
        \textcode{lutargci} & Input file unit with target eigenstates (default: 26). \\
        \textcode{lupropw} & Output file unit for moments (default: 624). \\
        \textcode{lucsf} & Two input file units containing ket and bra CSFs. \\
        \textcode{lucivec} & Two input file units containing ket and bra eigenstates. \\
        \textcode{nstat} & Numbers of ket and bra states to consider (0 = all). \\
        \textcode{nciset} & Datasets in \textcode{lucivec} to read states from. \\
        \textcode{numtgt} & Number of target states per space-spin symmetry as used \\
        & in SCATCI. \\
        \bottomrule
    \end{tabular}
    \label{tab:cdenprop-input}
\end{table}

On the other hand, \textprog{CDENPROP\_ALL} calculates matrix elements for transitions between all pairs of states in all symmetries provided, not limited to two of them. As such, it can act as an alternative to \textprog{DENPROP}, when applied on target eigenstates. When applied on scattering eigenstates, it may produce a massive amount of data, currently only needed by RMT (see below). The input namelist for \textprog{CDENPROP\_ALL} is identical to that of \textprog{DENPROP}, except that in the case of evaluation of (N+1)-electron properties, it also needs to contain the entry \textcode{numtgt} explained in Table~\ref{tab:cdenprop-input}.

\subsection{Interfaces} \label{sec:Interfaces}

The inner region wave functions obtained by \UKRmol+, or their properties, can be used to provide input for other suites or programs. The interfaces \textprog{SWINTERF} and \textprog{RMT\_INTERFACE} are responsible for extracting useful data from the inner region solutions that are passed further.


Once the interfaces have run, the output files from the target and inner region calculations can be deleted and further outer region/RMT calculations can be run using the interfaces output.

\subsubsection{\textprog{SWINTERF}}\label{sec:Swinterf}

\textprog{SWINTERF} interfaces with \UKRmol{}-out, the suite of outer region codes that perform the R-matrix propagation to obtain K-matrices and, from these, various scattering quantities: cross sections, eigenphase sums, resonance parameters, etc. It is also used by \textprog{RMT\_INTERFACE} (see the next section).

The information required by the program is: the target properties (generated by \textprog{DENPROP} and including the energies and permanent and transition moments), the raw boundary amplitudes (see \ref{app:bamps}) produced by \textprog{SCATCI\_INTEGRALS} and the inner region eigenpairs, generated by \textprog{SCATCI} or \textprog{MPI-SCATCI}.

The namelist \textcode{\&swintfin} is the only one needed by the program. The main parameters provided via this namelist are summarized in Table~\ref{tab:swinterf-input}.

\begin{table}[htbp]
      \caption{Selected parameters in the \textcode{\&swintfin} namelist in the input of \textprog{SWINTERF}.}
  \centering
    \begin{tabular}{ll}
        \toprule
        \textcode{\&swintfin} & \\
        \midrule
        \textcode{mgvn} &  Symmetry (irreducible representation) of the scattering \\ & wavefunction. \\ 
        \textcode{stot} & Spin multiplicity of the scattering wavefunction. \\
        \textcode{ntarg} & Number of target electronic states to be included in  \\
        	&  the outer region calculation \\
        \textcode{idtarg} & Array specifying which \textcode{ntarg} target states to select \\

	\textcode{luamp} & Input file unit containing the raw boundary amplitudes \\
	& (default:22)\\
	\textcode{luci} & Input file unit containing the  N+1 eigenpairs (default:25)\\
		\textcode{lutarg} & Input file unit containing the target data (default:24) \\
		\textcode{luchan} & Output file unit containing the channel information \\
		 & (default: 10) \\
		\textcode{lurmt} & Output file unit for the R-matrix poles, boundary  \\
		& (default: 21) \\
		& amplitudes and coefficients of the multipole potential \\
        \textcode{icform} &  Format flag for channel data file, 'F' for formatted, 'U' for \\
        &  unformatted (default) \\
        \textcode{irform} &  Format flag for R-matrix data file, 'F' for formatted, 'U'   \\
        &  for unformatted (default) \\
        \textcode{nvo} & Number of virtual orbitals used 'as continuum'  in \textprog{CONGEN}, \\
                       & also used to skip target states.\\
        \textcode{rmatr} & R-matrix radius \\
        \textcode{ismax} & Maximum multipole retained in expansion of long range\\
           &   potentials (default: 2)  \\
        \textcode{iposit} & Controls the charge sign for asymptotic potential  \\
		& interactions (=0, default, for electrons; =1 for positrons)  \\
		\textcode{last\textunderscore coeff\textunderscore saved} & Index of the last CI coefficient of all eigenvector  \\
		& saved by \textprog{SCATCI}/\textprog{MPI-SCATCI}. \\
        \bottomrule
    \end{tabular}
    \label{tab:swinterf-input}
\end{table}

As is the case for the construction and diagonalization of the N+1 Hamiltonian, the outer region calculation is performed separately for each space-spin symmetry of the N+1 system; this symmetry needs to be indicated on input using \textcode{mgvn} and \textcode{stot}.  The user should also indicate how many (\textcode{ntarg}) and which (\textcode{idtarg}) target states are to be considered (for the definition of the scattering channels) in the outer region: it is possible, and sometimes desirable, to use a smaller number than those included in the inner region calculation as this reduces the number of scattering channels and therefore the computational cost of the outer region calculation. It is inadvisable, however, to exclude energetically open states.

The radius of the R-matrix sphere (\textcode{rmatr}) needs to be provided too as it is used later on to calculate the R-matrix elements following Eq.~(\ref{eq:rmatrix}).  The user can also decide (using \textcode{ismax}) whether dipoles or both dipoles and quadrupoles are retained for the expansion of the projectile-target interaction potential in the outer region. In addition, if virtual orbitals are used 'as continuum'  in \textprog{CONGEN} (see Section~\ref{sec:congen}) the number of these has to be specified, for each target state, using the vector \textcode{nvo}. Finally, \textcode{iposit} indicates whether the projectile is an electron or a positron.

\textprog{SWINTERF} produces two output files: one containing the channel data and another containing the raw boundary  amplitudes (see \ref{app:bamps}) required for the construction of the R-matrix, and  the coefficients of the coupling potentials. We note that these two files also provide a convenient medium for archiving
the results of a calculation for use in possible future runs.

\subsubsection{\textprog{RMT\_INTERFACE}}\label{sec:Rmtinterface}

\begin{figure}[htbp]
    \centering
    \begin{tikzpicture}[auto, outer sep=3pt, node distance=1.5cm]

    \pgfmathsetmacro{\s}{0.5}  

    \draw[gray] (0.866*11.00*\s,-0.500*11.00*\s) arc (-30:30:11.00*\s);
    \draw[gray] (0.866*11.50*\s,-0.500*11.50*\s) arc (-30:30:11.50*\s);
    \draw[gray] (0.866*12.00*\s,-0.500*12.00*\s) arc (-30:30:12.00*\s);
    \draw[gray] (0.866*12.50*\s,-0.500*12.50*\s) arc (-30:30:12.50*\s);
    \draw[black,thick] (0.866*13.00*\s,-0.500*13.00*\s) arc (-30:30:13.00*\s);

    \draw[->,red,thick] (0,0) -- node[above] {\texttt{rmatr}} ++ (0.966*13.00*\s,0.259*13.00*\s);
    \draw[->,orange,thick] (0.966*11.50*\s,-0.259*11.50*\s) -- (0.966*12.50*\s,-0.259*12.50*\s);
    \draw[<-,orange,thick] (0.966*13.00*\s,-0.259*13.00*\s) -- (0.966*14.00*\s,-0.259*14.00*\s);
    \node[label=right:{\color{orange}{\texttt{delta\_r}}}] at (0.966*14.00*\s,-0.259*14.00*\s) {};

    \node[left,blue] at (0.866*11.00*\s,-0.500*11.00*\s) (N1) {\texttt{1}};
    \node[left,blue] at (0.866*11.50*\s,-0.500*11.50*\s) (N2) {\texttt{2}};
    \node[left,blue] at (0.866*12.00*\s,-0.500*12.00*\s) (N3) {\texttt{3}};
    \node[below,blue] at (0.866*12.00*\s,-0.500*12.00*\s) (Nd) {\texttt{\dots}};
    \node[below,blue] at (0.866*13.00*\s,-0.500*13.00*\s) (Nf) {\texttt{nfdm + 1}};

    \node[below] at (6.00*\s,0*\s) (Nin) {inner region};
    \node[above] at (16.00*\s,0*\s) (Nout) {outer region};

\end{tikzpicture}
    \caption{Parameters for finite-difference discretization of the inner wave function in RMT. Each number $j$ indicates a sphere of radius $r_j$.The physical R-matrix radius (i.e. fully containing the target orbitals) must be smaller than or equal to the radius 1.}
    \label{fig:nfdm}
\end{figure}
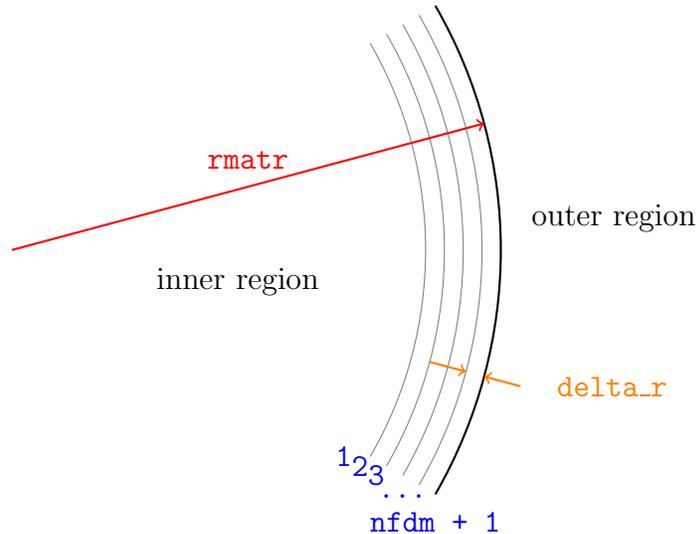

The \textprog{RMT\_INTERFACE} prepares the molecular input file for the RMT package~\cite{RMT}.  RMT solves the time-dependent Schr\"odinger equation of the molecule in a variable electric field, allowing for ionization of one electron into the continuum. Like \UKRmol+,  RMT splits the configuration space into an inner, fully correlated, region and an outer region where a one-electron channel expansion is used. The connection between these two regions is realized by an overlap that still counts as the inner region. However, this overlap of the two regions needs to be already free of bound  orbitals, which means that when preparing input for RMT the inner region (i.e. the R-matrix radius) has to be somewhat larger than the smallest physical one typically used when dealing with standard stationary scattering or photoionization. The continuity of the inner and outer wave function is maintained using the amplitudes \(w_{ik}(r_j)\) of the inner-region eigenstates \(i\) in the outer region channel \(k\); these are similar to the boundary amplitudes defined in Eq.~(\ref{eq:bamps}) but they are evaluated at several uniformly spaced radii \(r_j\) still within the R-matrix sphere, see Fig.~\ref{fig:nfdm}. The number of these extra evaluation radii (\textcode{nfdm} in the namelist \textcode{\&rmt\_interface\_inp}) and their spacing (\textcode{delta\_r}) must be consistent with the RMT setup. The parameter \textcode{nfdm}  is related to the finite difference order \(D\) used in the outer region of RMT and to the time propagation order \(Q\) as \textcode{nfdm} \(= D (Q + 1)\). While \(D\) is a compile-time constant in RMT and set to 2 (resulting in a 5-point discretization scheme), the propagation order \(Q\) is an input parameter to RMT. A typical value is \(Q = 8\), which results in the recommended \textcode{nfdm} = 18. The spacing between the evaluation radii needs to be equal to the finite difference discretization of the RMT outer region, and is usually in the \(0.1\,\Bohr\) range.

\textprog{RMT\_INTERFACE} also correctly transforms boundary amplitudes between the \UKRmol+ and RMT normalization conventions which differ by a factor of \(\sqrt2\); RMT expects \(w_{ik}\) such that Eq.~(\ref{eq:rmatrix}) becomes
\begin{equation}
    \label{eq:rmatrix_atom}
    R_{ij}(a,E) = \frac{1}{2a} \sum_k \frac{w_{ik}^{RMT}(a) w_{jk}^{RMT}(a)}{E_k - E}.
\end{equation}

\begin{table}[htbp]
    \caption{Parameters in the \textprog{RMT\_INTERFACE} input namelist \textcode{\&rmt\_interface\_inp}.}
    \centering
    \begin{tabular}{ll}
        \toprule
        \textcode{\&rmt\_interface\_inp} & \\
        \midrule
        \textcode{nfdm} & Number of in-sphere boundary evaluation radii... \\
        \textcode{delta\_r} & ... and spacing between them in atomic units. \\
        \textcode{lutarg} & Input file unit containing the target data (default: 24). \\
        \textcode{lunp1} & Input file unit containing the (N+1)-electron  \\ & moments (default: 667). \\
        \textcode{n\_symmetries} & Number of \textcode{\&swintfin} namelists to process. \\
        \bottomrule
    \end{tabular}
    \label{tab:rmt-interface}
\end{table}

The program \textprog{RMT\_INTERFACE} builds on \textprog{SWINTERF}, which is---under the hood---automatically called once for every (N+1)-electron spin-symmetry included in the calculation to obtain the channel information and the amplitudes at all requested radii. The input file follows this need and must contains enough copies of the \textprog{SWINTERF} input namelist \textcode{\&swintfin}, one for each space-spin symmetry required. Apart from these namelists, the input for \textprog{RMT\_INTERFACE} contains only a short additional namelist \textcode{\&rmt\_interface\_inp}, which is expected at the beginning of the standard input (before all \textcode{\&swintfin} namelists) and whose parameters are summarized in Table~\ref{tab:rmt-interface}.

The output of \textprog{RMT\_INTERFACE} is the binary stream file \textfile{molecular\_data} directly readable in RMT.

\section{Scattering calculations}
\label{sec:scatt}

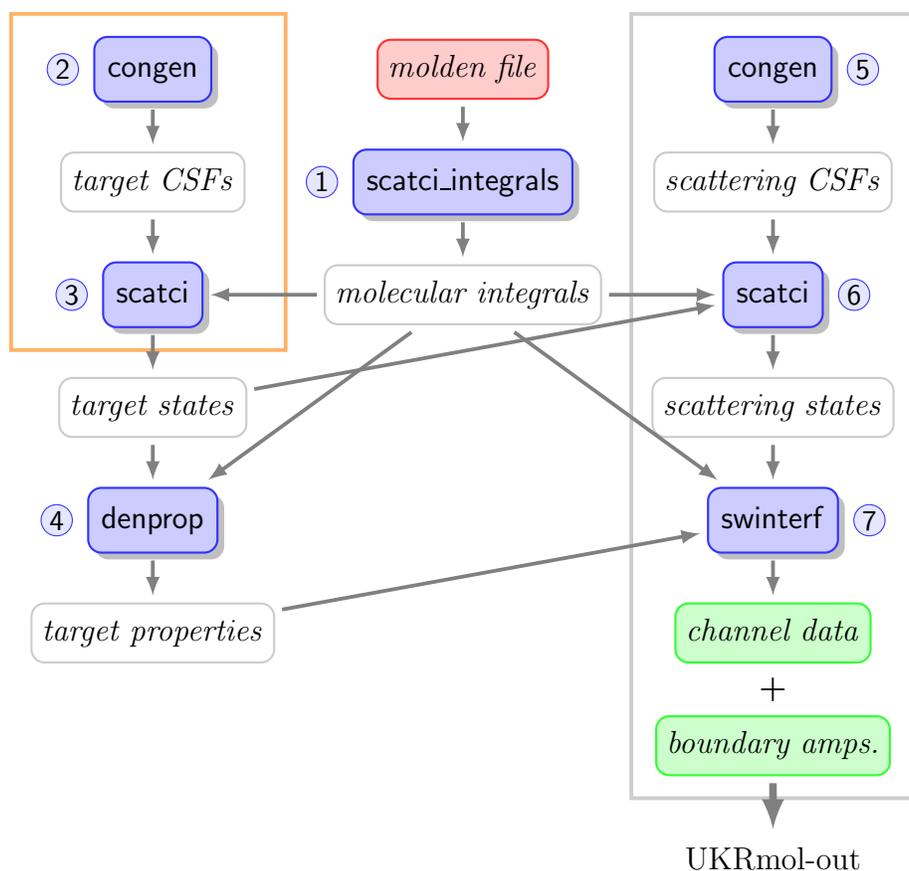
\begin{figure}[htbp]
    \centering
    \tikzstyle{program}=[rectangle, thick, minimum size=0.5cm, draw=blue!80, fill=blue!20, rounded corners = 5pt, inner sep = 5pt, drop shadow]
\tikzstyle{infile}=[rectangle, thick, minimum size=0.5cm, draw=red!80, fill=red!20, rounded corners = 5pt]
\tikzstyle{imtfile}=[rectangle, thick, minimum size=0.5cm, draw=black!20, fill=white!20, rounded corners = 5pt]
\tikzstyle{outfile}=[rectangle, thick, minimum size=0.5cm, draw=green!80, fill=green!20, rounded corners = 5pt]
\tikzstyle{number}=[circle, inner sep=1pt, minimum size=0mm, draw=blue!80, fill=blue!10]
\tikzstyle{inarrow}=[black!50, line width = 0.5mm, >=latex]
\tikzstyle{outarrow}=[black!50, line width = 0.5mm, >=latex]

\begin{tikzpicture}[auto, outer sep=3pt, node distance=1.5cm]
    \node [infile] (Nmol) {\strut\textit{molden file}};
    \node [program, below of = Nmol] (Nsci) {\strut\textsf{scatci\textunderscore integrals}};
    \node [imtfile, below of = Nsci] (Nmoi) {\strut\textit{molecular integrals}};
    \node [program, left = 2cm of Nmol] (Ncon) {\strut\textsf{congen}};
    \node [imtfile, below of = Ncon] (Ncsf) {\strut\textit{target CSFs}};
    \node [program, below of = Ncsf] (Nsca) {\strut\textsf{scatci}};
    \node [imtfile, below of = Nsca] (Neig) {\strut\textit{target states}};
    \node [program, right = 2cm of Nmol] (Ncons) {\strut\textsf{congen}};
    \node [imtfile, below of = Ncons] (Ncsfs) {\strut\textit{scattering CSFs}};
    \node [program, below of = Ncsfs] (Nscas) {\strut\textsf{scatci}};
    \node [imtfile, below of = Nscas] (Neigs) {\strut\textit{scattering states}};
    \node [program, below of = Neig] (Nden) {\strut\textsf{denprop}};
    \node [imtfile, below of = Nden] (Nprop) {\strut\textit{target properties}};
    \node [program, below of = Neigs] (Nswi) {\strut\textsf{swinterf}};
    \node [outfile, below of = Nswi] (Nchan) {\strut\textit{channel data}};
    \node [outfile, below of = Nchan] (Namps) {\strut\textit{boundary amps.}};
    \node [number, left = 0mm of Nsci] (N1) {\textsf{\small 1}};
    \node [number, left = 0mm of Ncon] (N2) {\textsf{\small 2}};
    \node [number, left = 0mm of Nsca] (N3) {\textsf{\small 3}};
    \node [number, left = 0mm of Nden] (N4) {\textsf{\small 4}};
    \node [number, right = 0mm of Ncons] (N5) {\textsf{\small 5}};
    \node [number, right = 0mm of Nscas] (N6) {\textsf{\small 6}};
    \node [number, right = 0mm of Nswi] (N7) {\textsf{\small 7}};
    \draw[orange!60, line width = 0.5mm] ($(Ncon.north west)+(-1.0,0.2)$)  rectangle ($(Nsca.south east)+(1.0,-0.2)$);
    \draw[black!20, line width = 0.5mm] ($(Ncons.north west)+(-1.0,0.2)$)  rectangle ($(Namps.south east)+(0.3,-0.2)$);
    \draw [inarrow, ->] (Nmol) --  node { } (Nsci) ;
    \draw [outarrow, ->] (Nsci) --  node { } (Nmoi) ;
    \draw [outarrow, ->] (Ncon) --  node { } (Ncsf) ;
    \draw [inarrow, ->] (Nmoi) --  node { } (Nsca) ;
    \draw [inarrow, ->] (Ncsf) --  node { } (Nsca) ;
    \draw [outarrow, ->] (Nsca) --  node { } (Neig) ;
    \draw [outarrow, ->] (Ncons) --  node { } (Ncsfs) ;
    \draw [inarrow, ->] (Nmoi) --  node { } (Nscas) ;
    \draw [inarrow, ->] (Ncsfs) --  node { } (Nscas) ;
    \draw [outarrow, ->] (Nscas) --  node { } (Neigs) ;
    \draw [inarrow, ->] (Neig) --  node { } (Nscas) ;
    \draw [inarrow, ->] (Neig) --  node { } (Nden) ;
    \draw [inarrow, ->] (Nmoi) --  node { } (Nden) ;
    \draw [outarrow, ->] (Nden) --  node { } (Nprop) ;
    \draw [inarrow, ->] (Nmoi) --  node { } (Nswi) ;
    \draw [inarrow, ->] (Nprop) --  node { } (Nswi) ;
    \draw [inarrow, ->] (Neigs) --  node { } (Nswi) ;
    \draw [outarrow, ->] (Nswi) --  node { } (Nchan) ;
    \path (Namps) -- (Nchan) node[midway,anchor=center] (Nand) {\textbf{+}};
    \path (Nchan) -- node (pt1) [below=0.5cm] {} (Namps);
    \path (Nchan) -- node (pt2) [below=2.0cm] {} (Namps);
    \node [below of = Namps, draw = white, fill = white] (Nouter) {UKRmol-out};
    \draw [outarrow, line width = 1.0 mm, >=latex, ->] (Namps) -- node {} (Nouter);
\end{tikzpicture}
    \caption{Workflow for the inner region and interface parts of a scattering calculation. Red indicates the molecular orbital (Molden) file produced by a compatible quantum chemistry package, white are intermediate files, green are the final outputs. Programs in the orange  box need to be executed for all the required (target) space-spin symmetries and those in the grey box for all (scattering) space-spin symmetries. Each program needs also its own input namelist(s) not shown in the diagram. The numbers next to individual programs indicate  the typical execution order.}
    \label{fig:workflow-scat}
\end{figure}

A typical workflow for the part of a scattering calculation using the \UKRmol{}-in programs together with the interface is illustrated in Fig.~\ref{fig:workflow-scat}. 

For SE and SEP calculations, the determination of the target properties is straightforward and the only significant choice is that of the basis set. For the inner region part of the calculation, the main choice is that of the virtual orbitals to be used in the construction of the L$^2$ functions (see Section~\ref{sec:theory}). The R-matrix radius will be given by the orbitals used and can be checked with \textprog{SCATCI\_INTEGRALS} (using the option \textcode{calc\_radial\_densities}) and the choice of the type of continuum basis is linked to it and the scattering energies to be investigated. Since scattering calculations are normally performed for energies up to no more than few eV higher than the ionization threshold, the energies to be covered are normally up to around 20~eV. This means that, in general, for R-matrix radii smaller than 15~a$_0$, GTOs only will suffice (see, for example, \cite{alanine,tarana}). Use of BTOs only as well as mixed bases have been tested for diatomic and triatomic targets \cite{Darby_Lewis_2017,SuG}. The choice of type of continuum basis is independent of the scattering model being used. The interface (\textprog{SWINTERF}) calculation is also straightforward, with the only significant choice being whether to use only  dipoles or include quadrupole moments too in the expansion of the static multipole molecular potential.

For CC calculations, the choices in terms of describing the target are several, and more difficult to make.
The first choice is that of the orbitals to be used: it is now customary to use State-Averaged  CASSCF orbitals generated by an external Quantum Chemistry code. An active space, and the states to be averaged need to be specified in the input to those codes. The former choice is usually limited, due to the requirement for balance \cite{jt474} between the N and N+1 electron wavefunctions,  by the number of N+1 electron CSFs that a specific active space will generate; this is particularly the case for molecules with large polarizabilites that require the use of several types of L$^2$ functions. 

The states to be averaged will normally be a subset of those included in the scattering calculation: usually, a few active spaces and state-averaging schemes are tested and the excitation thresholds and the ground state energy and dipole moment (if the molecule is polar) of the target are checked to decide on the best choice. 

We note that the symmetry of the target molecules is critical here: for a molecule with no symmetry, the size of the Hamiltonian to diagonalize is given by the total number of CSFs. For a molecule belonging to the D$_{2h}$ this number will be split roughly equally between the 8 irreducible representations leading to 8, much smaller, Hamiltonian blocks that need to be diagonalized (for this reason, it can be convenient to study higher symmetry  molecules as models for more complex targets \cite{MaG12}).

For the inner region calculation, one needs to choose, once again, the R-matrix radius and continuum basis to use: the same considerations as for SE/SEP calculations apply here. Under certain conditions (e.g. large number of bound orbitals, high-quality continuum description involving BTOs, large R-matrix radii)  \textprog{SCATCI\_INTEGRALS} can become the computationally heaviest part of a calculation.

In terms of the target states to include in the construction of the N+1 electron wavefunctions, it is customary to use all those that are energetically accessible in the scatting energy range of interest (if pseudostates are used, other considerations, like a good description of polarization effects, should be taken into account).

Increasing the number of target states does not contribute to the size of the Hamiltonian as much as increasing the number of L$^2$ function (by, for example, choosing a bigger active space). This increase, however, will have an effect in the outer region calculation since the number of channels is linked to the number of target states. 

\subsection{\UKRmol{}-out}
\label{sec:ukrmol-out}

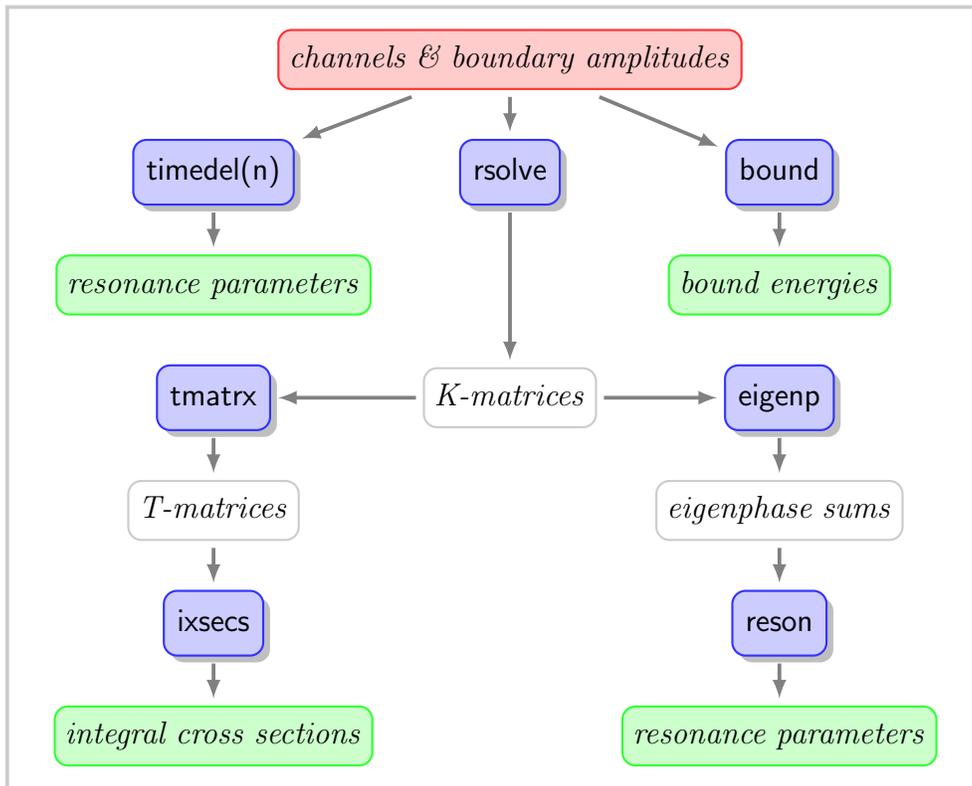
\begin{figure}[htbp]
    \centering
    \tikzstyle{program}=[rectangle, thick, minimum size=0.5cm, draw=blue!80, fill=blue!20, rounded corners = 5pt, inner sep = 5pt, drop shadow]
\tikzstyle{infile}=[rectangle, thick, minimum size=0.5cm, draw=red!80, fill=red!20, rounded corners = 5pt]
\tikzstyle{imtfile}=[rectangle, thick, minimum size=0.5cm, draw=black!20, fill=white!20, rounded corners = 5pt]
\tikzstyle{outfile}=[rectangle, thick, minimum size=0.5cm, draw=green!80, fill=green!20, rounded corners = 5pt]
\tikzstyle{number}=[circle, inner sep=1pt, minimum size=0mm, draw=blue!80, fill=blue!10]
\tikzstyle{inarrow}=[black!50, line width = 0.5mm, >=latex]
\tikzstyle{outarrow}=[black!50, line width = 0.5mm, >=latex]

\begin{tikzpicture}[auto, outer sep=3pt, node distance=1.5cm]
    \node [infile] (Namp) {\strut\textit{channels \& boundary amplitudes}};
    \node [program, below of = Namp] (Nrslv) {\strut\textsf{rsolve}};
    \node [program, left = 2cm of Nrslv] (Ntmdl) {\strut\textsf{timedel(n)}};
    \node [program, right = 2cm of Nrslv] (Nbnd) {\strut\textsf{bound}};
    \node [outfile, below of = Ntmdl] (Nrest) {\strut\textit{resonance parameters}};
    \node [below of = Nrslv, draw = none] (Ndummy) {};
    \node [imtfile, below of = Ndummy] (Nkmat) {\strut\textit{K-matrices}};
    \node [outfile, below of = Nbnd] (Nbene) {\strut\textit{bound energies}};
    \node [program, below of = Nrest] (Ntmx) {\strut\textsf{tmatrx}};
    \node [program, below of = Nbene] (Negp) {\strut\textsf{eigenp}};
    \node [imtfile, below of = Ntmx] (Ntmat) {\strut\textit{T-matrices}};
    \node [imtfile, below of = Negp] (Nesum) {\strut\textit{eigenphase sums}};
    \node [program, below of = Ntmat] (Nixs) {\strut\textsf{ixsecs}};
    \node [program, below of = Nesum] (Nrsn) {\strut\textsf{reson}};
    \node [outfile, below of = Nixs] (Ncs) {\strut\textit{integral cross sections}};
    \node [outfile, below of = Nrsn] (Nres) {\strut\textit{resonance parameters}};
    \draw[black!20, line width = 0.5mm] ($(Namp.north west)+(-3.5,0.2)$)  rectangle ($(Nres.south east)+(0.5,-0.2)$);
    \draw [inarrow, ->] (Namp) --  node { } (Nrslv) ;
    \draw [inarrow, ->] (Namp) --  node { } (Ntmdl) ;
    \draw [inarrow, ->] (Namp) --  node { } (Nbnd) ;
    \draw [outarrow, ->] (Ntmdl) --  node { } (Nrest) ;
    \draw [outarrow, ->] (Nrslv) --  node { } (Nkmat) ;
    \draw [outarrow, ->] (Nbnd) --  node { } (Nbene) ;
    \draw [inarrow, ->] (Nkmat) --  node { } (Ntmx) ;
    \draw [inarrow, ->] (Nkmat) --  node { } (Negp) ;
    \draw [outarrow, ->] (Ntmx) --  node { } (Ntmat) ;
    \draw [outarrow, ->] (Negp) --  node { } (Nesum) ;
    \draw [inarrow, ->] (Ntmat) --  node { } (Nixs) ;
    \draw [inarrow, ->] (Nesum) --  node { } (Nrsn) ;
    \draw [outarrow, ->] (Nixs) --  node { } (Ncs) ;
    \draw [outarrow, ->] (Nrsn) --  node { } (Nres) ;
\end{tikzpicture}
    \caption{Workflow for the outer region of a scattering calculation. The channel information and boundary amplitudes produced by the inner region (Fig.~\ref{fig:workflow-scat}) are used as input. This whole workflow (or its subset if not all output data are needed) is repeated for all space-spin symmetries relevant for the calculation. The color coding is as in Fig.~\ref{fig:workflow-scat}.}
    \label{fig:workflow-scat-out}
\end{figure}

A  workflow for the part of a scattering calculation using the \UKRmol{}-out programs is illustrated in Fig.~\ref{fig:workflow-scat-out}. Unlike the target and inner region part of the calculations, where most programs are run in a predetermined sequence, which \UKRmol{}-out programs are run depends on the scattering data the user requires.  

The first step in most  outer region calculations (the most significant exception is the study of bound states; see next section)  is the propagation of the R-matrix from the R-matrix boundary to an asymptotic region where, by matching to asymptotic expressions \cite{nn84}, the K-matrices can be determined. It is this step, performed by the program \textprog{RSOLVE}, that becomes resource consuming when a large number of channels are included in the calculation. For this reason, a parallelized version of it, \textprog{MPI-RSOLVE} is also available (it requires identical input to the serial version). In addition, the highly efficient, parallel program PFARM \cite{pfarm} can also be used for large calculations to perform the propagation and generate the K-matrices; the suite provides an interface program, \textprog{PFARM\_INTERFACE}, that ensures the inner region data is in the correct format for input to PFARM. The namelist of  \textprog{RSOLVE}/\textprog{MPI-RSOLVE}  is described in Table~\ref{tab:rsolve-input}. For large calculations it may be desirable to compute directly the T-matrices for the selected transitions of interest rather than full K-matrices: additional namelist variables available only in \textprog{MPI-RSOLVE} allow for this mode to be selected, see source code.

\begin{table}[htbp]
    \caption{Selected namelist variables in the namelist \textcode{\&rslvin} for \textprog{RSOLVE}.}
    \centering
    \begin{tabular}{ll}
        \toprule
        \textcode{\&rslvin} & \\
        \midrule
        \textcode{mgvn} &  Symmetry (irreducible representation) of scattering system  \\
        & (as in \textprog{SWINTERF})  \\
        \textcode{stot} & Spin multiplicity of the scattering system (as in \\ & \textprog{SWINTERF}) \\
        \textcode{nerang} &  Number of subranges of scattering energies \\
        \textcode{nescat} &  Number of input scattering energies in each subrange\\
        \textcode{einc}(1,$i$ ) & Initial energy in subrange $i$  \\
        \textcode{einc}(2,$i$ ) & Energy increment (step) in subrange $i$   \\
        \textcode{range ienut} &  Units in which  scattering energies are input; 1= Ryd \\
        & (default),  2= eV  \\
		\textcode{luchan} & Input file unit containing the channel information \\
		& (default: 10)\\
		\textcode{lurmt} & Input file unit for the R-matrix poles, boundary amplitudes \\
		& and coefficients of the multipole potential (default: 21)\\
			\textcode{lukmt} & Output file unit for the K-matrices \\
		\textcode{icform} &  Format flag for input channel data file, 'F' for formatted,  \\
        &  'U' for unformatted (default) \\
         \textcode{ikform} &  Format flag for output K-matrix data file, 'F' for formatted,   \\
        &  'U' for unformatted (default) \\
        \textcode{irform} &  Format flag for input R-matrix data file, 'F' for formatted,  \\
        &  'U' for  unformatted (default) \\
        \bottomrule
    \end{tabular}
    \label{tab:rsolve-input}
\end{table}

In order to perform the propagation, \textprog{RSOLVE} calls the propagator package \textprog{RPROP} \cite{mor84}. Input for this package is provided via the namelist \textcode{\&bprop} and includes the propagation radius \textcode{raf} and a flag \textcode{igail} that determines the type of asymptotic expansion used (0=Burke+Schey \cite{BSch}, 1=Gailitis \cite{gai76},  2=Bessel/Coulomb functions; default is 1).

\UKRmol{}-out contains a number of routines that use the K-matrices to generate scattering data: \textprog{TMATRX} calculates T-matrices that are then used to calculate integral cross sections using \textprog{IXSECS}. Cross section can be determined  for a set of chosen initial and final states. \textprog{EIGENP} calculates the eigenphase sum that can then be fitted by \textprog{RESON} \cite{RESON} to determine resonance positions and widths.   Another way of determining the energy and width of resonances is by fitting the largest eigenvalues of the time-delay matrix. \textprog{TIMEDEL} \cite{stibbe_timedel:program_1998} (and it's parallel implementation TIMEDELn \cite{TIMEDELn}, not distributed with the \UKRmol{+} suite, but available for download) can either use existing K-matrices or call \textprog{RSOLVE} to calculate K-matrices for an adaptive energy grid in order to calculate time-delays (an alternative version of a couple of subroutines is provided with the \UKRmol{}-out suite for use with TIMEDELn for this purpose).  Finally, for ionic targets the K-matrices generated can be fed into \textprog{MCQD} \cite{jt63} which computes 
(complex) multichannel quantum defects at each threshold in the calculation. Use  
\textprog{BOUND} to calculate bound states of the N+1 electron system is discussed below.

The CMake files provided will compile each program individually but will also generate an executable, called \textprog{OUTER}, that includes the most frequently run routines:   the interface \textprog{SWINTERF}, \textprog{RSOLVE},  \textprog{EIGENP},  \textprog{TMATRX}, \textprog{IXSEC} and  \textprog{RSOLVE}.

It is, of course, possible to use the K- or T-matrices obtained in a \UKRmol{} calculation as input for other programs. One that is regularly used is  POLYDCS \cite{polydcs} that enables the calculation of (electronically) elastic rotationally resolved  differential  and integral   cross sections. Use of this code also allows for the inclusion of a Born-approximation based correction \cite{gianturco_theory_1986} to the integral elastic cross sections, to account for the higher partial waves not included in the R-matrix calculation.

Elastic integral and differential (using an external program) cross sections can be calculated very accurately, although polar molecules present more of a problem \cite{RAW16,jt464}. The suite is able to model electronic excitation for small and mid-size molecules very accurately as well as describe their core-excited resonances \cite{RAMG16,thiophene}.
Among the largest targets studied (in terms of number of atoms in the system) are the molecular clusters pyridine-(H$_2$O)$_5$ and thymine-(H$_2$O)$_5$ for which SE and SEP calculations have been performed \cite{SiG17a,SiG17b}. 

\subsection{Bound states}
\label{sec:bound}

While \UKRmol+ is primarily a scattering code it also has the capability to find bound states. This can be useful for photoionization calculations (see next section) which
require (bound) target wavefunctions computed using the same model as the ionized
wavefunctions. However, given that one is most often interested in computing the photoionization cross sections for compact, deeply bound (ground) states, it is usually not necessary to consider the outer region contribution to their wavefunction. Conversely, highly-excited states of molecules become increasingly diffuse and Rydberg-like. Such states can be much more easily characterised using a negative energy scattering procedure than by standard quantum chemistry methods;
for example Little and Tennyson computed potential energy
curves of N$_2$ up to states with $n=10$ when standard electronic
structure procedures struggled to give full results even for
the $n=4$ states \cite{jt560}.

The original implementation of bound state finding, in outer
region module \textprog{BOUND}, was by Sarpal {\it et al.} \cite{jt106} which detected
bound states using the method of Seaton \cite{Seaton85}.
This implementation has been improved by both the use of a quantum-defect-based search grid \cite{jt184} and improved computation of outer-region wavefunctions \cite{jt491} using
a Runge-Kutta-Nystrom procedure \cite{RKN}. The need to compute
wavefunctions in the outer region makes bound state finding
numerically less stable than standard scattering calculations.
For this reason it is recommended that the range of the R-matrix
propagation is restricted to values below 50~a$_0$; $a=30$~a$_0$ is
typical whereas in calculations aimed at determining cross sections and resonance parameters it is more usual to propagate out to 100~a$_0$, sometimes further.

There is one further use for bound state finding which is important
for positron collisions. Positrons can annihilate with target electrons giving an effective annihilation parameter $Z_{\rm eff}$ which can be determined experimentally. Bound state wavefunctions
are required for the computation of $Z_{\rm eff}$ which was done
in the UKRmol outer region module \textprog{ZEFF}, see Zhang {\it et al.} \cite{jt491,jt510} for further details (\textprog{ZEFF} will be implemented in the suite in a forthcoming release).

\section{Photoionization calculations}
\label{sec:photo}
The capacity to calculate single photon ionization/recombination was added to the \UKRmol{} suite \cite{hbm14} since the last publication detailing its code base in 2012 \cite{epjd_ukrmol} and was used, via the Quantemol implementation \cite{jt601}, for a number of photodetachment studies  \cite{Slava16}.
The common need to treat higher photoelectron energies in photoionization/recombination applications provided one of the motivations for the development of \UKRmol+.
The first application of \UKRmol+ to photoionization was for NO$_2$ \cite{brambila2015}, where parallelization and the use of quad-precision allowed the photoelectron energy range to be doubled.
In a following study \cite{brambila2017} this calculation was extended to many different molecular geometries, and was then one of the key components used to model time-resolved laser controlled non-adiabatic dynamics in the vicinity of a conical intersection \cite{richter2019}. 
The \UKRmol{} CO$_2$ photoionization calculation \cite{hbm14,rouzee2014} were revisited, and extended to higher energies and the inclusion of many more target states, again made possible by parallelization and the use of quad-precision in \UKRmol+; the CO$_2$ recombination dipole matrix elements and Dyson orbitals (produced by \textprog{CDENPROP}) were then used to model the recombination step in high harmonic generation studies \cite{bruner2016,masin2018}.
Finally, recombination dipoles and Dyson orbitals for substituted benzenes were calculated for use in a study on the role of tunnel ionizaton in high harmonic generation \cite{austin2016}.
Tunnel ionization is sensitive to the exponential tail of the bound state wave function and the inclusion of diffuse continuum orbitals in the description of the Dyson orbitals was key for an accurate representation of this exponential decay. 

One-photon photoionization (in the perturbative regime) requires the calculation of transition dipole matrix elements between an initial N+1~electron bound state, $\Psineut{i}$, and a final continuum state, $\Psiscat{f}(\kfinalMOL;\Omega)$, with photoelectron momentum $\kfinalMOL$, and resultant ion in state $f$, 
\begin{equation} \label{e:photodipole1}
   \boldsymbol{d}_{fi}(\kfinalMOL;\Omega)=\braopket{\Psiscat{f}(\kfinalMOL;\Omega)}{\opdipole}{\Psineut{i}}.
\end{equation}
where the dipole operator, $\opdipole$, has 3 components, $d_p$, corresponding to the possible polarizations of the photon. 
The \UKRmol+ suite uses a molecular frame angular momentum basis for the continuum electron, so Eq.~(\ref{e:photodipole1}) has to be rewritten in the form
\begin{eqnarray}\label{e:photodipole2}
\boldsymbol{d}_{fi}(\kfinalLAB;\Omega)&=&\sum_{l}i^{-l}{e^{i\coulombphase{l}}}\MTXspharm{l}(\kdirLAB)
\MTXrotmat{l}(\Omega) \boldsymbol{d}_{fl,i}(E)\MTXrotmat{1\dagger}(\Omega),
\end{eqnarray} 
where the molecular frame partial wave dipoles (for a photoelectron with energy $E$), $\boldsymbol{d}_{fl,i}(E)$, have components
\begin{eqnarray}\label{e:partial_wave_dipole}
d_{p,flm,i}(E)=\sum_{k} A^{(-)*}_{flm,k}(E)\braopket{\psik{}{}}{d_p}{\Psineut{i}},
\end{eqnarray} 
and we have made explicit the representation of the final scattering state in terms of the inner region wave functions $\psik{}{}$ and the expansion coefficients of the scattering wave function in the inner region basis $A^{(-)}_{flm,k}(E)$ (see Eqs~(\ref{eq:inner_wfn_expansion}) and (\ref{e:rmat})). $\coulombphase{l}$ is the Coulomb phase, and $\MTXspharm{l}(\kdirLAB)$ is a vector of spherical harmonics $\spharm{l}{m}(\kdirLAB)$. $\MTXrotmat{l}(\Omega)$ are the Wigner rotation matrices and $\Omega=\euler$ are the Euler angles that relate lab and molecular frame (our conventions for these follow Brink and Satchler \cite{brink93}) and hence the orientation of the molecule. 

\begin{figure}[htbp]
    \centering
    \tikzstyle{program}=[rectangle, thick, minimum size=0.5cm, draw=blue!80, fill=blue!20, rounded corners = 5pt, inner sep = 5pt, drop shadow]
\tikzstyle{infile}=[rectangle, thick, minimum size=0.5cm, draw=red!80, fill=red!20, rounded corners = 5pt]
\tikzstyle{imtfile}=[rectangle, thick, minimum size=0.5cm, draw=black!20, fill=white!20, rounded corners = 5pt]
\tikzstyle{outfile}=[rectangle, thick, minimum size=0.5cm, draw=green!80, fill=green!20, rounded corners = 5pt]
\tikzstyle{number}=[circle, inner sep=1pt, minimum size=0mm, draw=blue!80, fill=blue!10]
\tikzstyle{inarrow}=[black!50, line width = 0.5mm, >=latex]
\tikzstyle{outarrow}=[black!50, line width = 0.5mm, >=latex]

\begin{tikzpicture}[auto, outer sep=3pt, node distance=1.5cm]
    \node [infile] (Nmol) {\strut\textit{molden file}};
    \node [program, below of = Nmol] (Nsci) {\strut\textsf{scatci\textunderscore integrals}};
    \node [imtfile, below of = Nsci] (Nmoi) {\strut\textit{molecular integrals}};
    \node [program, left = 2cm of Nmol] (Ncon) {\strut\textsf{congen}};
    \node [imtfile, below of = Ncon] (Ncsf) {\strut\textit{target CSFs}};
    \node [program, below of = Ncsf] (Nsca) {\strut\textsf{scatci}};
    \node [imtfile, below of = Nsca] (Neig) {\strut\textit{target states}};
    \node [program, right = 2cm of Nmol] (Ncons) {\strut\textsf{congen}};
    \node [imtfile, below of = Ncons] (Ncsfs) {\strut\textit{scattering CSFs}};
    \node [program, below of = Ncsfs] (Nscas) {\strut\textsf{scatci}};
    \node [imtfile, below of = Nscas] (Neigs) {\strut\textit{scattering states}};
    \node [below of = Nmoi, draw = none] (Ndummy1) {};
    \node [program, below of = Neig] (Nden) {\strut\textsf{denprop}};
    \node [imtfile, below of = Nden] (Nprop) {\strut\textit{target properties}};
    \node [program, below of = Ndummy1] (Ncden) {\strut\textsf{cdenprop}};
    \node [below of = Nprop, draw = none] (Ndummy2) {};
    \node [imtfile, below of = Ncden] (Ndips) {\strut\textit{inner dipoles}};
    \node [program, below of = Neigs] (Nswi) {\strut\textsf{swinterf}};
    \node [imtfile, below of = Nswi] (Namps) {\strut\textit{boundary amps.}};
    \node [imtfile, below of = Namps] (Nchan) {\strut\textit{channel data}};
    \node [program, below of = Ndips] (Nrslv) {\strut\textsf{rsolve}};
    \node [below of = Nchan, draw = none] (Ndummy3) {};
    \node [imtfile, below of = Nrslv] (Npwdip) {\strut\textit{outer dipoles}};
    \node [program, below of = Npwdip] (Ndipelm) {\strut\textsf{dipelm}};
    \node [below of = Nprop, draw = none] (Ndummy4) {};
    \node [below of = Ndummy4, draw = none] (Ndummy5) {};
    \node [below of = Ndummy5, draw = none] (Ndummy6) {};
    \node [below of = Ndummy3, draw = none] (Ndummy7) {};
    \node [outfile, below of = Ndummy6] (Ncs) {\strut\textit{cross sections}};
    \node [outfile, below of = Ndummy7] (Nbeta) {\strut\textit{asymmetry parameter}};
    \node [number, left = 0mm of Nsci] (N1) {\textsf{\small 1}};
    \node [number, left = 0mm of Ncon] (N2) {\textsf{\small 2}};
    \node [number, left = 0mm of Nsca] (N3) {\textsf{\small 3}};
    \node [number, left = 0mm of Nden] (N4) {\textsf{\small 4}};
    \node [number, right = 0mm of Ncons] (N5) {\textsf{\small 5}};
    \node [number, right = 0mm of Nscas] (N6) {\textsf{\small 6}};
    \node [number, right = 0mm of Nswi] (N7) {\textsf{\small 7}};
    \node [number, left = 0mm of Ncden] (N8) {\textsf{\small 8}};
    \node [number, left = 0mm of Nrslv] (N9) {\textsf{\small 9}};
    \node [number, left = 0mm of Ndipelm] (N10) {\textsf{\small 10}};
    \draw[orange!60, line width = 0.5mm] ($(Ncon.north west)+(-1.0,0.2)$)  rectangle ($(Nsca.south east)+(1.0,-0.2)$);
    \draw[purple!60, line width = 0.5mm] ($(Ncons.north west)+(-1.0,0.2)$)  rectangle ($(Nchan.south east)+(0.6,-0.2)$);
    \draw[blue!60, line width = 0.5mm] ($(Ncden.north west)+(-1.0,0.2)$)  rectangle ($(Npwdip.south east)+(0.6,-0.2)$);
    \draw [inarrow, ->] (Nmol) --  node { } (Nsci) ;
    \draw [outarrow, ->] (Nsci) --  node { } (Nmoi) ;
    \draw [outarrow, ->] (Ncon) --  node { } (Ncsf) ;
    \draw [inarrow, ->] (Nmoi) --  node { } (Nsca) ;
    \draw [inarrow, ->] (Ncsf) --  node { } (Nsca) ;
    \draw [outarrow, ->] (Nsca) --  node { } (Neig) ;
    \draw [outarrow, ->] (Ncons) --  node { } (Ncsfs) ;
    \draw [inarrow, ->] (Nmoi) --  node { } (Nscas) ;
    \draw [inarrow, ->] (Ncsfs) --  node { } (Nscas) ;
    \draw [outarrow, ->] (Nscas) --  node { } (Neigs) ;
    \draw [inarrow, ->] (Neig) --  node { } (Nscas) ;
    \draw [inarrow, ->] (Neig) --  node { } (Nden) ;
    \draw [inarrow, ->] (Nmoi) --  node { } (Nden.east) ;
    \draw [outarrow, ->] (Nden) --  node { } (Nprop) ;
    \draw [inarrow, ->] (Neigs.west) --  node { } (Ncden) ;
    \draw [inarrow, ->] (Neig.east) --  node { } (Ncden) ;
    \draw [inarrow, ->] (Nprop) --  node { } (Ncden) ;
    \draw [inarrow, ->] (Nmoi) --  node { } (Ncden) ;
    \draw [outarrow, ->] (Ncden) --  node { } (Ndips) ;
    \draw [inarrow, ->] (Nmoi) --  node { } (Nswi.west) ;
    \draw [inarrow, ->] (Nprop) --  node { } (Nswi) ;
    \draw [inarrow, ->] (Neigs) --  node { } (Nswi) ;
    \draw [outarrow, ->] (Nswi) --  node { } (Namps) ;
    \draw [inarrow, ->] (Nchan) --  node { } (Nrslv) ;
    \draw [inarrow, ->] (Namps) --  node { } (Nrslv) ;
    \draw [inarrow, ->] (Ndips) --  node { } (Nrslv) ;
    \draw [outarrow, ->] (Nrslv) --  node { } (Npwdip) ;
    \draw [inarrow, ->] (Npwdip) --  node { } (Ndipelm) ;
    \draw [outarrow, ->] (Ndipelm) --  node { } (Ncs) ;
    \draw [outarrow, ->] (Ndipelm) --  node { } (Nbeta) ;
    \path (Namps) -- (Nchan) node[midway,anchor=center] (Nand) {\textbf{+}};
\end{tikzpicture}
    \caption{Workflow for a photoionization calculation. Red indicates the molecular orbital (Molden) file produced by a compatible quantum chemistry package, white are intermediate files, green are the final outputs. Programs in the orange box need to be executed for all required (residual ion) spin-symmetries, while programs in the pink box need to be run for the initial neutral state  and all (final) spin-symmetries coupled to it by a component of the dipole operator. Programs in the blue box are run for each of the final spin-symmetries. Each program also needs  its own namelist input not shown in the diagram. The numbers next to individual programs indicate the typical execution order.}
    \label{fig:workflow-photo}
\end{figure}
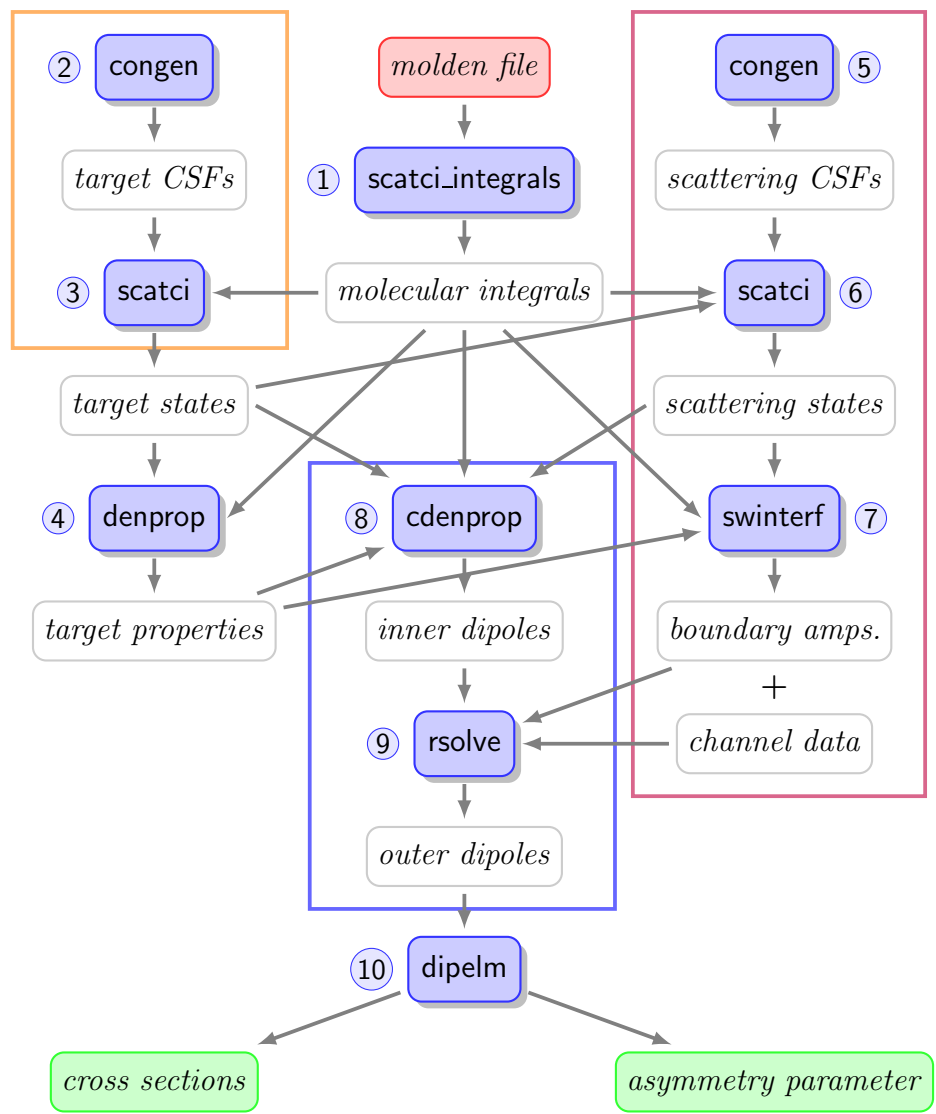

A typical workflow for photoionization is illustrated in Fig.~\ref{fig:workflow-photo}. Two programs belonging to the UKRmol+ suite are included: \textprog{RSOLVE} and \textprog{DIPELM}. The former  contains   routines  to determine the $A^{(-)}_{flm,k}(E)$ for each photoelectron energy $E$, and then use the transition dipoles from \textprog{CDENPROP}, $\boldsymbol{d}_{ki}=\braopket{\psik{}{}}{\opdipole}{\Psineut{i}}$ to construct the partial wave photoionization dipoles of Eq.~(\ref{e:partial_wave_dipole}) for input to \textprog{DIPELM}.
In the current implementation we consider the initial state to be fully contained within the R-matrix sphere.
This is a good approximation for the ground and low-lying excited states of many cations.

The additional variables in the  namelist input of \textprog{RSOLVE} needed for photoionization calculations are described in Table~\ref{tab:rsolve-photo-input}.

\begin{table}[htbp]
    \caption{Additional namelist variables for \textprog{RSOLVE} to be included in the namelist \textcode{\&rslvin}.}
    \centering
    \begin{tabular}{ll}
        \toprule
        \textcode{\&rslvin} & \\
        \midrule
        \textcode{calcak} & Number of final (ionic) states, for which to calculate scattering \\
         & wavefunction coefficients (default: 0). \\
        \textcode{calcdip} & Controls calculation of partial wave photoionization dipoles.\\
         & \textcode{calcdip(1)}: Calculate dipoles (= 1) or not (= 0, default). \\
         & \textcode{calcdip(2)}: The number of parent (neutral) states. \\
         & \textcode{calcdip(3)}: The number of final (ionic) states to consider.\\
        \bottomrule
    \end{tabular}
    \label{tab:rsolve-photo-input}
\end{table}

\textprog{DIPELM} calculates photoionization cross sections and asymmetry parameters.
These can be resolved for the final ionic state, molecular orientation, and photoelectron direction.
For photoionization from oriented molecules, the transformation from the angular momentum basis to the momentum basis (see Eq.~(\ref{e:photodipole2})) needs to be performed.
The photoelectron angular distribution is given by:
%
%
%
\begin{equation} \label{e:dcs_oriented}
   \frac{\mathrm{d}\sigma_{fi}}{\mathrm{d}\kfinalMOL}(\Omega)=C(
   \omega)|\boldsymbol{d}_{fi}(\kfinalMOL;\Omega)\cdot \boldsymbol{\hat \epsilon}|^2.
\end{equation}
where $C(\omega)$ is a photon energy dependent coefficient whose precise form depends on the gauge of the dipole operator (the current implementation uses the length gauge, $C(\omega)=4\pi^2\alpha a_0^2\omega$, where $\alpha$ is the fine structure constant and $a_0$ the Bohr radius), and $\boldsymbol{\hat \epsilon}$ is the polarization vector of the ionizing light.

For randomly oriented molecules, the energy dependent partial photoionization cross sections $\sigma_{fi}(E)$ and asymmetry parameters ($\beta_1(E)$ and $\beta_2(E)$) can be obtained from the partial wave dipoles of Eq.~(\ref{e:partial_wave_dipole}) (see, for example \cite{ritchie1976}). 
The photoelectron angular distribution is then:
\begin{equation} \label{e:dcs_averaged}
   \frac{\mathrm{d}\sigma_{fi}}{\mathrm{d}\theta}(E)=\frac{\sigma_{fi}(E)}{4\pi}[1+p\beta_1(E)\sin(\theta)+ \beta_2(E)\cos(\theta) ],
\end{equation}
where $p=0$ for linear polarization and $p=\pm 1$ for circular polarization. $\theta$ is the angle between the photoelectron emission direction and the photon polarization(/propagation) direction in the case of linear(/circular) polarization.
We note that $\beta_1(E)$ (also known as the dichroic parameter) only exists for the case of the photoionization of chiral molecules by circularly polarized light.

The default outputs of DIPELM are the partial photoionization cross sections and asymmetry parameters of Eq.~(\ref{e:dcs_averaged}).
Optionally, DIPELM also outputs the molecular (obtained by setting $\Omega=0$) and lab frame dipoles and photoelectron angular distributions of Eq.~(\ref{e:photodipole2}) and  Eq.~(\ref{e:dcs_oriented}).

The namelists controlling \textprog{DIPELM} are described in Table~\ref{tab:dipelm-input}. If only the default output is required (i.e. orientationally averaged cross sections and asymmetry parameters) then only the namelist \textcode{\&dipelminp} is needed.

\begin{table}[htbp]
    \caption{Selected parameters in the \textprog{DIPELM} input namelists:  \textcode{\&dipelminp} specifies the input file data,  
    \textcode{\&diptrans} controls the calculation of oriented photoionization observables and 
     \textcode{\&smooth}  allows for smoothing of the partial wave dipoles to remove narrow pseudoresonances (this will also remove narrow real resonances). }
    \centering
    \begin{tabular}{ll}
        \toprule
        \textcode{\&dipelminp} & \\
        \midrule
        \textcode{lu\_pw\_dipoles} & File unit(s) containing partial wave dipoles. \\
        \textcode{nset\_pw\_dipoles} & Set number(s) containing partial wave dipoles in \\ 
         & \textcode{lu\_pw\_dipoles}. \\
        \toprule
        \textcode{\&diptrans} & \\
        \midrule
        \textcode{ngrdproj} & No. of angular grid points (electron emission  \\ 
        & direction  $(N_\theta,N_\phi)$).  \\
        \textcode{ngrdalign} & No. of angular grid points (orientation $(N_\alpha,N_\beta,N_\gamma)$).  \\
        \textcode{scat\_angle\_limits} & Grid limits for photoelectron emission  \\
                                     & direction $(\theta_\mathrm{min},\phi_\mathrm{min}, \theta_\mathrm{max},\phi_\mathrm{max})$; default =\\
                                     & (0,0,180,360) \\
        \textcode{euler\_angle\_limits} & Grid limits for molecular orientation.\\
                                      & $(\alpha_\mathrm{min},\beta_\mathrm{min},\gamma_\mathrm{min}, \alpha_\mathrm{max},\beta_\mathrm{max},\gamma_\mathrm{max} )$; default =   \\
                                      & (0,0,0,360,180,360) \\
        %
        \toprule
        \textcode{\&smooth} & \\
        \midrule
        \textcode{ismooth} & Controls various smoothing options. \\
                         & 0 = No smoothing. \\
                         & 1 = Gaussian smoothing (no interpolation). \\
                         & 2 = Gaussian smoothing (with interpolation).\\
        \textcode{eleft} & Photoelectron energy to start smoothing from. \\
        
        \bottomrule
    \end{tabular}
    \label{tab:dipelm-input}
\end{table}

\subsection{\textprog{DIPOLE\_TOOLS}}

This program allows to post-process the partial wave dipoles, bound state dipoles (for the initial and the final states),  Dyson orbitals and package all this data in a formatted file for use in other programs. This program was used to generate data for the calculations in Refs.~\cite{richter_NO2_2019} and~\cite{bruner2016}.

The main functionality of this program lies in transforming the partial wave dipoles into momentum space in the molecular frame using the equation:
\begin{eqnarray}\label{e:molframe_dip}
d_{fi}^{q}(\mathbf{k}_{f}) = \sum_{l,m}X_{l,m}(\hat{\mathbf{k}}_{f}) d_{flm,i}(E) \imath^{-l} e^{\imath \sigma_{l}},
\end{eqnarray}
where $i$ and $f$ are indices of the initial and the final states, $q$ are spherical components of the photon polarization in the molecular frame, see Eq~(\ref{e:partial_wave_dipole}), $E$ is the total energy, $\sigma_{l}$ is the Coulomb phase and $\mathbf{k}_{f}$ is the final momentum of the photoelectron. In input for \textprog{dipole\_tools} the user specifies the initial and the final states, the angular grid of photoelectron directions and the range of photoelectron energies. 

The angular grid can be one of three types:
\begin{itemize}
    \item Lebedev grid of order $n$.
    \item Regular angular grid of $n$ points lying in one of zy, zx, xy planes.
    \item Custom grid of $n$ Cartesian points.
\end{itemize}{}

The program  also processes the Dyson orbitals generated by \textprog{CDENPROP}: it evaluates (a) the signs of the Dyson orbitals for the given radius and angular grid and optionally (b)  the orbital amplitudes in one of the chosen planes (xy, zy, zx).

Finally, the program allows phase corrections to be applied to  the initial (bound) states: this affects the phases of the final set of photoelectron dipoles, Dyson orbitals and the initial-state transition dipoles.

The input data for the program consists of
\begin{itemize}
    \item Partial wave dipole files for each irreducible representation dipole-coupled to the initial state as produced by \UKRmol{}-out (\textprog{RSOLVE}).
    \item Dyson orbitals saved in the \UKRmol+ format as output by \textprog{CDENPROP} (typically \textfile{dyson\_orbitals.ukrmolp}).
    \item Dyson orbitals saved in the \textprog{CDENPROP} format (typically \textfile{fort.123}).
    \item Property file for the final (ionic) states as produced by DENPROP (typically \textfile{fort.24}).
    \item Property file for the initial (bound) states as produced by CDENPROP (typically \textfile{fort.667}).
\end{itemize}

The program requires three namelists on input:
\begin{itemize}
    \item \textcode{\&dipelminp}: controls the input of the partial wave dipoles and is the same as the one required by \textprog{DIPELM}, see Table~\ref{tab:dipelm-input}.
    \item \textcode{\&input}: controls the input of the Dyson orbital files, property files, the selection of the initial and final states, phase correction factors for the bound states, the photoelectron energy grid and the precise format of the output data. The numerous but straightforward parameters are described in the source file \textfile{dipole\_tools.f90}.
    \item \textcode{\&anggrid}: specifies one of the three angular grids described above. The input parameters are described in the source file \textfile{dipole\_tools.f90}.
\end{itemize}

\subsection{\textprog{PHASE\_MATCH} and \textprog{PHASE\_MATCH\_ORBITALS}}

These programs are not part of the standard workflow of the scattering or photoionization calculation. Instead, they have been developed as a tool that can enable phase-matching of the partial wave photoionization matrix elements, see Eq.~(\ref{e:molframe_dip}), calculated for different geometries of the molecule. There are two arbitrary phases that can vary randomly with  the molecular geometry that manifest themselves in the partial wave photoionization amplitudes:
\begin{enumerate}
    \item Phase of the final (ionic) state(s) of the molecule.
    \item Phase of the initial (neutral) state of the molecule.
\end{enumerate}
When both of these phases are fixed, the photoionization amplitudes are a smooth function of geometry and can be used in various applications, e.g. in time-resolved photoelectron spectroscopy studies of ultrafast nuclear dynamics~\cite{richter_NO2_2019}. A concrete example of using these programs can be found in \cite{richter_NO2_2019}.

The program \textprog{PHASE\_MATCH} can be used to fix the phase of the final states by calculating the overlap (relative phase) between the final state wavefunctions of different geometries. The namelist \textcode{\&input} contains all the input parameters required  and is described in Table~\ref{tab:phase_match-input}. The program requires the output of the target calculations (molecular integrals file, the CSFs file, the CI vectors file -- i.e.  unit \textcode{nftw}  defined in Table~\ref{tab:scatci-cinorn} containing the eigenvectors) calculated for each geometry and supplied in the order in which the phase-matching will be done. It is recommended that the order in which these files are listed corresponds to the nearest-neighbour geometries. This choice typically maximises the reliability of the phase-matching process.

\begin{table}[htbp]
    \caption{Namelist variables for the program \textprog{PHASE\_MATCH}. 
    }
    \centering
    \begin{tabular}{ll}
        \toprule
        \textcode{\&input} & \\
        \midrule
        \textcode{n\_geom} & Number of geometries.\\
        \textcode{moints}  & List of the \textcode{n\_geom} file names of the molecular integral\\
                           & files produced by \textprog{SCATCI\_INTEGRALS}.\\
        \textcode{lucivec} & List of the \textcode{n\_geom} unit numbers containing the CI\\
                           & vectors in the format produced by \textprog{SCATCI}. \\
        \textcode{nciset } & List of the \textcode{n\_geom} set numbers on \textcode{lucivec}\\
                           & corresponding to sets of CI vectors to match.\\
        \textcode{nciset}  & List of the \textcode{n\_geom} unit numbers corresponding\\
                           & to the CSF files produced by \textprog{CONGEN}.\\
        \textcode{energies\_file} & Name of the file where the energies of the\\
                                  & phase-matched states will be saved.\\
        \textcode{replace\_with\_phase} &  if set to \textcode{.true.} the phase-matched CI vectors will   \\
        \textcode{\_corrected\_vectors}  & replace the original eigenvectors saved on \textcode{lucivec} \\
        \bottomrule
    \end{tabular}
    \label{tab:phase_match-input}
\end{table}

The program makes use of GBTOlib to evaluate the overlap integrals between the molecular orbitals from the neighbouring geometries. This overlap matrix is then used, together with the Slater rules for non-orthogonal orbital sets~\cite[p.~140-141]{mayer2003}, to calculate overlap integrals between the multiconfigurational wavefunctions representing the final states. The sign of the overlap is used to phase correct the wavefunction for the geometry currently processed. Optionally, see Table~\ref{tab:phase_match-input}, the phase-matched CI vectors can be saved back to disk replacing the original set of CI vectors. It is highly recommended that the user carefully inspects the generated file (named using the variable \textcode{energies\_file} in the namelist) which lists the energies of the phase-matched states: smooth energy curves for the states whose phases are required signifies that no mismatch has  occurred.

The program \textprog{PHASE\_MATCH\_ORBITALS} can be used to phase-match orbitals from different geometries. This program has been used to phase-match the Dyson orbitals (i.e. single particle orbitals obtained calculating the overlap integrals between the initial (N+1)-electron and the final N-electron wavefunctions) generated by \textprog{CDENPROP} from photoionization calculations which are using the phase-consistent final state wavefunctions obtained by \textprog{PHASE\_MATCH}.  

The input for \textprog{PHASE\_MATCH\_ORBITALS} is straightforward and is summarized in Table~\ref{tab:phase_match_orbitals-input}. The program operates similarly to \textprog{PHASE\_MATCH} by calculating overlap integrals between the orbitals from neighbouring geometries saved on separate integral files produced by \textprog{SCATCI\_INTEGRALS}. However, unlike \textprog{PHASE\_MATCH}, the program by default does not perform automatic matching of the orbitals from the neighbouring geometries: instead the user provides, via the namelist \textcode{\&input}, a list of orbitals for each geometry which will be compared directly against the set from the previous geometry. The reason for this set-up is that in case of Dyson orbitals their norm can vary from geometry to geometry which could potentially lead to a mismatch of orbitals since the (optional) matching orbital is found as the one with the largest magnitude of the overlap integral. A sign change of the overlap of the Dyson orbitals from two neighbouring geometries can be used to phase-correct the corresponding initial neutral state (if the ionic states have been phase-matched before). The output of the program is the list of phase-correction factors for each orbital and geometry. The orbitals themselves are not modified.

\begin{table}[htbp]
    \caption{Namelist variables for the program \textprog{PHASE\_MATCH\_ORBITALS}. The linear arrays \textcode{orbital\_num} and \textcode{orbital\_sym}  emulate 2D arrays with dimensions (\textcode{n\_orbitals\_to\_match},\textcode{n\_geom}).}
    \centering
    \begin{tabular}{ll}
        \toprule
        \textcode{\&input} & \\
        \midrule
        \textcode{n\_geom} & Number of geometries\\
        \textcode{n\_orbitals\_to\_match} & Number of orbitals to be matched.\\
        \textcode{moints} & List of \textcode{n\_geom} file names of the molecular integral\\
        &files produced by \textprog{SCATCI\_INTEGRALS}.\\
        \textcode{orbital\_num} & Linear array containing indices, for each geometry,\\
        & of the orbitals selected for phase-matching.\\
        \textcode{orbital\_sym} & Linear array containing the symmetries,\\
        &for each geometry, of the orbitals indicated by \\
        & \textcode{orbital\_num}.\\
        \textcode{find\_matching\_orbitals} & Selects automatic matching of orbitals from \\
        & the neighbouring geometries, see text.\\
        \bottomrule
    \end{tabular}
    \label{tab:phase_match_orbitals-input}
\end{table}

\section{Producing input for R-matrix with time (RMT) calculations}
\label{sec:rmt}
The RMT (R-matrix with time) code \cite{RMT} solves the time-dependent Schr\"odinger equation to describe molecules in intense, ultrashort, arbitrarily-polarized laser pulses. It requires the following data from \UKRmol+:
\begin{itemize}
    \item elements of the dipole operator between all pairs of inner region, (N+1)-electron states (produced by \textprog{CDENPROP\_ALL}),
    \item elements of the dipole operator, as produced by \textprog{DENPROP}, between all pairs of final (N-electron) states,
    \item energies of both the N+1-electron and N-electron states (produced by \textprog{SCATCI}/\textprog{MPI-SCATCI}),
    \item the list of outer region channels and coefficients for reconstruction of the long-range potentials that couple the channels (produced by \textprog{SWINTERF}; see Appendix~A in \cite{RMT} for a detailed description of these),
    \item the (boundary) amplitudes \(w_{ik}(r_j)\) of the inner eigenstate wave functions at several radii smaller than and equal to the radius of the R-matrix sphere (produced by \textprog{SWINTERF}) for maintaining the continuity between inner and outer region
\end{itemize}
and some other auxiliary data computed directly by the \textprog{RMT\_INTERFACE} program itself. In order to generate the  RMT data, one must perform a ``target'' run, where eigenstates and properties of the ionized (N-electron) molecule are evaluated, and a ``scattering'' run, where similar data are obtained for the neutral (N+1-electron) system.  As discussed in Section 3.6.2 the R-matrix radius needs to be larger than in similar scattering/photoionization calculations.

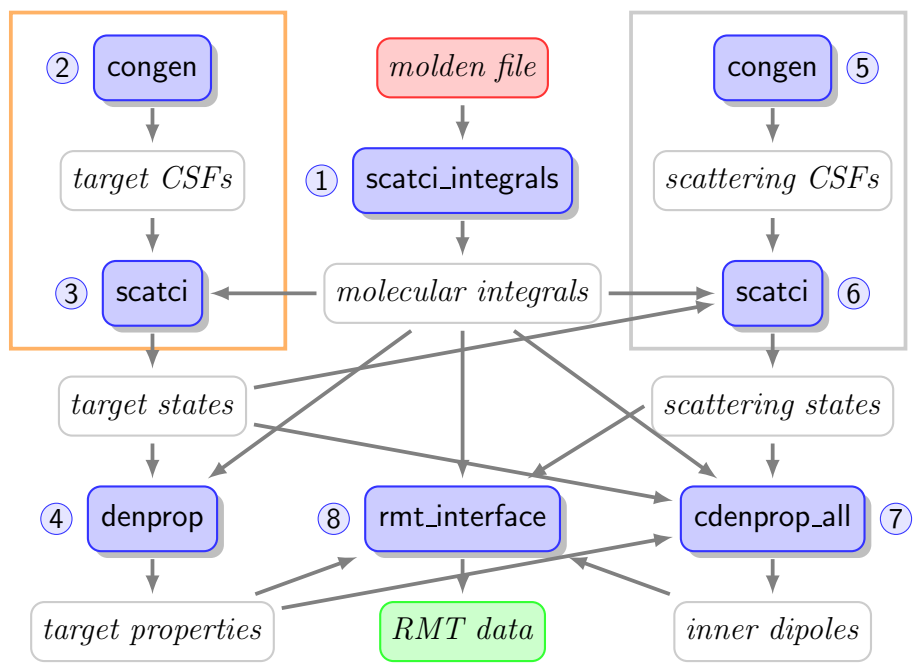
\begin{figure}[htbp]
    \centering
    \tikzstyle{program}=[rectangle, thick, minimum size=0.5cm, draw=blue!80, fill=blue!20, rounded corners = 5pt, inner sep = 5pt, drop shadow]
\tikzstyle{infile}=[rectangle, thick, minimum size=0.5cm, draw=red!80, fill=red!20, rounded corners = 5pt]
\tikzstyle{imtfile}=[rectangle, thick, minimum size=0.5cm, draw=black!20, fill=white!20, rounded corners = 5pt]
\tikzstyle{outfile}=[rectangle, thick, minimum size=0.5cm, draw=green!80, fill=green!20, rounded corners = 5pt]
\tikzstyle{number}=[circle, inner sep=1pt, minimum size=0mm, draw=blue!80, fill=blue!10]
\tikzstyle{inarrow}=[black!50, line width = 0.5mm, >=latex]
\tikzstyle{outarrow}=[black!50, line width = 0.5mm, >=latex]

\begin{tikzpicture}[auto, outer sep=3pt, node distance=1.5cm]
    \node [infile] (Nmol) {\strut\textit{molden file}};
    \node [program, below of = Nmol] (Nsci) {\strut\textsf{scatci\textunderscore integrals}};
    \node [imtfile, below of = Nsci] (Nmoi) {\strut\textit{molecular integrals}};
    \node [program, left = 2cm of Nmol] (Ncon) {\strut\textsf{congen}};
    \node [imtfile, below of = Ncon] (Ncsf) {\strut\textit{target CSFs}};
    \node [program, below of = Ncsf] (Nsca) {\strut\textsf{scatci}};
    \node [imtfile, below of = Nsca] (Neig) {\strut\textit{target states}};
    \node [program, right = 2cm of Nmol] (Ncons) {\strut\textsf{congen}};
    \node [imtfile, below of = Ncons] (Ncsfs) {\strut\textit{scattering CSFs}};
    \node [program, below of = Ncsfs] (Nscas) {\strut\strut\textsf{scatci}};
    \node [imtfile, below of = Nscas] (Neigs) {\strut\textit{scattering states}};
    \node [program, below of = Neig] (Nden) {\strut\textsf{denprop}};
    \node [below of = Nmoi, draw = none] (Ndummy) {};
    \node [below of = Ndummy, draw = none] (Ndummy2) {};
    \node [program, below of = Ndummy] (Nrmt) {\strut\textsf{rmt\textunderscore interface}};
    \node [imtfile, below of = Nden] (Nprop) {\strut\textit{target properties}};
    \node [program, below of = Neigs] (Ncden) {\strut\textsf{cdenprop\textunderscore all}};
    \node [imtfile, below of = Ncden] (Ndips) {\strut\textit{inner dipoles}};
    \node [outfile, below of = Nrmt] (Ndata) {\strut\textit{RMT data}};
    \node [number, left = 0mm of Nsci] (N1) {\textsf{\small 1}};
    \node [number, left = 0mm of Ncon] (N2) {\textsf{\small 2}};
    \node [number, left = 0mm of Nsca] (N3) {\textsf{\small 3}};
    \node [number, left = 0mm of Nden] (N4) {\textsf{\small 4}};
    \node [number, right = 0mm of Ncons] (N5) {\textsf{\small 5}};
    \node [number, right = 0mm of Nscas] (N6) {\textsf{\small 6}};
    \node [number, right = 0mm of Ncden] (N7) {\textsf{\small 7}};
    \node [number, left = 0mm of Nrmt] (N8) {\textsf{\small 8}};
    \draw[orange!60, thick, line width = 0.5mm] ($(Ncon.north west)+(-1.0,0.2)$)  rectangle ($(Nsca.south east)+(1.0,-0.2)$);
    \draw[black!20, thick, line width = 0.5mm] ($(Ncons.north west)+(-1.0,0.2)$)  rectangle ($(Nscas.south east)+(1.0,-0.2)$);
    \draw [inarrow,->] (Nmol) --  node { } (Nsci) ;
    \draw [outarrow, ->] (Nsci) --  node { } (Nmoi) ;
    \draw [outarrow, ->] (Ncon) --  node { } (Ncsf) ;
    \draw [inarrow, ->] (Ncsf) --  node { } (Nsca) ;
    \draw [outarrow, ->] (Ncons) --  node { } (Ncsfs) ;
    \draw [inarrow, ->] (Nmoi) --  node { } (Nsca) ;
    \draw [outarrow, ->] (Nsca) --  node { } (Neig) ;
    \draw [inarrow, ->] (Nmoi) --  node { } (Nscas) ;
    \draw [inarrow, ->] (Ncsfs) --  node { } (Nscas) ;
    \draw [outarrow, ->] (Nscas) --  node { } (Neigs) ;
    \draw [inarrow, ->] (Neig) --  node { } (Nscas) ;
    \draw [inarrow, ->] (Neig) --  node { } (Nden) ;
    \draw [inarrow, ->] (Nmoi) --  node { } (Nden) ;
    \draw [outarrow, ->] (Nden) --  node { } (Nprop) ;
    \draw [inarrow, ->] (Neigs) --  node { } (Ncden) ;
    \draw [inarrow, ->] (Neig) --  node { } (Ncden) ;
    \draw [inarrow, ->] (Nprop) --  node { } (Ncden) ;
    \draw [inarrow, ->] (Nmoi) --  node { } (Ncden) ;
    \draw [outarrow, ->] (Ncden) --  node { } (Ndips) ;
    \draw [inarrow, ->] (Nmoi) --  node { } (Nrmt) ;
    \draw [inarrow, ->] (Nprop) --  node { } (Nrmt) ;
    \draw [inarrow, ->] (Ndips) --  node { } (Nrmt) ;
    \draw [inarrow, ->] (Neigs.west) --  node { } (Nrmt) ;
    \draw [outarrow, ->] (Nrmt) --  node { } (Ndata) ;
\end{tikzpicture}
    \caption{Workflow for production of the RMT input file. Red indicates the molecular orbital (Molden) file produced by a compatible quantum chemistry package, white are intermediate files, green is the output file needed for input to RMT. Programs in the orange  box need to be executed for all the required (target) space-spin symmetries and those in the grey box for all (scattering) space-spin symmetries. (\textprog{RMT\_INTERFACE} requires all transition moments irrespective of whether only some are actually needed in RMT). Each program also needs  its own namelist input not shown in the diagram. The numbers next to individual programs indicate the typical execution order.}
    \label{fig:workflow-rmt}
\end{figure}

\begin{figure}[htbp]
    \centering
    \tikzstyle{program}=[rectangle, thick, minimum size=0.5cm, draw=blue!80, fill=blue!20, rounded corners = 5pt, inner sep = 5pt, drop shadow]
\tikzstyle{infile}=[rectangle, thick, minimum size=0.5cm, draw=red!80, fill=red!20, rounded corners = 5pt]
\tikzstyle{imtfile}=[rectangle, thick, minimum size=0.5cm, draw=black!20, fill=white!20, rounded corners = 5pt]
\tikzstyle{outfile}=[rectangle, thick, minimum size=0.5cm, draw=green!80, fill=green!20, rounded corners = 5pt]
\tikzstyle{number}=[circle, inner sep=1pt, minimum size=0mm, draw=blue!80, fill=blue!10]
\tikzstyle{inarrow}=[black!50, line width = 0.5mm, >=latex]
\tikzstyle{outarrow}=[black!50, line width = 0.5mm, >=latex]

\begin{tikzpicture}[auto, outer sep=3pt, node distance=1.5cm]
    \node [infile] (Nmol) {\strut\textit{molden file}};
    \node [program, below of = Nmol] (Nsci) {\strut\textsf{scatci\textunderscore integrals}};
    \node [imtfile, below of = Nsci] (Nmoi) {\strut\textit{molecular integrals}};
    \node [program, left = 2cm of Nmol] (Ncon) {\strut\textsf{congen}};
    \node [imtfile, below of = Ncon] (Ncsf) {\strut\textit{target CSFs}};
    \node [program, below of = Ncsf] (Nsca) {\strut\textsf{mpi-scatci}};
    \node [imtfile, below of = Nsca] (Neig) {\strut\textit{target states}};
    \node [program, right = 2cm of Nmol] (Ncons) {\strut\textsf{congen}};
    \node [imtfile, below of = Ncons] (Ncsfs) {\strut\textit{scattering CSFs}};
    \node [program, below of = Ncsfs] (Nscas) {\strut\textsf{mpi-scatci}};
    \node [outfile, below of = Nscas] (Ndata) {\strut\textit{RMT data}};
    \node [program, below of = Nmoi] (Nden) {\strut\textsf{denprop}};
    \node [imtfile, below of = Nden] (Nprop) {\strut\textit{target properties}};
    \node [number, left = 0mm of Nsci] (N1) {\textsf{\small 1}};
    \node [number, left = 0mm of Ncon] (N2) {\textsf{\small 2}};
    \node [number, left = 0mm of Nsca] (N3) {\textsf{\small 3}};
    \node [number, right = 0mm of Nden] (N4) {\textsf{\small 4}};
    \node [number, right = 0mm of Ncons] (N5) {\textsf{\small 5}};
    \node [number, right = 0mm of Nscas] (N6) {\textsf{\small 6}};
    \draw[orange!60, line width = 0.5mm] ($(Ncon.north west)+(-1.0,0.2)$)  rectangle ($(Ncsf.south east)+(0.5,-0.2)$);
    \draw[black!20, line width = 0.5mm] ($(Ncons.north west)+(-1.0,0.2)$)  rectangle ($(Ncsfs.south east)+(0.5,-0.2)$);
    \draw [inarrow, ->] (Nmol) --  node { } (Nsci) ;
    \draw [outarrow, ->] (Nsci) --  node { } (Nmoi) ;
    \draw [outarrow, ->] (Ncon) --  node { } (Ncsf) ;
    \draw [inarrow, ->] (Nmoi) --  node { } (Nsca) ;
    \draw [outarrow, ->] (Nsca) --  node { } (Neig) ;
    \draw [outarrow, ->] (Ncons) --  node { } (Ncsfs) ;
    \draw [inarrow, ->] (Ncsf) --  node { } (Nsca) ;
    \draw [inarrow, ->] (Nmoi) --  node { } (Nscas) ;
    \draw [inarrow, ->] (Ncsfs) --  node { } (Nscas) ;
    \draw [outarrow, ->] (Nscas) --  node { } (Ndata) ;
    \draw [inarrow, ->] (Neig) --  node { } (Nscas) ;
    \draw [inarrow, ->] (Neig) --  node { } (Nden) ;
    \draw [inarrow, ->] (Nmoi) --  node { } (Nden) ;
    \draw [outarrow, ->] (Nden) --  node { } (Nprop) ;
    \draw [inarrow, ->] (Nprop) --  node { } (Nscas) ;
\end{tikzpicture}
    \caption{Alternative workflow for production of the RMT input file using the outer interface built into \textprog{MPI-SCATCI}. The colour-coding and numbering are as in Fig.~\ref{fig:workflow-rmt}.}
    \label{fig:workflow-rmt-alternative}
\end{figure}
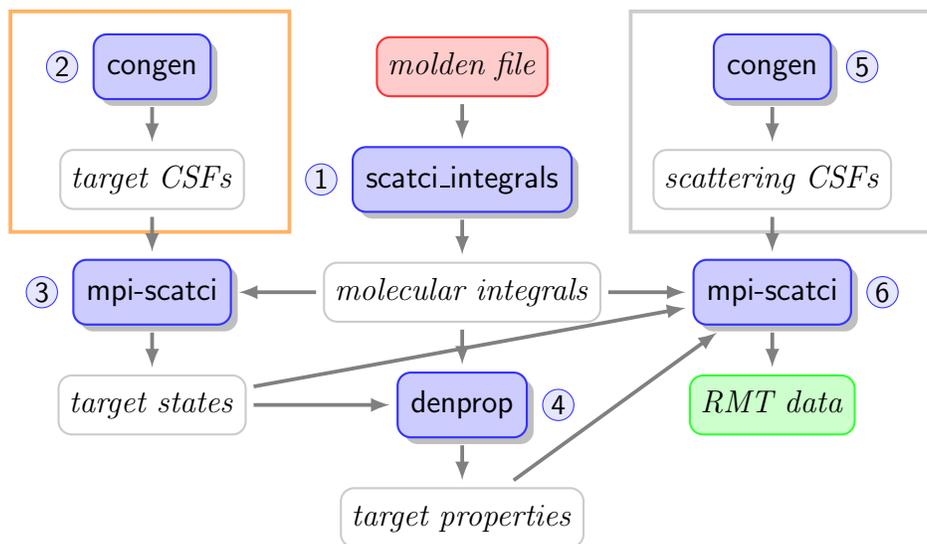

The two alternative workflows for generating RMT data are illustrated in the diagrams in Figs.~\ref{fig:workflow-rmt} and~\ref{fig:workflow-rmt-alternative}. The first of the two methods allows greater control over input parameters by exposing the full \textprog{SWINTERF} input namelists to the user for each scattering symmetry. The other method, on the other hand, has the advantage of not needing to write out the inner region eigenstates and their properties to disk, which can save considerable  storage space and speed up the full task. Nevertheless,  the resulting file \textfile{molecular\_data} tends to be large even for modest physical systems, so to achieve efficient writing even in distributed mode \textprog{MPI-SCATCI} uses MPI-IO facilities, where each process writes its own part of the distributed data directly to the proper location in the output file.

In the alternative method shown in Fig.~\ref{fig:workflow-rmt-alternative}, \textprog{MPI-SCATCI} is executed just twice: once for all ionic, N-electron, space-spin symmetries and once for all neutral, N+1-electron, space-spin symmetries. The input for each of these runs consists of directly concatenated inputs like those used for the corresponding \textprog{SCATCI} executions in Fig.~\ref{fig:workflow-rmt}, without any changes. Attention needs to be paid, though, to the numbering of the scratch file units, so that different irreducible representations data are not written to the same ones. In the ``scattering'' (neutral) calculation, the additional namelist \textcode{\&outer\_interface} is required, as explained earlier, to provide RMT-related inputs (see Tab.~\ref{tab:mpi-scatci-outer-interface}).

\section{Test suite}
\label{sec:test}

A test suite is provided with the \UKRmol+ release. It contains:

\begin{itemize}
    \item an extensive  set of electron scattering tests that cover all point groups for which calculations can be run 
    \item a photoionization test for H$_2$
    \item an H$_2$ run that generates input for RMT.
\end{itemize}

The electron scattering calculations can be run for all or each point group and from beginning to end or in stages: target, inner region and outer region. Both HF and CAS target model inputs are provided as well as  SEP and CC scattering inputs. In addition, when possible, the target calculations are performed for different orientations of the molecule with respect to the coordinate frame. All tests can be run using Molden files provided; the scattering tests  use GTO-only continuum basis sets whereas the photoionization and RMT ones use BTOs. The tests can either be run serially or as a parallel job (two processes). In the latter case, \textprog{SCATCI\_INTEGRALS} is run in parallel, \textprog{MPI-SCATCI} is used instead of \textprog{SCATCI} and \textprog{MPI-RSOLVE}, an  MPI-enabled version of \textprog{RSOLVE} is used in the outer region.

Details of how to use the test suite are provided with the release. Executing the suite will generate a set of output files that summarize the main data produced. A set of benchmark outputs is provided for comparison; it contains eigenvalues of the Hamiltonian for the N and N+1 electron systems as well as cross sections and eigenphase sums for each irreducible representation. The expected accuracy for the different data is also explained in the release. 



\section{Results}
\label{sec:results}

Here we present some  illustrative results of both scattering and photoionization calculations. For RMT results, we refer the reader to Brown \textit{et al}.~\cite{RMT}.

\subsection{Electronic excitation and core-excited resonances: thiophene}

\begin{figure}[htbp]
    \centering
    \includegraphics[width=10cm]{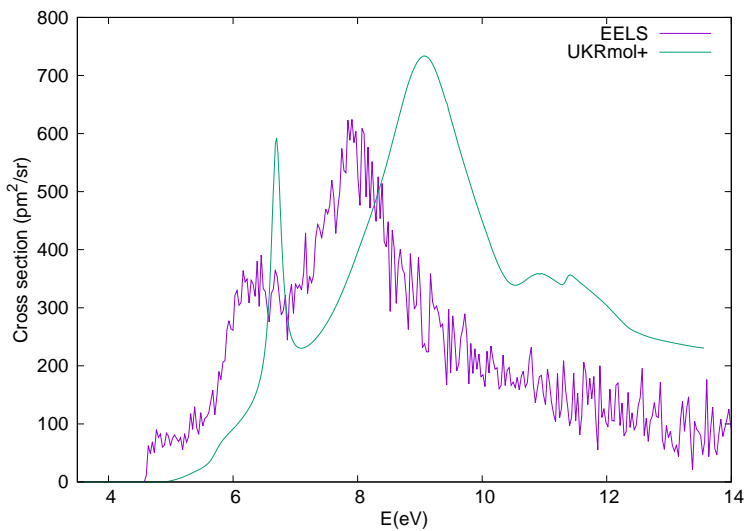}
    \caption{Cross sections for excitation  into the second triplet state of thiophene for an electron scattering angle of 90$^{\circ}$. The EEL spectrum was measured for an energy loss of 4.61~eV \cite{thiophene} and. Two core-excited resonances are clearly visible as peaks below 10~eV in both results. }
    \label{fig:thiophene}
\end{figure}

Cross sections for electronic excitation have been calculated with the \UKRmol{}/\UKRmol{+}  for a variety of small and medium-sized molecules: hardly any other software is available for this purpose in the low-energy (a few eV up to ionization threshold) range. Figure~\ref{fig:thiophene} shows an example for thiophene, C$_4$H$_4$S (details of the calculation can be found in \cite{thiophene}): the cross section for excitation into the second excited state of the target (1$^3$A$_1$)  for a projectile scattering angle  of 90$^{\circ}$ (these angle-resolved cross sections are  normally called excitation functions). The agreement  with results from electron energy loss (EELS)  measurements is excellent \cite{thiophene}. We note that the experimental cross sections were not normalized to calculated values. The figure also shows that two core-excited resonances (of symmetry $^2$B$_1$ and $^2$A$_2$) are  visible in both cross sections: the shift to higher energies of the calculated resonances is a well known effect due to an incomplete description of the polarization effects \cite{thiophene}.

This calculations was performed using  \textprog{MPI-SCATCI} on 40 nodes (960 cores) as the Hamiltonians for each symmetry were built from around 250~000 configurations each.  The option to store  only the continuum coefficients was used (\textcode{vecstore = 1}). 

\subsection{Inclusion of BTOs in the continuum: electron scattering from BeH}

\begin{figure}[htbp]
    \centering
    \includegraphics[angle=-90,width=\textwidth]{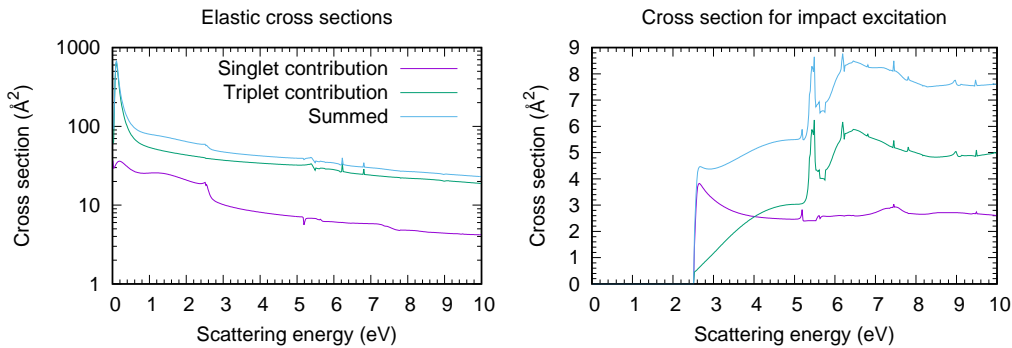}
    \caption{Cross sections for electron scattering from BeH (doublet ground state). Left panel: elastic cross sections. Right panel: cross sections for electron impact electronic excitation of the ground electronic state. No Born correction for dipolar scattering was added.}
    \label{fig:beh}
\end{figure}

To illustrate the capability of the codes to represent very diffuse target electronic states we show in Fig.~\ref{fig:beh} cross sections for electron scattering from BeH. The calculations and the results are described in detail in~\cite{Darby_Lewis_2017}. 

The calculations used R-matrix radius of \(35\,\Bohr\) (the largest used so far in any published molecular R-matrix calculation) and partial waves for the continuum up to $l=6$. The continuum was represented using the mixed GTO/BTO scheme as shown in Fig.~\ref{fig:continuum} with the B-spline basis starting at radius \(3.5\,\Bohr\). The total number of atomic functions in the basis was $1277$. The calculation included $50$ electronic states of the target reaching up to approx.\ $12$~eV above the ground state: this led to construction of up to $669$ channels per symmetry in the outer region. The dimension of the inner region Hamilonians was approx. $45000$ per symmetry. In order to speed-up the outer region calculation for a fine grid of $1000$ energies we used \textprog{MPI\_RSOLVE}, the parallel version of the \textprog{RSOLVE} outer region propagator: employing $60$ cores the propagation took approximately 1 hour per symmetry.

\subsection{Positron scattering from H$_2$: use of pseudostates}
A number of positron-molecule collision calculations have been performed using the \UKRmol{} codes, e.g.\ \cite{jt411,SFB13+}, without inclusion of a pseudocontinuum basis set and pseudostates. However,
polarisation effects are even more important in positron-molecule collisions than they are in
electron-molecule collisions. To address this issue, studies on positron-H$_2$ \cite{jt491}
and positron-acetylene \cite{jt510} used pseudostate expansions to improve the treatment of polarisation
effects which are often referred to as virtual positronium formation.
These studies particularly focused on the calculation of the positron annihilation cross section as characterised by the parameter
$Z_{\rm eff}$, the effective number of electrons available for
positron annihilation. Use of pseudostates was found to greatly increase the computed values for $Z_{\rm eff}$, but it was found that convergence
could only be achieved using pseudostate expansions which go
to high $l$; in practice extrapolation formulae were used to
achieve convergence.

\begin{figure}
    \centering
    \includegraphics[width=0.75\textwidth]{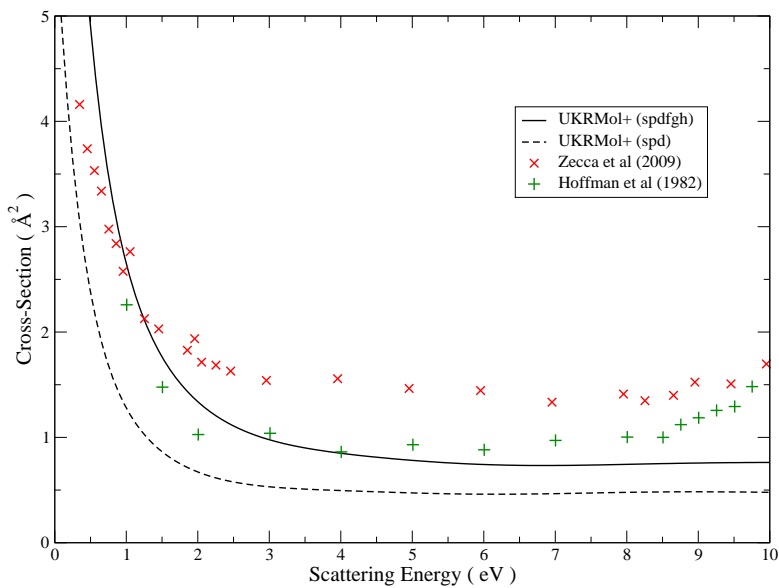}
    \caption{Total cross sections for positron-H$_2$ collisions. Calculations including pseudocontinuum orbitals with angular momenta up to 2 ('spd') and up to 5 ('spdfgh') are compared with  experimental results \cite{PhysRevA.25.1393,PhysRevA.80.032702}. }
    \label{fig:e+h2}
\end{figure}

As an illustration, Figure~\ref{fig:e+h2} shows the total cross section for positron-H$_2$ calculated with the \UKRmol{+} suite. The models used is that of Zhang et al. \cite{jt491}. The results obtained with the new suite are, as expected,  similar to those of the older suite. It is also clear from the figure that increasing the number of angular momenta used for the pseudocontinuum orbitals from 3 to 6 (and also increasing the number of pseudostates from 15 to 31) improves the description of polarization effects, making the cross section larger and closer to the experimental values. The efficiency improvements implemented in the suite (in particular \textprog{MPI-SCATCI}) will enable both electron and positron calculations using pseudostates for larger targets: this will be particularly useful for biological molecules that tend to have big polarizabilities.

\subsection{Photoionization of benzene}

The recently implemented capability of the \UKRmol+ suite to generate photoionization observables, Dyson orbitals and partial wave dipoles has enabled several applications, see e.g.~\cite{bruner2016,richter_NO2_2019}. Most of these applications used high quality R-matrix data for smaller polyatomic molecules and photon energies up to $100$~eV. Nevertheless, the suite is capable of generating data for larger molecules too~\cite{austin2016} as we demonstrate below in Fig.~\ref{fig:benzene_photo} for the case of photoionization of benzene.

These calculations were carried out using the simple HF (i.e. Static-Exchange) approximation, the cc-pVDZ atomic GTO basis set and used $20$~virtual orbitals in the $L^2$ expansion, see Section~\ref{sec:rmat_models}. The purpose of the calculations is not to present accurate observables but to illustrate the flexibility of the suite in representing the continuum and to provide an elementary guidance to the user on the current capability of the suite in representing the continuum. The size of the molecule and the wide range of photon energies typically used in experiments presents a challenge for accurate description of the continuum. Here we compare the results of calculations that used three different choices of the continuum basis:
\begin{itemize}
    \item GTO-only continuum and integrals calculated in double precision (deletion thresholds $10^{-7}$, see Table~\ref{tab:sintegrals-input-continuum}).
    \item GTO-only continuum and integrals calculated in quad precision (deletion thresholds $10^{-14}$, i.e. no continuum orbitals removed from the basis).
    \item Mixed GTO/BTO continuum (double precision) with B-spline basis starting at $r=3.5$~\(\Bohr\) and two choices of the parameters\\ \textcode{max\_l\_legendre\_1el} and \textcode{max\_l\_legendre\_2el}, see Table~\ref{tab:sintegrals-input-process}.
\end{itemize}
In all cases listed above the calculations used R-matrix radius \(13\,\Bohr\) and maximum continuum angular momentum $l_{max}=6$. We have intentionally chosen a demanding example from the point of view of the size of the molecule and the corresponding mixed integral evaluation. Nevertheless, all the integral calculations presented were carried out using a single MPI task on a 20-core node equipped with 256~GB memory.

Figure~\ref{fig:benzene_photo} shows the photoionization cross sections and asymmetry parameters for the first two ionic (final) states of benzene ($^{2}E_{1g}$ and $^{2}E_{2g}$). As expected the double precision GTO-only continuum performs the worst. It starts to develop unphysical oscillations in the results at approximately $25$~eV and breaks down completely at around $50$~eV. In the quad precision GTO-only continuum the basis is reliable up to approximately $50$~eV, i.e. twice that energy. However, the continuum GTO basis set used was optimized for use in scattering from a neutral target rather than in photoionization calculations and therefore the performance of this calculation could be improved employing a GTO continuum basis optimized using \textprog{NUMCBAS} and \textprog{GTOBAS} for photoionization calculations.

\begin{figure}[htbp]
    \centering
    \includegraphics[angle=-90,width=\textwidth]{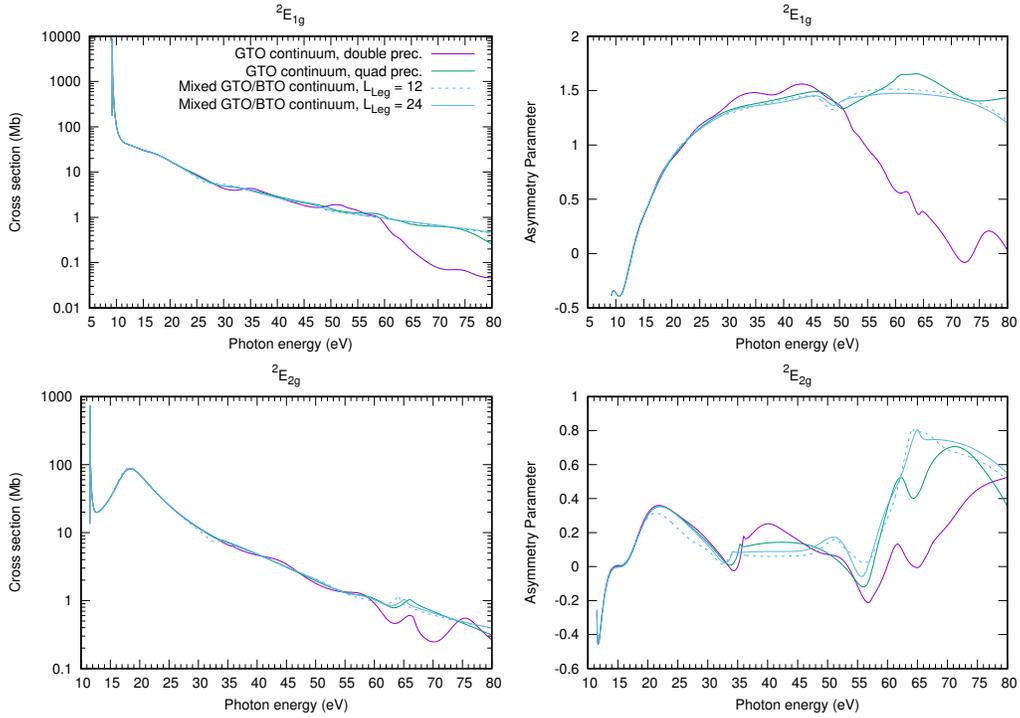}
    \caption{Cross sections (on log-scale) and $\beta$ asymmetry parameters (angular distributions) for photoionization of benzene into the two lowest-lying states ($^{2}E_{1g}$ and $^{2}E_{2g}$) as calculated using GTO-only continua ($l_{max}=6$, double and quad precision) and mixed GTO/BTO continua ($l_{max}=6$ and two values of the maximum angular momentum used in the evaluation of the mixed Coulomb integrals: $L_{Leg}=12$, $L_{Leg}=24$). The ionization potentials of the final states have been shifted to their accurate values.}
    \label{fig:benzene_photo}
\end{figure}

The quality of the mixed GTO/BTO continuum basis set is much higher, as is apparent from the asymmetry parameter for the $^{2}E_{1g}$ state which is smooth throughout the whole energy range, showing that the continuum basis accurately represents the highly oscillatory character of the scattering wavefunction. However, all the calculations at energies above approximately $50$~eV are not fully converged with respect to the continuum angular momentum. The figure also compares the effect of the parameters \textcode{max\_l\_legendre\_1el} and \textcode{max\_l\_legendre\_2el} (here denoted collectively as $L_{Leg}$) on the integral calculations in the mixed GTO/BTO basis. We can see that, as expected, with increasing electron energy the results are more sensitive to the convergence of the integrals involving continuum orbitals representing higher angular momentum electrons which in turn require higher $L_{Leg}$ to converge. In this case the minimum sensible value of these parameters ($L_{Leg}=12$) is insufficient for energies above approx. $25$~eV and a higher value ($L_{Leg}=24$) must be used to achieve more accurate results. Using higher values of $L_{Leg}$ requires more memory for evaluation of the mixed atomic integrals. Therefore convergence of the calculation with respect to this parameter should be carefully checked especially if results for high photon/electron energies are required.

Finally, the radial B-spline basis typically generates larger number of orbitals than the GTO basis with the same angular momentum. This includes continuum orbitals with near-zero amplitudes on the boundary. Therefore we can expect that the mixed continuum basis is also effectively contributing somewhat to the $L^2$ expansion.

\section{Conclusions}

The \UKRmol+ suite is a completely reengineered and extended version of the UK polyatomic molecular R-matrix codes. The suite enables the calculation of low energy electron and positron (excluding positronium formation) scattering from molecules and molecular clusters, photoionization and also the production of input for the RMT suite of codes \cite{RMT}.  Use of the GBTOlib library (included in the releases but developed as a stand-alone entity) means that both GTOs and BTOs can be used to describe the unbound particle. The suite is maintained  and developed on GitLab (developers-only access) ensuring sustainability.

The codes can be downloaded from Zenodo: provision of set of CMake files greatly simplifies the compilation of the codes; provision of a test suite enables users to test their compilation.  A set of perl scripts (\textprog{ukrmol\textunderscore scripts}, not included in the release) is available to greatly simplify the production of input for the suite, execution of the different programs and gathering of the relevant output data. These scripts were originally developed by Dr. K. Houfek and will soon be made available  for general use. 

Alternatively,  Quantemol-N \cite{jt416} provides a graphical expert system for running the \UKRmol{} codes; a new expert system, Quantemol-EC \cite{jtQEC}, performs
the same service for the \UKRmol+ codes described in this paper.

Future releases of UKRmol+ will include a new version of the GBTOlib library currently in development that will make the evaluation of the integrals more efficient, and the re-implementation of both the program  for the evaluation of $Z_{\rm eff}$ in  positron scattering and the partitioned R-matrix approach.

\section*{Acknowledgement}
We would like to dedicate this paper to the memory of Phillip Burke who created, inspired and led the development
of the R-matrix method for atomic, molecular and optical physics.
We thank the many people who have contributed to the development of the UK Molecular R-matrix codes over many years, in particular  Ahmed Al-Refaie, Rui Zhang, Daniel Darby-Lewis (who also performed the positron calculations included in this paper), Dermot Madden,  Martin Plummer, Andrew Sunderland, Jo Carr, Michael Lysaght, and Paul Roberts. Earlier versions of the suite involved the contribution of many other researchers, in particular Cliff Noble, Lesley Morgan and Charles Gillan. JDG, ZM and JB also acknowledge Andrew Brown, Greg Armstrong and Hugo van der Hart for the stimulating R-MADAM collaboration that has been the impulse for some of the more recent developments as well as Karel Houfek for developing the Perl scripts that are now regularly used in many UKRmol+ calculations. ZM further acknowledges Roman {\v C}urík for suggesting the idea of the free-potential scattering test.
The development of UKRmol+ has been supported by EPSRC under grants EP/G055556/1, EP/G055599/1, EP/P022146/1 and EP/R029342/1.
and eCSE projects eCSE01-13, eCSE08-7 and eCSE13-14. ZM acknowledges partial support by OP RDE project No.CZ.02.2.69/0.0/0.0/16\_027/0008495, International Mobility of Researchers at Charles University.

\appendix

\section{Expressions for the boundary amplitudes}\label{app:bamps}

Equation~(\ref{eq:bamps}) defines the boundary amplitudes in terms of projections of the R-matrix basis functions on the channel wavefunctions:
\begin{eqnarray}
w_{pk}(a) = \frac{1}{\sqrt{2}}\bigg\langle\chnlshort{p}{}\frac{1}{r}\bigg\vert\psik{}{}\bigg\rangle\bigg\vert_{r=a} = \frac{1}{\sqrt{2}}\bigg\langle\Phi_{p_i}^{N}\frac{1}{r}X_{l_{p},m_{p}}(\mathbf{r}_{N+1})\bigg\vert\psik{}{}\bigg\rangle\bigg\vert_{r=a}.
\end{eqnarray}
In the UKRmol+ implementation the R-matrix basis functions are given by Eq.~(\ref{e:rmat}):
\begin{equation}
    \psi_{k}^{N+1} = \hat A \sum_{i, j} c_{ijk} \Phi^{N}_{i}(\mathbf{x}_{1},\dots,\mathbf{x}_{N}) \eta_{ij}(\mathbf{x}_{N+1}) + \sum_{m} b_{m k} \chi^{N+1}_{m}(\mathbf{x}_{1},\dots,\mathbf{x}_{N+1}).
\end{equation}
Using this equation and considering that only the first sum on the right-hand side contributes to the boundary amplitude we obtain:
\begin{eqnarray}
w_{pk}(a) &=& \sum_{i, j} c_{ijk} \frac{1}{\sqrt{2}}\bigg\langle\chnlshort{p}{}\frac{1}{r}\bigg\vert\phi_{i}^{N}\eta_{ij} \bigg\rangle\bigg\vert_{r=a}=\\
&=&\sum_{i, j} c_{ijk} \frac{1}{\sqrt{2}}\delta_{i,p_{i}}\bigg\langle\frac{1}{r}X_{l_{p},m_{p}}\bigg\vert\eta_{ij}\bigg\rangle\bigg\vert_{r=a},
\end{eqnarray}
where we used orthonormality of the target electronic states. The matrix elements on the second line are called the raw boundary amplitudes. They are evaluated with help of expansion of the continuum orbitals $\eta_{ij}(\mathbf{x}_{N+1})$ in the single-particle basis of target, pseudocontinuum and continuum functions. However at $r=a$ only the continuum functions which have the form $f(r)X_{l,m}(\Omega)$ can contribute to the raw boundary amplitude:
\begin{eqnarray}
\bigg\langle\frac{1}{r}X_{l_{p},m_{p}}\bigg\vert\eta_{ij}\bigg\rangle\bigg\vert_{r=a} &=& \delta_{l_{p},l_{ij}}\delta_{m_{p},m_{ij}}a\sum_{s} q_{ij,s}\gamma(a).
\end{eqnarray}
Here $q_{ij,s}$ are the coefficients in expansion of the continuum orbital $\eta_{ij}(\mathbf{x}_{N+1})$ in the basis of the single particle functions and the sum over $s$ is assumed to be only over the continuum part of the basis. The factor of $a$ is the result of the projection $\frac{1}{r}$ on the reduced radial part of the continuum orbital. Finally, $\gamma(a)$ are the amplitudes of the radial parts of the continuum functions which can be GTOs and/or BTOs (cf.~Section~\ref{sec:sintegrals}) centered on the center of mass:
\begin{eqnarray}
\gamma_{GTO,s}(a) &=& N_{\alpha,l_{s}}^{GTO}\sqrt{\frac{4\pi}{2l_{s}+1}}a^{l_{s}}\exp[-\alpha_{s} a^2],\\
\gamma_{BTO,s}(a) &=& N_{i_{s}}\frac{B_{i_{s}}(a)}{a}.
\end{eqnarray}
In UKRmol+ the raw boundary amplitudes are generated by GBTOlib at the interfacing stage, see Section~\ref{sec:Interfaces}.

\section{Irreducible representations and multiplication table}
\label{app:irr-tables}

The order (and index number) of the irreducible representations in the \UKRmol{}-in input is listed in Table~\ref{tab:M-values}. This mirrors the order of irreducible representations used in Molpro~\cite{MOLPRO} and enables use of a single multiplication table for all groups (see Table~\ref{tab:Abel-mul-table}). However, this convention is not universal and use of other compatible software (Psi4~\cite{Psi4}, for instance) together with \UKRmol+ requires additional attention to the ordering when writing input for \textprog{SCATCI\_INTEGRALS} and other programs.

\begin{table}[htbp]
    \caption{Association of ``$M$-values'' with irreducible representations of the supported finite groups in \UKRmol+.}
    \centering
    \begin{tabular}{lllllllll}
        \toprule
              & 0   & 1 & 2 & 3 & 4 & 5 & 6 & 7 \\
        \midrule
        C$_1$ & $A$ &   &   &   &   &   &   &   \\
        C$_2$ & $A$ & $B$ & & &  &   &   &   \\
        C$_i$ & $A_g$ & $A_u$ &   &   &   &   &   &   \\
        C$_s$ & $A'$ & $A''$ &   &   &   &   &   &   \\
        C$_{2h}$ &  $A_g$  & $A_u$  & $B_u$  & $B_g$  &   &   &   &   \\
        C$_{2v}$ & $A_1$ & $B_1$ & $B_2$ & $A_2$ &   &   &   &   \\
        D$_{2h}$ & $A_g$ & $B_{3u}$ & $B_{2u}$ & $B_{1g}$ & $B_{1u}$ & $B_{2g}$ & $B_{3g}$ & $A_u$ \\
        \bottomrule
    \end{tabular}
     \label{tab:M-values}
\end{table}

Note that in some of the output of the calculations, for example, the target property file produced by \textprog{DENPROP}, the irreducible representations are numbered starting from 1.

\begin{table}[htbp]
    \caption{Multiplication table for combinations of irreducible representations of the Abelian groups supported with \UKRmol+. Numbers are the $M$-values introduced in Table~\ref{tab:M-values}.}
   \centering
    \begin{tabular}{c|cccccccc}
        \toprule
            & 0 & 1 & 2 & 3 & 4 & 5 & 6 & 7 \\
        \midrule
         0  & 0 & 1 & 2 & 3 & 4 & 5 & 6 & 7 \\
         1  & 1 & 0 & 3 & 2 & 5 & 4 & 7 & 6 \\    
         2  & 2 & 3 & 0 & 1 & 6 & 7 & 4 & 5 \\
         3  & 3 & 2 & 1 & 0 & 7 & 6 & 5 & 4 \\
         4  & 4 & 5 & 6 & 7 & 0 & 1 & 2 & 3 \\
         5  & 5 & 4 & 7 & 6 & 1 & 0 & 3 & 2 \\
         6  & 6 & 7 & 4 & 5 & 2 & 3 & 0 & 1 \\
         7  & 7 & 6 & 5 & 4 & 3 & 2 & 1 & 0 \\
        \bottomrule
    \end{tabular}
     \label{tab:Abel-mul-table}
\end{table}

\section{Dipole and quadrupole operators}
\label{app:properties}

The \((l,m)\) labeling of the matrix elements of the multipole operator produced by
\textprog{(C)DENPROP} are subject to a convention.
In \UKRmol+, the electronic multipoles are written without the (negative) charge factor. When the projectile is a positron, the values receive an additional minus sign.
Tab.~\ref{tab:denprop-properties} lists explicit expressions for the multipole operators used
in \UKRmol+.

\begin{table}[htbp]
    \centering
    \begin{tabular}{lrll}
        \toprule
         \(l\) & \(m\) & operator & in terms of Molpro properties \\
        \midrule
           1   & \(-1\) & \(y\)   & DMY \\
           1   & 0      & \(z\)   & DMZ \\
           1   & \(+1\) & \(x\)   & DMX \\
        \midrule
           2   & \(-2\) & \(\sqrt{3}xy\)                     & 2 QMXY / \(\sqrt{3}\) \\
           2   & \(-1\) & \(\sqrt{3}yz\)                     & 2 QMYZ / \(\sqrt{3}\) \\
           2   & 0      & \(\frac{1}{2} (3z^2 - r^2)\)       & QMZZ \\
           2   & \(+1\) & \(\sqrt{3}xz\)                     & 2 QMXZ / \(\sqrt{3}\) \\
           2   & \(+2\) & \(\frac{\sqrt{3}}{2} (x^2 - y^2)\) & (QMXX \(-\) QMYY) / \(\sqrt{3}\) \\
        \bottomrule
    \end{tabular}
    \caption{Multipole operators in \UKRmol+ and relation to properties calculated by Molpro.}
    \label{tab:denprop-properties}
\end{table}



\bibliographystyle{elsarticle-num}
\bibliography{sample,rmat,PhotoRefs}







\end{document}